\documentclass[a4paper,11pt]{article}
\pdfoutput=1 

\usepackage{booktabs}
\usepackage{bm}         
\usepackage{array}
\usepackage{float}

\usepackage[T1]{fontenc}
\usepackage{jheppub} 
\newcommand{\gev}{\,\textrm{GeV }}
\newcommand{\BL}{L+R}
\newcommand{\Extraone}{G}
\newcommand{\Extratwo}{B}

\renewcommand{\d}{\mathrm d}
\renewcommand{\i}{\mathrm i}

\newcommand{\U}[1]{\mathrm{U}(1)_{\mathrm{#1}}}			
\newcommand{\SU}[2]{\mathrm{SU}(#1)_{\mathrm{#2}}}		
\newcommand{\T}[2]{T_{\mathrm{#1}}^{#2}}				
\newcommand{\g}[1]{g_{\mathrm{#1}}}						

\newcommand{\Sl}[3]{(\tilde{L}^{ #1} )^{ #2 }_{ #3 }}
\newcommand{\SlS}[3]{(\tilde{L}^*_{ #1} )_{ #2 }^{ #3 }}

\newcommand{\SQL}[3]{({\tilde{Q}_\mathrm{L}}{}^{ #1} )^{ #2 }_{ #3 }}
\newcommand{\SQR}[3]{({\tilde{Q}_\mathrm{R}}{}^{ #1} )^{ #2 }_{ #3 }}
\newcommand{\SQLS}[3]{({\tilde{Q}^*_\mathrm{L}} {}_{ #1} )_{ #2 }^{ #3 }}
\newcommand{\SQRS}[3]{({\tilde{Q}^*_\mathrm{R}} {}_{ #1} )_{ #2 }^{ #3 }}

\newcommand{\LL}[3]{(L^{ #1} )^{ #2 }_{ #3 }}
\newcommand{\QL}[3]{({Q_\mathrm{L}}{}^{ #1} )^{ #2 }_{ #3 }}
\newcommand{\QR}[3]{({Q_\mathrm{R}}{}^{ #1} )^{ #2 }_{ #3 }}
\newcommand{\QRd}[3]{({Q_\mathrm{R}^\dagger}{}_{ #1} )_{ #2 }^{ #3 }}

\newcommand{\QLE}[3]{({\mathcal{Q}_\mathrm{L}}^{ #1} )^{ #2 }_{ #3 }}
\newcommand{\QRE}[3]{({\mathcal{Q}_\mathrm{R}}^{ #1} )^{ #2 }_{ #3 }}

\newcommand{\QREd}[3]{({\mathcal{Q}_\mathrm{R}^\dagger}^{ #1} )^{ #2 }_{ #3 }}

\newcommand{\Llf}[2]{(l_\mathrm{L}{}^{ #1} )^{ #2 }}
\newcommand{\Lrf}[2]{(l_\mathrm{R}{}^{ #1} )_{ #2 }}
\newcommand{\Llfs}[1]{l_\mathrm{L}^{s} {}^{ #1 }}
\newcommand{\Lrfs}[1]{l_\mathrm{R}^{s} {}_{ #1 }}

\newcommand{\HH}[3]{(\tilde{H}^{ #1} )^{ #2 }_{ #3 }}
\newcommand{\Ll}[2]{({\tilde{l}_\mathrm{L}}{}^{ #1} )^{ #2 }}
\newcommand{\Lr}[2]{({\tilde{l}_\mathrm{R}}{}^{ #1} )_{ #2 }}
\newcommand{\Lrt}{\tilde{l}_\mathrm{R}}
\newcommand{\hh}[2]{\tilde{h}^{ #1}_{ #2 }{}}

\newcommand{\Lrs}[2]{({\tilde{l}^{*}_\mathrm{R}{}_{ #1} )}^{ #2 }}
\newcommand{\hhs}[2]{\tilde{h}^*{}_{ #1}^{ #2 }}

\renewcommand\Re{\mathrm{Re}}
\renewcommand\Im{\mathrm{Im}}

\newcommand{\HHf}[3]{(H^{ #1} )^{ #2 }_{ #3 }}

\preprint{
\begin{minipage}{4cm}
\begin{flushright}
LU TP 16-32\\
July 2016
\end{flushright}
\end{minipage}}

\title{On a radiative origin of the Standard Model from Trinification}

\makeatletter
\def\@fpheader{\relax}
\makeatother

\author[a]{Jos\'e~Eliel Camargo-Molina}
\author[a,b]{Ant\'onio~P.~Morais}
\author[a]{Roman~Pasechnik}
\author[a]{Jonas~Wess\'en}

\affiliation[a]{
Department of Astronomy and Theoretical Physics, Lund
University, S\"olvegatan 14A, SE-223 62 Lund, Sweden}
\affiliation[b]{
Departamento de F\'\i sica, Universidade de Aveiro and CIDMA,\\ 
Campus de Santiago, 3810-183 Aveiro, Portugal}

\emailAdd{Eliel@thep.lu.se}
\emailAdd{Aapmorais@ua.pt}
\emailAdd{Roman.Pasechnik@thep.lu.se}
\emailAdd{Jonas.Wessen@thep.lu.se}

\abstract{In this work, we present a trinification-based grand unified theory incorporating a global $\SU{3}{}$ family symmetry 
that after a spontaneous breaking leads to a left-right symmetric model. Already at the classical level, this model can accommodate 
the matter content and the quark Cabbibo mixing in the Standard Model (SM) with only one Yukawa coupling at the unification scale.
Considering the minimal low-energy scenario with the least amount of light states, we show that the resulting effective theory enables 
dynamical breaking of its gauge group down to that of the SM by means of radiative corrections accounted for by the renormalisation 
group evolution at one loop. This result paves the way for a consistent explanation of the SM breaking scale and fermion mass hierarchies. }

\begin{document} 
\maketitle
\flushbottom

\section{Introduction} 

The discovery of the Higgs boson at the Large Hadron Collider (LHC) \cite{Aad:2012tfa,Chatrchyan:2012xdj} and continuous studies of its properties have revealed an intriguing consistency of experimental results with the Standard Model (SM) predictions. This highlights yet another major step in precision verification of the SM structure. Besides being a big phenomenological success as a fundamental theory of particle physics, the SM as an effective theory still allows for possibilities for new interactions and particles (such as Dark Matter, right-handed neutrinos, heavy Higgs boson partners, new gauge interactions, among others) at energy scales much larger then the electroweak (EW) scale. This kind of new physics might bring answers to current open questions and could be detected already at the LHC. In addition, the origin of the large set of measured (not predicted!) fermion mass and mixing parameters as well as the Higgs boson mass and self-couplings still remains as one of the most interesting open questions to date. Furthermore, there is still no explanation for the characteristic hierarchies in the measured fermion mass spectrum.

In order to get a better understanding of these long-standing issues in the framework of quantum field theory, one naturally considers the SM as a low-energy approximation of a bigger and more symmetric (unified) theory whose dynamics at high energies is implicitly encoded in the observed structure of the SM.

An important example of such a grand unified theory (GUT) based upon the trinified gauge group $\SU{3}{L} \times \SU{3}{R} \times 
\SU{3}{C}\equiv [\SU{3}{}]^3$ (also known as trinification) was proposed by De R\'ujula, Georgi and Glashow (RGG-model) back in 1984 \cite{original}. 
Since then, trinified extensions of the SM have been traditionally considered as good bets for a GUT, both with and without 
supersymmetry (SUSY) \cite{Babu:1985gi,Lazarides:1993uw,Lazarides:1994px,Kim:2003cha,Willenbrock:2003ca,Carone:2004rp}, due to many 
attractive features (for a good introduction into trinification GUTs, see \cite{Hetzel:2015cca} and references therein).

The gauge trinification $[\SU{3}{}]^3 \ltimes \mathbb{Z}_3$ is a maximal subgroup of $\mathrm{E}_6$, where $\mathbb{Z}_3$ is the group of cyclic (L,R,C)-permutations
(for a comprehensive discussion of $\mathrm{E}_6$-inspired GUT scenarios, see e.g.~Refs.~\cite{Gursey:1975ki,Achiman:1978vg,Shafi:1978gg,Barbieri:1981yy,Stech:2003sb,Stech:2010tv,King:2005my,King:2005jy,King:2007uj,Braam:2010sy,Athron:2008np,King:2008qb,Athron:2009ue,Athron:2009bs,Athron:2011wu,Hall:2010ix,Athron:2012sq,Nevzorov:2012hs,Nevzorov:2013tta,Nevzorov:2013ixa,Nevzorov:2015sha,Athron:2015vxg,King:2016wep,Athron:2012pw,Kawamura:2013rj,Rizzo:2012rf,Reuter:2010nx}). 
Typically, this model is considered to be a low-energy limit of the heteroic $\mathrm{E}_8 \times \mathrm{E}_8'$ string theory \cite{Gross:1984dd} as well as the ${\cal N}=4$ supergravity 
\cite{Cremmer:1979uq}. It naturally incorporates the left-right (LR) symmetric gauge interactions \cite{Ma:1986we} as well as the gauge couplings' unification 
at a GUT scale. All matter fields (including the Higgs fields) can be elegantly arranged into bi-fundamental representations where each family belongs to a $\mathbf{27}$-plet,
$ \mathbf{27} = (\mathbf{ 3},\mathbf{\bar 3},\mathbf{1})\oplus (\mathbf{1},\mathbf{3}, \mathbf{\bar 3}) \oplus (\mathbf{\bar 3},\mathbf{1},\mathbf{ 3}) \,, $
being the fundamental representation of the $\mathrm{E}_6$ group \cite{Gursey:1975ki,Ma:1995xk}. Remarkably, no adjoint Higgs fields are needed to break 
$\SU{3}{L} \times \SU{3}{R}$ down to the electroweak (EW) symmetry group of the SM, $\SU{2}{L} \times \U{Y}$. The spontaneous breaking of trinification 
with at least two Higgs $\mathbf{27}$-plets yields the standard GUT-scale prediction for the weak mixing angle, $\sin^2\theta_W=3/8$, which leads to quantization of the 
$\U{Y}$ hypercharge in the SM (e.g.~resulting in electron charge being exactly opposite to the proton charge) and provides a consistent explanation of parity 
violation in the SM. As was shown in Refs.~\cite{Kim:2004pe,Cauet:2010ng} it is possible to achieve naturally light neutrinos via a seesaw mechanism as well. 
Moreover, in the RGG formulation, the model accommodates any quark mixing angles \cite{Sayre:2006ma} and a natural suppression of proton decay 
\cite{Babu:1985gi,Willenbrock:2003ca}.

However, many existing realizations of the RGG model suffer from severe issues with phenomenology, a considerable amount of particles in 
its spectrum and many (e.g.~Yukawa) parameters. One particular issue, common to most of the well-known GUTs, is an unmotivated strong hierarchy 
between the trinification and the EW symmetry breaking scales as well as hierarchies in the SM fermion mass spectrum. In addition, the existing minimal SUSY-based trinification \cite{Sayre:2006ma} 
has problems in avoiding TeV-scale lepton masses without imposing higher-dimensional operators, large Higgs representations or an artificial and 
simultaneous fine tuning of many parameters. At the same time, realistic calculations including quantum corrections are cumbersome due to 
a large number of scalar particles and gauge bosons in any $[\SU{3}{}]^3$-symmetric theory. These issues left the trinification-based models among the least-developed GUT scenarios so far.

To be consistent with SM phenomenology, a number of additional $\U{}$ groups emerging in $\mathrm{E}_6$ (or $\mathrm{E}_8$) breaking \cite{Huang:2001bm} should be consistently 
broken at intermediate steps by the conventional Higgs mechanism. Having a few $\mathbf{27}$ Higgs multiplets coupled to fermions which acquire several 
low-scale vacuum expectation values (VEVs), may resolve this issue. However, those interactions induce potentially large flavour-changing neutral current
processes which are severely restricted by experiment, and a large degree of fine tuning is required. 

Due to a huge hierarchy in the mass spectra, at low energy scales heavy d.o.f.'s have to be integrated out at each intermediate symmetry 
breaking scale giving rise to a new effective model having a fewer amount of light fields in the spectrum. Depending on the symmetry breaking scheme and the hierarchy 
in the initial $[\SU{3}{}]^3$ GUT model parameters, one may end up with a few possible low-energy effective models having different light particle content. 

One possible development would be to consider a mechanism for Yukawa couplings unification, severely reducing the number of free parameters at the GUT scale \cite{Stech:2003sb,Stech:2010tv}. Similarly to the gauge couplings, the unified Yukawa coupling would then give rise to several different couplings by means of radiative corrections via the renormalisation group (RG) evolution and loop-induced operators, which may reproduce the SM fermion mass and mixing hierarchies at low energies. In this way, family symmetries acting in the space of fermion generations \cite{Georgi:1979md} are known to provide a convenient tool for generating the necessary patterns in fermion spectra \cite{Stech:2010tv,Stech:2014tla}. In particular, such symmetries help to avoid GUT-scale lepton masses in trinification-type models. 

An example of an effective LR-symmetric scenario with very interesting phenomenology has been discussed in Ref.~\cite{Hetzel:2015bla}. There the authors introduce the gauge group $\SU{3}{C} \times \SU{2}{L} \times \SU{2}{R} \times \U{B-L}$ as originating at lower energies from the trinification model with two $\mathbb{Z}_2$-even and odd Higgs $\mathbf{27}$-plets. However, the properties of the Yukawa sector in this model rely on additional higher-dimensional representations of $\mathrm{E}_6$ such as (anti)symmetric $\mathbf{351}$ reps.

In this work, we consider an alternative non-SUSY trinification model $[\SU{3}{}]^3 \ltimes \mathbb{Z}_3$ augmented by a $\SU{3}{F}$ global family symmetry which acts both on fermion and scalar multiplets. The latter are thus incorporated in a symmetric way essentially inspired by SUSY. Our scenario is however manifestly non-supersymmetric and it does not invoke any higher-dimensional reps or extra singlets besides lowest $\mathbf{27}$-plets of $\mathrm{E}_6$. The scenario we present is naturally inspired by a reduction $\mathrm{E}_8\to \mathrm{E}_6 \times \SU{3}{}$ where the remnant $\SU{3}{}$ is identified with a global family symmetry $\SU{3}{F}$ at the trinification breaking scale. The symmetry group is spontaneously broken down to a LR-symmetric model with an extra $\SU{2}{F} \times \U{X} \times \U{Z} \times \U{B}$ global symmetry that is a remnant of the $\SU{3}{F}$ and an accidental $\U{A}\times\U{B}$ symmetry in the high-scale trinification theory. 

As we show in the present work, this model inherits all the important features of trinified GUTs and resolves some of their known difficulties. In the considered implementation 
of the family symmetry together with the trinification model, all Yukawa couplings are manifestly unified into a single coupling at the GUT scale, and the number of free scalar self-couplings in the scalar potential is remarkably low, making a complete RG analysis of this model feasible. Many of the relevant interactions in the low-energy effective theory emerge radiatively at one (or higher) loop level, bringing a potential explanation to a variety of hierarchies in the SM parameters. However, a detailed calculation of such quantities is beyond the scope of this paper and left for future work. Simultaneously, the family symmetry 
forbids proton decay due to an appearance of an accidental $\U{B}$ symmetry and protects the light SM fermion sector from large radiative corrections offering potentially 
interesting phenomenological consequences. Another feature of this model, that we show in this work, is that the $\SU{2}{R} \times \U{\BL}$ subgroup gets broken radiatively 
to $\U{Y}$ at a much lower scale in a natural way for a large region of the parameter space of the GUT-scale trinification model.

In Sect.~\ref{sec:TriniMod}, we introduce the high-scale trinification model augmented by the family symmetry. In Sect.~\ref{sec:TriBreaking}, we discuss in detail 
the first symmetry breaking stage down to a low-energy LR-symmetric effective theory. In Sect.~\ref{sec:LR}, we describe the effective model and the matching of effective 
couplings in order to study, in Sect.~\ref{sec:RadBreak}, under which circumstances the effective theory shows radiative breaking of the $\SU{2}{R} \times \U{\BL}$ 
symmetry. In Sect.~\ref{sec:Results}, we perform a parameter space scan to find the regions where the radiative symmetry breaking happens in the simplest 
feasible scenario. In Sect.~\ref{sec:Discussion} we discuss, in the light of our results, under which conditions it could be possible to reproduce the SM mass spectra.
Concluding remarks are given in Sect.~\ref{sec:Conclusions}.

\subsection{A quick note on notations}

In the text that follows, we employ the following notations:

\begin{itemize}
\item Fundamental representations carry superscript indices while anti-fundamental representations carry subscript indices.
\item Fundamental and anti-fundamental indices of $\SU{3}{}$ groups are denoted with lower case letters, while fundamental 
and anti-fundamental indices under $\SU{2}{}$ groups are denoted with upper case letters.
\item $\SU{3}{K}$ and $\SU{2}{K}$ (anti-)fundamental indices are denoted by $k,k',k_1,k_2 \dots$ and $K,K',K_1,K_2 \dots$ for $K=L,R,C$, respectively.
\item Indices belonging to (anti-)fundamental representations of $\SU{3}{F}$ and $\SU{2}{F}$ are denoted by $i,j,k \dots$ and $I,J,K \dots$ respectively. 
\item If a field transforms both under gauge and global symmetry groups, the index corresponding to the global one is placed within the parenthesis around the field, while the indices corresponding to the gauge symmetries are placed outside.
\item Global symmetry groups will be indicated by $\{\dots\}$. 
\end{itemize} 

For example, $\LL{i}{l}{r}$ is a $\bm{3} \otimes \bm{3} \otimes \bar{\bm{3}}$ representation of $\SU{3}{L} \times \SU{3}{R} \times \{\SU{3}{F}\}$, 
and $\Lr{I}{R}$ is a $\bm{2} \otimes \bar{\bm{2}}$ representation of $ \SU{2}{R} \times \{\SU{2}{F}\}$, where $\SU{3}{F}$ and $\SU{2}{F}$ 
are global family symmetry groups.
           
\section{The GUT-scale $[\SU{3}{}]^3 \ltimes \mathbb{Z}_3 \times \lbrace \SU{3}{F} \rbrace$ model} \label{sec:TriniMod}

The fields in the high-scale trinification model form representations of the symmetry group
\begin{equation}\label{eq:trinigaugegroup}
\left[ \SU{3}{L} \times \SU{3}{R} \times \SU{3}{C} \right] \ltimes \mathbb{Z}_3 \times \{\SU{3}{F}\} , 
\end{equation}
as shown in Tab.~\ref{tab:HSmodelT}, and consist of three Weyl fermion multiplets ($L$, $Q_\mathrm{L}$, $Q_\mathrm{R}$), 
three scalar multiplets ($\tilde{L}$,  $\tilde{Q}_\mathrm{L}$, $\tilde{Q}_\mathrm{R}$) and gauge bosons 
($G_\mathrm{L}$, $G_\mathrm{R}$, $G_\mathrm{C}$). Here, $\SU{3}{F}$ is a global family symmetry acting on the space of fermion and scalar field generations, while 
$ \SU{3}{L} \times \SU{3}{R} \times \SU{3}{C}$ is the standard trinification gauge group. Although our model 
is not supersymmetric, we employ a notation inspired by SUSY, since we have the same group 
representations in the scalar and fermion sectors. The fermions and scalars both form bi-triplet representations 
under the  gauge group, but \textit{tri}-triplets under the full symmetry group (including the $\SU{3}{F}$)\footnote{Gauging the family $\SU{3}{F}$
in an $E_8$ inspired scenario would effectively mean the doubling the number of chiral and scalar multiplets as is required by the anomaly cancellation condition. 
In this paper, however, we avoid such a huge complication by treating the family symmetry as a global one, as a first step.}. 
\begin{table}[tbp]
  \begin{center}
    \begin{tabular}{ccccc}
     \toprule                     
                        				 &$\SU{3}{L}$ & $\SU{3}{R}$ & $\SU{3}{C}$ & $ \{\SU{3}{F}\}$ \\      
      \midrule
      fermions &  & & & \\
      \midrule
       $L$      			    & $\bm{3}$				& $\bar{\bm{3}}$	& $\bm{1}$				& $\bm{3}$				 \\
       $Q_\mathrm{L}$ 			& $\bar{\bm{3}}$		& $\bm{1}$			& $\bm{3}$				& $\bm{3}$ 				 \\
       $Q_\mathrm{R}$ 	 		& $\bm{1}$				& $\bm{3}$ 			& $\bar{\bm{3}}$		& $\bm{3}$		 \\
      \midrule
      scalars & & & &    \\
      \midrule       	
       $\tilde{L}$         	    & $\bm{3}$				& $\bar{\bm{3}}$	& $\bm{1}$				& $\bm{3}$				 \\
       $\tilde{Q}_\mathrm{L}$  	& $\bar{\bm{3}}$		& $\bm{1}$			& $\bm{3}$				& $\bm{3}$ 				 \\
       $\tilde{Q}_\mathrm{R}$	& $\bm{1}$				& $\bm{3}$ 			& $\bar{\bm{3}}$		& $\bm{3}$		 \\
      \midrule
      gauge bosons & & & & \\
      \midrule
      $G_{\mathrm{L}}$	    		& $\bm{8}$ 				& $\bm{1}$ 			& $\bm{1}$ 				& $\bm{1}$ \\
      $G_{\mathrm{R}}$            & $\bm{1}$ 				& $\bm{8}$ 			& $\bm{1}$ 				& $\bm{1}$ \\
      $G_{\mathrm{C}}$            & $\bm{1}$ 				& $\bm{1}$ 			& $\bm{8}$ 				& $\bm{1}$ \\
     \bottomrule
    \end{tabular}
    \caption{Field content of the GUT-scale trinification model. The fermionic fields are left-handed Weyl fermions.}
    \label{tab:HSmodelT}
  \end{center}
\end{table}

The $\mathbb{Z}_3$ symmetry refers to the cyclic permutation of the fields
\begin{equation}
\begin{aligned}
G_\mathrm{L} \overset{\mathbb{Z}_3}{\rightarrow} G_\mathrm{C},				\\ 
G_\mathrm{C} \overset{\mathbb{Z}_3}{\rightarrow} G_\mathrm{R},				\\ 
G_\mathrm{R} \overset{\mathbb{Z}_3}{\rightarrow} G_\mathrm{L},			
\end{aligned}
\qquad
\begin{aligned}
L \overset{\mathbb{Z}_3}{\rightarrow} Q_\mathrm{L},				\\
Q_\mathrm{L} \overset{\mathbb{Z}_3}{\rightarrow} Q_\mathrm{R}	,			\\ 
Q_\mathrm{R} \overset{\mathbb{Z}_3}{\rightarrow} L	,		
\end{aligned}
\qquad
\begin{aligned}
\tilde{L} \overset{\mathbb{Z}_3}{\rightarrow} \tilde{Q}_\mathrm{L},				\\ 
\tilde{Q}_\mathrm{L} \overset{\mathbb{Z}_3}{\rightarrow} \tilde{Q}_\mathrm{R}	,			\\
\tilde{Q}_\mathrm{R} \overset{\mathbb{Z}_3}{\rightarrow} \tilde{L}	.			
\end{aligned}
\end{equation}
which in turn enforces the gauge coupling unification. This symmetry 
combined with the global $\SU{3}{F}$ also dramatically reduces the number of possible terms 
in the scalar potential as well as in the fermion sector of the theory. 

The most general renormalizable scalar potential for the trinification model reads
\begin{equation}\label{eq:TLpot}
V = V_{1} +V_{2}+V_{3}
\end{equation}
where
\begin{equation}
\begin{aligned}
V_{1} &= -\mu^2 \Sl{i}{l}{r} \, \SlS{i}{l}{r}	 + \lambda_1 \left[ \Sl{i}{l}{r} \, \SlS{i}{l}{r} \right]^2	\\
   &+ \lambda_2 \Sl{i}{l}{r} \, \Sl{j}{l'}{r'} \, \SlS{j}{l}{r} \, \SlS{i}{l'}{r'} \\
   &+ \lambda_3 \Sl{i}{l}{r} \, \Sl{j}{l'}{r'} \, \SlS{i}{l}{r'} \, \SlS{j}{l'}{r} \\
   &+ \lambda_4 \, \Sl{i}{l}{r} \, \Sl{j}{l'}{r'} \, \SlS{j}{l}{r'} \, \SlS{i}{l'}{r} \\
   &+ \,\,\,\, (\text{$\mathbb{Z}_3$ permutations}),	\\
V_{2}  &= \alpha_1 \, \Sl{i}{l}{r} \, \SlS{i}{l}{r} \, \SQL{j}{c}{l'} \, \SQLS{j}{c}{l'}		\\
	 &+ \alpha_2 \, \Sl{i}{l}{r} \, \SlS{j}{l}{r} \, \SQL{j}{c}{l'} \, \SQLS{i}{c}{l'}				\\
	 &+ \alpha_3 \, \Sl{i}{l}{r} \, \SlS{i}{l'}{r} \, \SQL{j}{c}{l} \, \SQLS{j}{c}{l'}				\\
	 &+ \alpha_4 \, \Sl{i}{l}{r} \, \SlS{j}{l'}{r} \, \SQL{j}{c}{l} \, \SQLS{i}{c}{l'}				\\
	 &+ \,\,\,\, (\text{$\mathbb{Z}_3$ permutations}),	
\end{aligned}
\end{equation}
and
\begin{equation}\label{eq:Gamma}
V_{3} = \gamma \, \epsilon_{i j k} \, \Sl{i}{l}{r} \, \SQL{j}{c}{l} \, \SQR{k}{r}{c} + \mathrm{c.c.}
\end{equation}
The scalar potential thus contains two dimensionfull parameters, mass parameter $\mu$ and trilinear coupling $\gamma$, 
and eight quartic couplings $\lambda_{1,\dots,4}$ and $\alpha_{1,\dots,4}$ which can be taken to be real without loss of generality.

Due to the interplay between $\SU{3}{F}$ and $\mathbb{Z}_3$, combined with the trinification gauge group, the fermion sector in the model only contains one single Yukawa coupling,
\begin{equation} \label{eq:TriYukawa}
\begin{aligned}
\mathcal{L}_{\mathrm{Fermion}} &=& - y \, \epsilon_{ijk} \, \Sl{i}{l}{r} \, \QL{j}{c}{l} \, \QR{k}{r}{c} + \mathrm{c.c.}		\\
& & + \,\,\,\, (\text{$\mathbb{Z}_3$ permutations})	.
\end{aligned}
\end{equation}
The trinification Yukawa coupling $y$ can be taken to be real since any complex phase may 
be absorbed into the definition of the fermion fields. 

Once all the renormalizable terms invariant under the trinification gauge group and the global $\SU{3}{F}$ symmetry are written, one can notice that 
the terms are also invariant under an accidental $\{\U{A} \times \U{B}\}$ symmetry. A convenient charge assignment under the accidental $\U{}$ 
groups is shown in Tab.~\ref{tab:AccSym}, where one immediately recognizes $\U{B}$ as giving rise to a conserved baryon number. Furthermore, 
$\U{B}$ will stay unbroken at lower scales (including the SM), since $\tilde{L}$ is uncharged under $\U{B}$ and no other fields will develop VEVs throughout 
the evolution to the EW scale. Though the symmetry would still allow for the proton to decay into coloured scalars $\tilde{Q}_\mathrm{L,R}$, 
we will see that all the coloured scalar states acquire their masses of the order of the unification scale (i.e. much larger than the proton mass), making 
such a proton decay kinematically impossible. Meanwhile, the heavy coloured scalars are relevant for generation of loop-induced lepton mass terms 
at the matching scale in the low-energy effective model as will be discussed in more detail below.
\begin{table}[tbp]
  \begin{center}
    \begin{tabular}{ccc}
     \toprule                     
                        				 				&$\U{A}$ 	& $\U{B}$  	\\    
      \midrule
       $L$, $\tilde{L}$     			    			& $+1$		& $0$		\\
       $Q_\mathrm{L}$, $\tilde{Q}_\mathrm{L}$  			& $-1/2$	& $+1/3$	\\
       $Q_\mathrm{R}$, $\tilde{Q}_\mathrm{R}$ 	 		& $-1/2$	& $-1/3$ 	\\
      \midrule
      $G_{\mathrm{L,R,C}}$								& $0$		& $0$ \\
     \bottomrule
    \end{tabular}
    \caption{Charge assignment under the accidental symmetries.}
    \label{tab:AccSym}  
  \end{center}
\end{table}

\section{Spontaneous trinification breaking down to a LR-symmetric model} \label{sec:TriBreaking}

In this paper, we would like to explore whether after the spontaneous symmetry breaking (SSB) of the trinification gauge symmetry, 
it will be possible for the effective LR-symmetric model to break down to the SM gauge group by means of the RG evolution of the 
corresponding couplings, in particular, mass parameters. In order to do that, one has to explore first the SSB of the group in 
Eq.~\eqref{eq:trinigaugegroup} (also taking into account the accidental $\{\U{A}\times\U{B}\}$ symmetry). The most straightforward 
way to break trinification is when only one component in $\tilde{L}$ acquires a real non-zero VEV, namely,
\begin{equation} \label{eq:vevStructure}
\langle \Sl{i}{l}{r} \rangle = \delta^i_3 \delta^l_3 \delta^3_r \frac{v_3}{\sqrt{2}} = \left( \begin{array}{ccc}
0 & 0 & 0 \\
0 & 0 & 0 \\
0 & 0 & \frac{v_3}{\sqrt{2}}
\end{array} \right)^{i=3} \,,
\end{equation}
where the $l$ ($r$) index labels the rows (columns), while $\langle \tilde{Q}_L \rangle = \langle \tilde{Q}_R \rangle = 0$. 
As will be shown in Sect.~\ref{sec:Homotopy}, this often corresponds to the global minimum of the potential in Eq.~\eqref{eq:TLpot}, assuming no $\SU{3}{C}$ breaking VEVs. 
The extremal conditions (i.e. the requirement that the first derivatives of the scalar potential vanish in the minimum) allow us to rewrite $\mu$ in terms of $v_3$ as follows
\begin{equation} \label{eq:TriTadpole}
\mu^2 = (\lambda_1 + \lambda_2 + \lambda_3 + \lambda_4) \,  v_3^2 \,.
\end{equation} 
By applying a general infinitesimal gauge transformation on $\langle \Sl{i}{l}{r} \rangle$, we find that the following subset of the 
trinification gauge symmetry generators leaves that vacuum invariant:
\begin{equation} \label{eq:TriUnbrokenGaugeGenerators}
\T{C}{1,\dots,8} \,, \qquad \T{L}{1,2,3} \,, \qquad \T{R}{1,2,3}	 \,, \qquad \T{\BL}{} \equiv \frac{2}{\sqrt{3}} \left( \T{L}{8} + \T{R}{8} \right) \,.
\end{equation}
Therefore, the vacuum (\ref{eq:vevStructure}) spontaneously breaks the trinification gauge group to the LR-symmetric gauge group 
$\SU{3}{C} \times \SU{2}{L} \times \SU{2}{R} \times \U{\BL}$. 
Before the SSB, the global symmetry group is the full symmetry group $\SU{3}{C} \times \SU{3}{L} \times \SU{3}{R} \times \SU{3}{F} \times \U{A} \times \U{B}$. When applying a general infinitesimal global symmetry transformation on the vacuum given by Eq.~\eqref{eq:vevStructure}, we find that the following generators leave it invariant
\begin{equation} \label{eq:TriUnbrokenGlobalGenerators}
 \T{F}{1,2,3} \,, \qquad \T{X}{} \equiv \frac{2}{\sqrt{3}} \left(\T{L}{8} - \T{R}{8} - 2 \T{F}{8} \right) \,, \qquad 
 \T{Z}{} \equiv \frac{2}{3} \left( \T{A}{} + \sqrt{3} \T{F}{8} \right) \,,
\end{equation}
in addition to $\T{B}{}$ and the generators in Eq.~\eqref{eq:TriUnbrokenGaugeGenerators}. Here, we have constructed $\T{X,Z}{}$ such that they 
are orthogonal to $\T{\BL}{}$ and chosen their normalisation for convenience. However, any other two linearly independent combinations of 
$\T{X,Z}{}$ and $\T{\BL}{}$ that are also linearly independent of $\T{\BL}{}$, generate an unbroken $\{\U{} \times \U{}\}$ symmetry. 
Therefore, after the SSB, in addition to the unbroken gauge group, the symmetry $\{\SU{2}{F} \times \U{X} \times \U{Z} \times \U{B} \}$ remains unbroken as well. In summary, the VEV setting \eqref{eq:vevStructure} leads to the SSB pattern
\begin{equation}\label{eq:triniBreakingPattern}
\begin{gathered}
\SU{3}{L} \times \SU{3}{R} \times \SU{3}{C} \times \{\SU{3}{F} \times \U{A} \times \U{B}\} \\
\downarrow	\\
\SU{3}{C} \times \SU{2}{L} \times \SU{2}{R} \times \U{\BL} \times \{\SU{2}{F} \times \U{X} \times \U{Z} \times \U{B} \} \,,
\end{gathered}
\end{equation}
and the basic properties of the resulting effective LR-symmetric model will be studied below in detail.

\subsection{Colour-singlet scalar sector}

The colour-singlet scalars (CSS) are contained in $\tilde{L}$ which is the tri-triplet representation $\bm{3}\otimes \bar{\bm{3}} \otimes \bm{3}$ of 
$\SU{3}{L} \times \SU{3}{R} \times \{\SU{3}{F}\}$. It therefore contains $54$ real degrees of freedom. The VEV structure (\ref{eq:vevStructure}) breaks 
nine gauge symmetry generators, meaning that one identifies nine massless real d.o.f.'s in the CSS mass spectrum that become the longitudinal 
polarisation states of nine massive gauge bosons. In addition, the non-gauge part of the symmetry group is reduced from $\SU{3}{F}\times \U{A} \times \U{B}$ down to $\SU{2}{F} \times \U{X} \times \U{Z} \times \U{B}$ 
so that the CSS spectrum, after the SSB, also contains four corresponding Goldstone d.o.f.'s. These so-called ``global Goldstone'' bosons remain as physical massless 
scalar d.o.f.'s. However, at energy scales much lower than $v_3$, these are effectively decoupled from all other light fields (including the SM fields) 
since their interactions are always suppressed by powers of $v_3$. The decoupling of global Goldstone bosons is further discussed in Sect.~\ref{sec:GoldstoneDecoupling}.
\begin{table}[tbp]
  \begin{center}
    \begin{tabular}{llll}
     \toprule                     
       Fields    						& (Mass)$^2$ 													& $(\BL,X,Z)$ & Comment					\\      
      \midrule
       $\Sl{I}{L}{R}$  					& $- ( \lambda_2 + \lambda_3 + \lambda_4)\,  v_3^2$				& $(0,0,+1)$	&							\\
       $\Sl{I}{3}{R}$  					& $m_R^2 \equiv - ( \lambda_2 + \lambda_3) \, v_3^2 $			& $(-1,-1,+1)$	&							\\
       $\Sl{3}{L}{R}$ 		    		& $m_h^2 \equiv - ( \lambda_3 + \lambda_4) \, v_3^2 $			& $(0,+2,0)$	&							\\
       $\Sl{I}{L}{3}$   				& $- ( \lambda_2 + \lambda_4) \, v_3^2 $						& $(+1,-1,+1)$	&							\\
       $\mathrm{Re}[\Sl{3}{3}{3}]$  	& $2 (\lambda_1 + \lambda_2 + \lambda_3 + \lambda_4) \, v_3^2$	& $(0,0,0)$		&							\\
       $\mathrm{Im}[\Sl{3}{3}{3}]$ 		& 0																& $(0,0,0)$ 	&	Gauge Goldstone			\\
       $\Sl{3}{L}{3}$			   		& 0																& $(+1,+1,0)$ 	&	Gauge Goldstone			\\
       $\Sl{3}{3}{R}$			   		& 0																& $(-1,+1,0)$ 	&	Gauge Goldstone			\\
       $\Sl{I}{3}{3}$   				& 0																& $(0,-2,+1)$	&	Global Goldstone		\\
     \bottomrule
    \end{tabular}
    \caption{Mass eigenstates in $\tilde{L}$ after the SSB of the trinification group, and the corresponding tree-level squared masses and $\U{}$ charges. 
    All states have zero baryon number here. In Sect.~\ref{sec:LR}, we consider the LR-symmetric low-energy effective model with $\Sl{I}{3}{R}\equiv \Lr{I}{R}$ 
    and $\Sl{3}{L}{R} \equiv \hh{L}{R}$ assuming that $m^2_{R,h} \ll v_3^2$ while all other masses are heavy, i.e. $\sim v_3$. }
 \label{tab:CSSMassStates}  
  \end{center}
\end{table}
The mass eigenstates and the corresponding squared masses of the CSS after $\Sl{3}{3}{3}$ develops a VEV, are listed in Tab.~\ref{tab:CSSMassStates}. 
Local stability of the minima in the CSS sector is obtained when these squared masses are non-negative. Combined with the requirement that $\mu^2>0$, 
this is ensured when 
\begin{equation}\label{eq:conditionsLam} 
\lambda_1 + \lambda_2 + \lambda_3 + \lambda_4 > 0 \,, \qquad \lambda_2 + \lambda_3 \leq 0	 \,, \qquad
\lambda_2 + \lambda_4 \leq 0	 \,, \qquad \lambda_3 + \lambda_4 \leq 0	\,.
\end{equation}

\subsection{Coloured scalar sector}

\begin{table}[tbp]
  \begin{center}
    \begin{tabular}{lll}
     \toprule                     
       Fields    						& (Mass)$^2$ 													&  $(\BL,X,Z,B)$ \\      
      \midrule
       $\SQL{I}{c}{L}$	& $\frac{1}{2} \left[ \alpha_1 - 2 (\lambda_1 + \lambda_2 + \lambda_3 + \lambda_4) \right] v_3^2$									& 	$(-1/3,-1,0,+1/3)$	\\
       $\SQR{I}{R}{c}$	& $\frac{1}{2} \left[ \alpha_1 - 2 (\lambda_1 + \lambda_2 + \lambda_3 + \lambda_4) \right] v_3^2$									& 	$(+1/3,-1,0,-1/3)$	\\
       $\SQL{3}{c}{L}$ 	& $\frac{1}{2} \left[ \alpha_1+\alpha_2 - 2 (\lambda_1 + \lambda_2 + \lambda_3 + \lambda_4) \right] v_3^2$							& 	$(-1/3,+1,-1,+1/3)$ \\
       $\SQR{3}{R}{c}$ 	& $\frac{1}{2} \left[ \alpha_1+\alpha_2 - 2 (\lambda_1 + \lambda_2 + \lambda_3 + \lambda_4) \right] v_3^2$							& 	$(+1/3,+1,-1,-1/3)$ \\
        $(\tilde{Q}^I_{\mathrm{LR}\pm})^c$ & $\frac{1}{2} \left[ \alpha_1+\alpha_3 \pm \frac{\gamma}{\sqrt{2} v_3} - 2 (\lambda_1 + \lambda_2 + \lambda_3 + \lambda_4) \right]  v_3^2$																														& 	$(+2/3,0,+1/3)$ \\
       $\SQL{3}{c}{3}$ 	& $\frac{1}{2} \left[ \alpha_1+\alpha_2 + \alpha_3 + \alpha_4 - 2 (\lambda_1 + \lambda_2 + \lambda_3 + \lambda_4) \right]  v_3^2$	& 	$(+2/3,+2,-1,+1/3)$ \\
       $\SQR{3}{3}{c}$ & $\frac{1}{2} \left[ \alpha_1+\alpha_2 + \alpha_3 + \alpha_4 - 2 (\lambda_1 + \lambda_2 + \lambda_3 + \lambda_4) \right]  v_3^2$	& 	$(-2/3,+2,-1,-1/3)$ \\
     \bottomrule
    \end{tabular}
    \caption{Mass eigenstates in $\tilde{Q}_\mathrm{L}$ and $\tilde{Q}_\mathrm{R} $ after the SSB of the trinification symmetry group, and 
    the corresponding tree-level squared masses and $\U{}$ charges. Here, $(\tilde{Q}^I_{\mathrm{LR}\pm})^c \equiv \frac{1}{\sqrt{2}} \left[\SQL{I}{c}{3} 
    \pm \epsilon^{I J} \SQRS{J}{3}{c} \right]$.}
    \label{tab:CSMassStates}
  \end{center}
\end{table}
When $\tilde{L}$ acquires a VEV according to Eq.~\eqref{eq:vevStructure}, all coloured scalar (CS) d.o.f.'s become massive. The mass eigenstates and masses are listed in Tab.~\ref{tab:CSMassStates}. Requiring that the squared masses must be non-negative constrains the parameters in $V_2$ and $V_{3}$ as
\begin{equation}\label{eq:conditionsAlpha}
\begin{aligned}
\alpha_1 &\geq 2 (\lambda_1 + \lambda_2 + \lambda_3 + \lambda_4)	\,,\\
\alpha_1 + \alpha_2 &\geq 2 (\lambda_1 + \lambda_2 + \lambda_3 + \lambda_4) \,,\\
\alpha_1 + \alpha_3 - \frac{|\gamma|}{\sqrt{2} v_3}&\geq 2 (\lambda_1 + \lambda_2 + \lambda_3 + \lambda_4) \,,\\
\alpha_1+\alpha_2 + \alpha_3 + \alpha_4 &\geq 2 (\lambda_1 + \lambda_2 + \lambda_3 + \lambda_4) \,.
\end{aligned}
\end{equation}
%

\subsection{Gauge boson sector}

After the SSB, nine gauge bosons become massive. Their masses are determined by the trinification gauge 
coupling $g$ as indicated in Tab.~\ref{tab:TriBosons}. They can be conveniently grouped into two doublets 
(one for each $\SU{2}{L,R}$)
\begin{equation}
{V_\mathrm{L}}^L_\mu \equiv \frac{1}{\sqrt{2}}\left( \begin{array}{c}
{G_\mathrm{L}}^6 + \mathrm{i} {G_\mathrm{L}}^7 	\\
 {G_\mathrm{L}}^4 + \mathrm{i} {G_\mathrm{L}}^5 	
\end{array} \right)_\mu \, , \quad {V_\mathrm{R}}^R_\mu \equiv \frac{1}{\sqrt{2}}\left( \begin{array}{c}
{G_\mathrm{R}}^6 + \mathrm{i} {G_\mathrm{R}}^7 	\\
 {G_\mathrm{R}}^4 + \mathrm{i} {G_\mathrm{R}}^5 	
\end{array} \right)_\mu \, ,
\end{equation}
and one singlet 
\begin{equation}
{V_\mathrm{s}}_\mu \equiv \frac{1}{\sqrt{2}} \left( {G_\mathrm{L}}^8 -  {G_\mathrm{R}}^8 \right)_\mu \,.
\end{equation}

\begin{table}[tbp]
  \begin{center}
    \begin{tabular}{llll}
     \toprule                     
       Fields    												& (Mass)$^2$ 			&  $(\BL,X)$ 	& Comment	\\      
      \midrule
       ${G_\mathrm{C}}^{1\dots 8}_\mu$								& $ 0 $					& 	$(0,0)$		&	Gauge field of $\SU{3}{C}$	\\
       ${G_\mathrm{L}}^{1 \dots 3}_\mu$								& $ 0 $					& 	$(0,0)$		&	Gauge field of $\SU{2}{L}$	\\
       ${G_\mathrm{R}}^{1 \dots 3}_\mu$								& $ 0 $					& 	$(0,0)$		&	Gauge field of $\SU{2}{R}$	\\
       $\frac{1}{\sqrt{2}}\left(  {G_\mathrm{L}}^8 + {G_\mathrm{R}}^8 \right)_\mu $		& $ 0 $	 &	$(0,0)$		&	Gauge field of $\U{\BL}$	\\
       ${V_\mathrm{L}}^L_\mu$										& $ \frac{1}{4} g^2 \, v_3^2 $		& 	$(+1,+1)$	&		\\
       ${V_\mathrm{R}}^R_\mu$										& $ \frac{1}{4} g^2 \, v_3^2 $		& 	$(+1,-1)$	&		\\
       ${V_\mathrm{s}}_\mu$											& $ \frac{2}{3} g^2 \, v_3^2 $		&	$(0,0)$				&	\\
     \bottomrule
    \end{tabular}
    \caption{Gauge boson states after the SSB of the trinification group. All gauge boson states are uncharged under $\{ \U{Z} \times \U{B} \}$.}
    \label{tab:TriBosons}
  \end{center}
\end{table}

\subsection{Fermion sector}

The fermions $\LL{i}{l}{r}$, $\QL{i}{c}{l}$ and $\QR{i}{r}{c}$ couple to the scalars $\Sl{i}{l}{r}$, $\SQL{i}{c}{l}$ and $\SQR{i}{r}{c}$ via 
the Yukawa interactions given in Eq.~\eqref{eq:TriYukawa}. In particular, the VEV leads to a Dirac mass term for one fermionic $\SU{2}{F}$ doublet. 
The corresponding down-type Dirac state is built out of $\QL{I}{c}{3}$ and $\QRd{I}{3}{c}$ as follows
\begin{equation}
({\mathcal{D}_\mathrm{H}}^I )^c \equiv \left( \begin{array}{c}
\QL{I}{c}{3} 	\\
\epsilon^{I J} \QRd{J}{3}{c} 	
\end{array} \right) \,.
\end{equation}
The $\U{}$ charges of $\mathcal{D}_\mathrm{H}$ are shown in Tab.~\ref{tab:TriFermions}.

\begin{table}[tbp]
  \begin{center}
    \begin{tabular}{llll}
     \toprule                     
       Fields    								& (Mass)$^2$ 												&  $(\BL,X,Z,B)$ 		& Comment	\\      
      \midrule
       $({\mathcal{D}_\mathrm{H}}^I)^c$							& $ \frac{1}{2} y^2 v_3^2 $									& 	$(+2/3,0,0,+1/3)$		&	Dirac fermion	\\
     \bottomrule
    \end{tabular}
    \caption{The first and second generation $\SU{2}{L} \times \SU{2}{R}$-singlet quarks make up an $\SU{2}{F}$-doublet Dirac fermion 
    that gets a tree-level mass at the trinification breaking scale. All other fermionic d.o.f.'s in $\LL{i}{l}{r}$, $\QL{i}{c}{l}$ 
    and $\QR{i}{r}{c}$ are massless at tree-level. }
\label{tab:TriFermions}  
  \end{center}
\end{table}

\subsection{Finding the global minimum through homotopy continuation} \label{sec:Homotopy}

If we restrict ourselves to the case of two generations of colour singlet scalars $\tilde{L}^i$ getting VEVs, 
the most general VEV setting after accounting for gauge \cite{Cauet:2010ng} and family symmetries can be 
written as\footnote{Using the $\SU{2}{F} \subset \SU{3}{F}$ family symmetry we can ``rotate away'' 
one of the VEVs of the general case shown in Ref.~\cite{Cauet:2010ng}.}
\begin{equation} \label{eq:MinimumTri}
\langle (\tilde{L}^1)^l_r \rangle= \frac{1}{\sqrt{2}}
\begin{pmatrix}
v^1 & 0 & 0 \\
0 & v^2 & 0 \\
0 &0 & v^3
\end{pmatrix}   \, ,
\qquad
\langle (\tilde{L}^2)^l_r \rangle=
\frac{1}{\sqrt{2}} \begin{pmatrix}
v^5 & 0 & v^6 \\
0 & 0 & 0 \\
v^7 &0 & v^8
\end{pmatrix} \,.
\end{equation}
Note that, due to $\mathbb{Z}_3$, this choice is physically equivalent to assuming only VEVs in two generations of either $\tilde{Q}_\mathrm{L}$ 
or $\tilde{Q}_\mathrm{R}$.

As discussed in the previous sections, we are interested in the case where only one real scalar field aqcuires a non-zero VEV. 
The question remains as to whether this is indeed the global minimum of the scalar potential or if the global minimum has a different 
set of non-zero VEVs and therefore a different symmetry breaking chain takes place. 

Using the homotopy continuation method through HOM4PS2 \cite{lee2008hom4ps}, we performed a random scan over 5000 
parameter points satisfying the conditions in Eqs.~\eqref{eq:conditionsLam} and \eqref{eq:conditionsAlpha}. The homotopy 
continuation method finds all the solutions of systems of polynomial equations, in this case the minimisation conditions 
of the tree-level potential \eqref{eq:TLpot} for the VEV setting in Eq.~\eqref{eq:Minimum}. 

For all the points in the scan, the global minimum was always the one for which $v^{3}\equiv v \not=0$ and $v^{i\neq 3}=0$, even 
for parameter points where other minima were present. In other words, for the model described here if we require that there exists 
a minimum with one real field acquiring a VEV then, excluding pathological cases that might have been missed in the numerical analysis, 
that minimum is the global one. 

The most general case where a third generation is also allowed to acquire VEVs could not be treated with HOM4PS2 
due to a complicated system of equations outpacing our computational resources. However, a purely numerical minimisation 
was performed over a second scan of parameter space leading to the same result as for the case of two generations. Given a parameter point that 
satisfies the positive scalar mass-squares condition in the one VEV minima, the numerical minimisation procedure was started in a random 
point in field space, whereby the minimum of the potential was found by a simple steepest-descent method. For the minima obtained 
in this way, we computed the gauge boson mass spectrum and observed that it numerically matched the masses in Tab.~\ref{tab:TriBosons}. By pretending 
that $\SU{3}{F}$ is gauged, we computed the number of unbroken global symmetry generators by counting the number of ``new'' massless gauge bosons, and 
could in all cases conclude that it matched the number of global symmetry generators in the effective LR-symmetric model. Therefore, we believe that the all 
of these minima are related to the one VEV minima in Eq.~\eqref{eq:vevStructure} by a symmetry transformation, and are hence physically equivalent.
\section{The low-scale effective LR-symmetric model} \label{sec:LR}

\subsection{Minimal particle content of the effective model}

The trinification group is spontaneously broken by the VEV $v_3$ in Eq.~\eqref{eq:vevStructure} to the following symmetry 
\begin{equation} \label{eq:LRgaugegroup}
\SU{3}{C} \times \SU{2}{L} \times \SU{2}{R} \times \U{\BL} \times \{ \SU{2}{F} \times \U{X} \times \U{Z} \times \U{B}\}.
\end{equation}
The decomposition of $\tilde{L}$ in terms of representations of the group \eqref{eq:LRgaugegroup} can be written as
\begin{equation}\label{eq:expansionL}
\begin{aligned}
\Sl{i}{l}{r} = &  \delta^i_I \left[ \delta_L^l \delta_r^R \, \HH{I}{L}{R} + \delta_L^l \delta_r^3 \, \Ll{I}{L}+\delta^l_3 \, \delta_r^R \Lr{I}{R} + \delta_3^l \delta_r^3 \, \tilde{\Phi}^I \right] 		\\
&  + \delta_3^i \left[ \delta_L^l \delta_r^R \, \hh{L}{R} + \delta_L^l \delta_r^3 \, \tilde{l}_\mathrm{L}^s {}^L + \delta_3^L \delta_r^R \, \tilde{l}_\mathrm{R}^s{}_R + \delta_3^l \delta_r^3 \, \left( \tilde{\Phi}^s + \frac{v_3}{\sqrt{2}} \right) \right].
\end{aligned}
\end{equation}
Here, $\tilde{l}_{\mathrm{L,R}}^s$ and $\mathrm{Im} [\tilde{\Phi}^s]$ are the gauge Goldstone d.o.f.'s 
that become the longitudinal polarisation states of the heavy vector bosons listed in Tab.~\ref{tab:TriBosons}, wheras $\tilde{\Phi}^I$ is the ``global'' Goldstone boson. Similarly, 
the fermion multiplet $L$ can be written in terms of reps of the new symmetry group as follows
\begin{equation}\label{eq:expansionLfermion}
\begin{aligned}
\LL{i}{l}{r} = &  \delta^i_I \left[ \delta_L^l \delta_r^R \, \HHf{I}{L}{R} + \delta_L^l \delta_r^3 \, \Llf{I}{L}+\delta^l_3 \delta_r^R \, \Lrf{I}{R} + \delta_3^l \delta_r^3 \, \Phi^I \right] 		\\
&  + \delta_3^i \left[ \delta_L^l \delta_r^R \, \HHf{s}{L}{R} + \delta_L^l \delta_r^3 \, \Llfs{L}+\delta^l_3 \delta_r^R \, \Lrfs{R} +\delta_3^l \delta_r^3 \, \Phi^s \right].
\end{aligned}
\end{equation}
Moreover, the decomposition of the trinification quark multiplets, $Q_\mathrm{L}$ and $Q_\mathrm{R}$, reads
\begin{equation}\label{eq:expansionQ}
\begin{aligned}
\QL{i}{c}{l} = &  \delta^i_I \left[ \delta^L_l \,  (\mathcal{Q}_\mathrm{L}{}^I)^c_L  + \delta_l^3 \, (D_\mathrm{L}{}^{I})^c \right] + \delta_3^i \left[  \delta_l^L \, \mathcal{Q}_\mathrm{L}^s{}_L^c + \delta_l^3 \, D_\mathrm{L}^s{}^c  \right],  \\
\QR{i}{r}{c} = &  \delta^i_I \left[ \delta_R^r \, (\mathcal{Q}_\mathrm{R}{}^I)^R_c  + \delta^r_3 (D_\mathrm{R}{}^{I})_c \right] + \delta_3^i \left[  \delta^r_R \, \mathcal{Q}_\mathrm{R}^s {}^R_c + \delta^r_3 \, D_\mathrm{R}^s{}_c  \right],  \\
\end{aligned}
\end{equation}
and similarly for $\tilde{Q}_\mathrm{L,R}$.

As mentioned in Sect.~\ref{sec:TriBreaking}, we want to explore whether a radiatively induced breaking 
down to the SM gauge group can happen for the proposed model. As will be discussed later in detail, in order for that to happen, 
we need to have at least one $\SU{2}{R}$ and one $\SU{2}{L}$ scalar doublet in the effective theory so that the $\SU{2}{R}$-doublet 
mass parameter can run negative. With this in mind, and considering the simplest possible scenario, we will focus our attention on a subset of the parameter space where 
after the trinification symmetry is broken, the scalar spectrum comprises two very light states, namely,
\begin{equation}\label{eq:lightfields}
\hh{L}{R} \equiv \Sl{3}{L}{R} \qquad \mbox{and} \qquad  \Lr{I}{R} \equiv \Sl{I}{3}{R} \, ,
\end{equation}
in addition to the global Goldstone field $\tilde{\Phi}^I \equiv \Sl{I}{3}{3}$. The remaining scalars either get masses of $\mathcal{O}(v_3)$ 
or get ``eaten'' by heavy gauge bosons. For this particular case, one can integrate out all the heavy scalars ending up with a simpler effective theory 
containing the light and massless fields only. In the fermion sector, the fields $\QL{I}{c}{3}$ and $\QR{I}{3}{c}$ can also be integrated out 
as they make up the heavy Dirac fermions shown in Tab.~\ref{tab:TriFermions}. The fields that are present in the effective theory are shown in Tab.~\ref{tab:LRmodel}.

In order to parametrize the relevant regions of parameter space giving rise to such a minimal particle content of the effective model, let us define small dimensionless $\delta$ and $\varepsilon$ parameters as follows
\begin{equation}
\varepsilon \equiv -\lambda_2  - \lambda_3   \, ,\quad	\delta \equiv - \lambda_3 - \lambda_4,
\end{equation}
such that $m_R^2 = \varepsilon v_3^2$ and $m_h^2 = \delta v_3^2$. We will then construct the effective 
LR-symmetric model assuming $\varepsilon\ll 1$ and $\delta \ll 1$. At higher orders in perturbation theory, other tree-level couplings (such as $\alpha_i$ and $\gamma$) will enter in the full expressions for $m_{h,R}^2$. To still keep these states sufficiently light at the matching scale will then further constrain the parameter space as the simple assumptions $\varepsilon, \delta \ll 1$ will not suffice. This is further discussed in Sec.~\ref{sec:LightScalars_loop}.

In the LR-symmetric effective model, the fields interact with gauge bosons according to their representations under the gauge groups as given in 
Tab.~\ref{tab:LRmodel}, with strengths determined by the gauge couplings $g_\mathrm{L}$, $g_\mathrm{R}$, $g_\mathrm{C}$ and 
$g_\mathrm{\BL}$. At the matching scale $v_3$, these are related to the trinification gauge coupling $g$ as
\begin{equation}
\g{L} = \g{R} = \g{C} = g \,,	\quad 	\g{\BL} = \sqrt{\frac{3}{8}} g \,.
\end{equation}

\subsection{Matching of the scalar potential parameters}

The GUT-scale scalar potential $V$ should be matched onto the most general renormalizable scalar 
potential for $\hh{L}{R}$, $\Lr{I}{R}$ and $\tilde{\Phi}^I$ in the low-energy LR-symmetric model:
\begin{equation}
\begin{aligned}
V_{LR} &= m_h^2 \, |\tilde{h}|^2 + m_R^2 \, |\tilde{l}_\mathrm{R}|^2 + m_{\tilde{\Phi}}^2 \, |\tilde{\Phi}|^2 		\\
& + \lambda_a \, |\tilde{h}|^4 + \lambda_b \, |\tilde{l}_\mathrm{R}|^4 + \lambda_c \, |\tilde{\Phi}|^4 + \lambda_d \, |\tilde{h}|^2 |\tilde{\Phi}|^2 + \lambda_e \, |\tilde{l}_\mathrm{R}|^2 |\tilde{\Phi}|^2 + \lambda_f \, | \tilde{l}_\mathrm{R}|^2 |\tilde{h}|^2		\\
& + \lambda_g \, \Lr{I}{R_1} \,\, \Lrs{I}{R_2} \,\, \hh{L}{R_1'} \,\, \hhs{L}{R_2'} \quad \epsilon^{R_1 R_1'} \, \epsilon_{R_2 R_2'} \\ 
& + \lambda_h \, \Lr{I_1}{R} \,\, \Lrs{J_1}{R} \,\, \tilde{\Phi}^{I_2} \,\, \tilde{\Phi}^*_{J_2} \quad \epsilon_{I_1 I_2} \, \epsilon^{J_1 J_2}	\\											
& + \lambda_i \, \Lr{I_1}{R_1} \,\, \Lr{I_2}{R_1'} \,\, \Lrs{J_1}{R_2} \,\, \Lrs{J_2}{R_2'} \quad \epsilon_{I_1 I_2} \, \epsilon^{J_1 J_2} \, \epsilon_{R_2 R_2'} \, \epsilon^{R_1 R'_1} \\
& + \lambda_j \, \hh{L_1}{R_1} \,\, \hh{L_1'}{R_1'} \,\, \hhs{L_2}{R_2} \,\, \hhs{L_2'}{R_2'}  \quad \epsilon_{L_1 L_1'} \, \epsilon^{L_2 L_2'} \, \epsilon_{R_2 R_2'} \, \epsilon^{R_1 R_1'}
\end{aligned}
\end{equation}

The tree-level matching conditions in the scalar sector are obtained by requiring that the $n$-point functions with external scalars in the high-scale and the low-scale theory coincide 
at tree-level at the matching scale $\mu_m$, which we take to be the trinification breaking scale $v_3$. In this case, we get all relations between the high scale parameters $\lbrace v_3,\lambda_{1,3},\epsilon,\delta\rbrace$, 
and the low-scale parameters $\lbrace m_{h,R,\tilde{\Phi}}^2,\lambda_{a,\dots,j}\rbrace$ from the two-point functions (i.e.~from the masses squared) and the four-point 
functions (quartic couplings). We compute the four-point functions taking the limit of small external momenta (compared to $v_3$ scale), since we are only interested in the matching of the renormalizable operators. Any momentum dependence in the four-point functions in the low-scale theory would instead be attributed to (non-renormalizable) derivative interactions. We do not take these higher order derivative operators into account since they presumably would have a negligible effect in the infrared behaviour of the theory. 
Thus, the matching conditions are
\begin{equation}\label{eq:matchingcond}
\begin{aligned}
m_h^2 &= \delta v^2	\, , \quad 	m_R^2 = \varepsilon v^2	\, , \quad 	m_{\tilde{\Phi}}^2 = 0	\, ,		\\									
\lambda_c &= \lambda_d = \lambda_e = \lambda_h = 0		\,	 ,												\\
\lambda_g &= - 2 \lambda_3 \, , \quad \lambda_i =  \frac{1}{2} \varepsilon	\, , \quad 	\lambda_j =  \frac{1}{2} \delta	\,	,				\\
\lambda_a &= \lambda_1-\lambda_3-\varepsilon-\delta - \frac{(\lambda_1-\lambda_3-\varepsilon)^2}{\lambda_1-\lambda_3-\varepsilon-\delta} \approx - 2 \delta	+ \mathcal{O}(\varepsilon^2,\varepsilon \delta, \delta^2)		\, ,\\
\lambda_b &= \lambda_1-\lambda_3-\varepsilon-\delta - \frac{(\lambda_1-\lambda_3-\delta)^2}{\lambda_1-\lambda_3-\varepsilon-\delta} \approx - 2 \varepsilon + \mathcal{O}(\varepsilon^2,\varepsilon \delta, \delta^2)		\, ,	\\
\lambda_f &= 2 \left[ \lambda_1+\lambda_3 - \frac{(\lambda_1-\lambda_3-\varepsilon)(\lambda_1-\lambda_3-\delta)}{\lambda_1-\lambda_3-\varepsilon-\delta} \right]	\approx 4 \lambda_3					+ \mathcal{O}(\varepsilon^2,\varepsilon \delta, \delta^2)		\, .
\end{aligned}
\end{equation}
Interestingly, all $\lambda_1$'s cancel out in the matching conditions, provided that $\varepsilon$ and $\delta$ are sufficiently small, which means that $\lambda_1$ does not affect the values of the couplings in the effective LR-symmetric model at tree-level. This can be seen as a consequence 
of $\hh{L}{R}$, $\Lr{I}{R}$ and $\tilde{\Phi}^I$ becoming Goldstone bosons of the $\mathrm{O}(54) \rightarrow \mathrm{O}(53)$ breaking that is induced in the CSS sector by the $v_3$ VEV in the limit $\lambda_2, \, \lambda_3, \, \lambda_4 \rightarrow 0$ (since they are Goldstone bosons in this limit, they must decouple from the scalar potential in the same limit). 
 
\subsection{Fermion sector}

Though the trinification theory only contains one Yukawa coupling, many terms are allowed by the symmetry group of the LR-symmetric effective model in the fermion sector:
\begin{equation}\label{eq:LRYukawa}
\begin{aligned}
\mathcal{L}_{\mathrm{Fermion}}^{(LR)} = \,\, &  Y_\alpha \,\, \Lrs{I}{R} \,\, \Lrf{I}{R} \,\, \Phi^s \,			  	\\
 +\,\, &  Y_\beta \,\, \Lrs{I}{R} \,\, \Lrfs{R} \,\, \Phi^I \,													\\
 +\,\, &  Y_\gamma \,\, \Lr{I}{R} \, \, \QRE{J}{R}{} \,\, D_\mathrm{L}^s \,\, \epsilon_{IJ}							\\
& &	&																												\\
 +\,\, & Y_\delta \,\, \hhs{L}{R} \,\, \HHf{s}{L}{R} \,\, \Phi^s													\\
 +\,\, & Y_\epsilon \,\, \hhs{L}{R} \,\, \Llfs{L} \,\, \Lrfs{R}	           										\\
 +\,\, & Y_\zeta \,\, \hh{L}{R}  \,\, \QLE{I}{}{L} \,\, \QRE{J}{R}{}  \,\, \epsilon_{I J}							\\
& &	&																												\\
 +\,\, &	Y_\eta \,\, \tilde{\Phi}^*_I \,\, \Phi^I \Phi^s															\\
& & &																												\\
 +\,\,& \frac{m_{\Phi^s}}{2} \, \Phi^s \Phi^s 			& + \mathrm{c.c.}&					
\end{aligned}
\end{equation} 

The matching conditions for these Yukawa couplings are rather easy at tree-level, as only two of them are found 
to have non-vanishing values at the matching scale. This does not mean that the other couplings are not present in the effective theory 
and it would be a subject of a future study to calculate the matching conditions at higher orders where they might not necessarily vanish. 
However, for the purpose of this work, which is to explore a potential for the radiative breaking of the LR symmetry down to the SM gauge group, the tree-level approximation is expected to be sufficient. In this case, the matching conditions are such that 
\begin{equation}\label{eq:matchingY}
Y_\zeta = -y \qquad \mbox{and} \qquad Y_\gamma = y \,,
\end{equation}
while $Y_{\alpha,\beta,\delta,\epsilon, \eta} = 0$. Furthermore, the $\beta$-functions in Appendix~\ref{sec:RGEs} indicate that the Yukawa 
couplings that are zero at the matching scale will also remain zero at lower scales, since $\beta_{Y_i} \propto Y_i$.

With the matching conditions defined above, the parameters of the effective LR-symmetric model can be determined from 
a reduced set of parameters of the GUT-scale theory. The vacuum stability constraints \eqref{eq:conditionsLam} and \eqref{eq:conditionsAlpha} 
can be then translated as well to reduce further the allowed parameter space for the effective theory. With this framework in mind, the question 
remains as to whether the remaining symmetries of the effective theory can be broken radiatively by one-loop RG running at a lower scale leading 
to an effective model which approaches the SM.

\subsection{Decoupling of ``global'' Goldstone bosons} \label{sec:GoldstoneDecoupling}

It can be shown, on very general grounds that Goldstone bosons appearing due to the spontaneous breaking of a global symmetry at a scale $v_3$, 
have negligibly small interactions at scales $\mu \ll v_3$ \cite{Burgess:1998kh, Burgess:1998ku}. This decoupling is obvious if one chooses 
a specific exponential representation of the Goldstone d.o.f.'s, which comes at expense of manifest renormalisability. In this work, we have instead 
chosen the simple (but equivalent) linear representation of the global Goldstone d.o.f.'s $\tilde{\Phi}^I$ such that renormalisability (but not decoupling) 
is manifest in the GUT-scale trinification theory. Nevertheless, with the results from the three previous sections, we can see explicitly how $\tilde{\Phi}^I$ 
decouple at scales well below $v_3$.

\begin{table}[tbp]
    \setlength{\tabcolsep}{3pt}
    \begin{tabular}{ccccccccc}
     \toprule                     
                        		&$\SU{2}{L}$ 		& $\SU{2}{R}$ 		& $\SU{3}{C}$ 		& $\U{\BL}$ 	& $\{\SU{2}{F}\}$	& $\{\U{X}\}$	& $\{\U{Z}\}$ & $\{\U{B}\}$ \\      
      \midrule
      fermions 					&  					& 					& 					& 				&				& 			&			 &			 \\
      \midrule
       $H$      		  		&	 $\bm{2}$		&	$\bar{\bm{2}}$	&	$\bm{1}$		&	$0$			&	$\bm{2}$ 	&	$0$		&	$+1$	 &	$0$		 \\
       $l_\mathrm{L}$	   		&	 $\bm{2}$		&	$\bm{1}$		&	$\bm{1}$		&	$+1$		&	$\bm{2}$ 	&	$-1$	&	$+1$	 &	$0$		 \\
       $l_\mathrm{R}$	   		&	 $\bm{1}$		&	$\bar{\bm{2}}$	&	$\bm{1}$		&	$-1$		&	$\bm{2}$ 	&	$-1$	&	$+1$	 &	$0$		 \\
       $\Phi$      		   		&	 $\bm{1}$		&	$\bm{1}$		&	$\bm{1}$		&	$0$			&	$\bm{2}$ 	&	$-2$	&	$+1$	 &	$0$	  	 \\
       $\mathcal{Q}_\mathrm{L}$	&	 $\bar{\bm{2}}$	&	$\bm{1}$		&	$\bm{3}$		&	$-1/3$		&	$\bm{2}$ 	&	$-1$	&	$0$		 &	$+1/3$	 \\
       $\mathcal{Q}_\mathrm{R}$	&	 $\bm{1}$		&	$\bm{2}$		&	$\bar{\bm{3}}$	&	$+1/3$		&	$\bm{2}$ 	&	$-1$	&	$0$		 &	$-1/3$	 \\
       $H^s$      		  		&	 $\bm{2}$		&	$\bar{\bm{2}}$	&	$\bm{1}$		&	$0$			&	$\bm{1}$ 	&	$+2$	&	$0$		 &	$0$		 \\
       $l_\mathrm{L}^s$			&	 $\bm{2}$		&	$\bm{1}$		&	$\bm{1}$		&	$+1$		&	$\bm{1}$ 	&	$+1$	& 	$0$		 &	$0$		 \\
       $l_\mathrm{R}^s$	   		&	 $\bm{1}$		&	$\bar{\bm{2}}$	&	$\bm{1}$		&	$-1$		&	$\bm{1}$ 	&	$+1$	&	$0$		 &	$0$		 \\
       $\Phi^s$      	  		&	 $\bm{1}$		&	$\bm{1}$		&	$\bm{1}$		&	$0$			&	$\bm{1}$ 	&	$0$		&	$0$		 &	$0$		 \\
       $\mathcal{Q}_\mathrm{L}^s$&	 $\bar{\bm{2}}$	&	$\bm{1}$		&	$\bm{3}$		&	$-1/3$		&	$\bm{1}$ 	&	$+1$	&	$-1$	 &	$+1/3$	 \\
       $\mathcal{Q}_\mathrm{R}^s$&	 $\bm{1}$		&	$\bm{2}$		&	$\bar{\bm{3}}$	&	$+1/3$		&	$\bm{1}$ 	&	$+1$	&	$-1$	 &	$-1/3$	 \\
       $D_\mathrm{L}^s$		    &	 $\bm{1}$ 		&	$\bm{1}$		&	$\bm{3}$		&	$+2/3$		&	$\bm{1}$ 	&	$+2$	&	$-1$	 &	$+1/3$	 \\
       $D_\mathrm{R}^s$		    &	 $\bm{1}$		&	$\bm{1}$		&	$\bar{\bm{3}}$	&	$-2/3$		&	$\bm{1}$ 	&	$+2$	&	$-1$	 &	$-1/3$	 \\
      \midrule
      scalars					& 					& 					& 					& 				&    			&			&			 &			 \\
      \midrule       
       $\tilde{h}$      		& $\bm{2}$			&	$\bar{\bm{2}}$	&	$\bm{1}$	 	&	$0$			& $\bm{1}$		&	$+2$	&	$0$		 &	$0$		 \\
       $\tilde{l}_\mathrm{R}$	& $\bm{1}$			&	$\bar{\bm{2}}$	&   $\bm{1}$		& 	$-1$ 		& $\bm{2}$		&	$-1$	&	$+1$	 &	$0$		 \\
       $\tilde{\Phi}$			& $\bm{1}$			&	$\bm{1}$		&	$\bm{1}$		&	$0$			& $\bm{2}$		&	$-2$	&	$+1$	 &	$0$		 \\
      \midrule
      gauge bosons 				& 					& 					& 					& 				& 				&			&			 &			\\
      \midrule
      $G_\mathrm{L}$			& $\bm{3}$	 		& $\bm{1}$ 			& $\bm{1}$ 			& $0$			& $\bm{1}$ 		&	$0$		&	$0$		 &	$0$		\\
      $G_\mathrm{R}$			& $\bm{1}$ 			& $\bm{3}$ 			& $\bm{1}$ 			& $0$			& $\bm{1}$  	&	$0$		&	$0$		 &	$0$		\\
      $G_\mathrm{C}$			& $\bm{1}$ 			& $\bm{1}$ 			& $\bm{8}$ 			& $0$			& $\bm{1}$ 		&	$0$		&	$0$		 &	$0$		\\
      $G_{\mathrm{\BL}}$		& $\bm{1}$ 			& $\bm{1}$ 			& $\bm{1}$ 			& $0$			& $\bm{1}$ 		&	$0$		&	$0$		 &	$0$		\\
     \bottomrule
    \end{tabular}
    \caption{Field content of the effective LR-symmetric model.}
    \label{tab:LRmodel}
\end{table}

Firstly, we notice that $\tilde{\Phi}^I$ decouples from the scalar potential at the matching scale since $m_{\tilde{\Phi}}=0$ and 
$\lambda_c=\lambda_d=\lambda_e=\lambda_h=0$. 
Instead, one finds that for the matching to 
agree at non-zero Goldstone boson momenta, one has to introduce derivative interactions among the Goldstone bosons as well as between the Goldstone 
bosons and non-Goldstone fields. These derivative interactions necessarily have $\mathrm{dim}>4$ and hence must be suppressed by the trinification 
breaking scale $\sim v_3$. Since these operators presumably will be increasingly irrelevant in the infrared, we simply omit them from our LR-symmetric 
effective Lagrangian. 

In the trinification fermion sector shown in Eq.~\eqref{eq:TriYukawa}, one can check that the only Yukawa interactions involving $\tilde{\Phi}^I$ that are 
non-zero at tree-level also involve the heavy quark fields $\QL{I}{c}{3}$ and $\QR{I}{3}{c}$ (which are integrated out at the trinification breaking scale). 
In addition, in the effective theory, one new Yukawa interaction with $\tilde{\Phi}^I$ is allowed by symmetry ($Y_{\eta}$ in Eq.~\eqref{eq:LRYukawa}), 
but it vanishes at the matching scale. The corresponding $\beta$-function, $\beta_{Y_\eta}$, is proportional to $Y_{\eta}$, meaning that the vanishing 
matching condition also forces $Y_{\eta}=0$ at lower scales. 

Having shown the disappearance of scalar and Yukawa interactions with $\tilde{\Phi}^I$, only gauge interactions remain. However, since $\tilde{\Phi}^I$ 
is a gauge singlet, it will neither interact via gauge interactions nor contribute to the running of the gauge couplings in the effective LR-symmetric model.

\section{Breaking of $\SU{2}{R} \times \U{\BL}$ in the effective model} \label{sec:RadBreak}

In order to reproduce the phenomenology of the SM at low energies, the gauge $\SU{2}{R} \times \U{\BL}$ subgroup needs to be broken to the SM hypercharge group $\U{Y}$. One of the persistent issues in high energy models is the fact that the vastly different energy scales associated have to be given through input parameters. One way of dealing with this issue is to introduce the possibility of radiative symmetry breaking, i.e SSB triggered by the RG evolution of the model. This is a standard way of understanding, for example, EW symmetry breaking in the MSSM where the running of $m_{H_u}^2$ drives the breaking of $\SU{2}{L} \times \U{Y}$ \cite{Dedes:1995sb,Gamberini:1989jw, Carena:1993ag}. The question remains as to whether this model offers a possibility of breaking $\SU{2}{R} \times \U{\BL}$ through the RG running of $m_R^2$ and the rest of parameters of the effective model with initial conditions coming from tree-level matching of the high energy theory. This will give us the possibility of checking under which conditions (for the high-energy input parameters) this radiative breaking can be induced. 

For this purpose, we consider two separate scenarios. In scenario I, we study the properties of a minimum in the scalar potential of the effective LR-symmetric model where $\SU{2}{R} \times \U{\BL}$ is broken to the analogous of $\U{Y}$ in the SM, (i.e.~the EW gauge group \mbox{$\SU{2}{L} \times \U{Y}$} is unbroken in this minimum). To study the model at the electro-weak scale, where $\SU{2}{L} \times \U{Y}$ is broken, a second step of matching and running would have to be performed to discover whether there is a sign change of the Higgs squared mass parameter inducing the electro-weak symmetry breaking.
In scenario II, we instead study minima with a more complicated VEV structure such that $\SU{2}{L} \times \SU{2}{R} \times \U{\BL}$ is directly broken down to $\U{E.M.}$. This is accomplished by suitable VEVs both in $\tilde{h}$ and in $\tilde{l}_\mathrm{R}$.  The VEV in $\tilde{l}_\mathrm{R}$ needs to be larger than the Higgs VEV to keep the $W'$ and $Z'$ bosons heavy, but not too large as to ruin the convergence the perturbative expansion through large logarithms of the ratio between the two VEVs. If this ratio would become too large, Scenario I is more appropriate.

\subsection{Scenario I: Breaking to the SM gauge group}

Let us first understand what are the conditions necessary for $\SU{2}{R} \times \U{\BL}$ breaking through non-vanishing VEV for the scalar field $\tilde{l}_\mathrm{R}$, 
\begin{equation}
\langle \Lr{I}{R} \rangle = \delta^I_2 \delta_R^2 \frac{w}{\sqrt{2}}	= \begin{pmatrix}
0   & 0      \\
0   & \frac{w}{\sqrt{2}} 
\end{pmatrix}  ,	\label{LRvev}
\end{equation}
where $w$ is taken to be real. The extremal condition for such a  VEV setting reads 
\begin{equation}
-m_R^2 =  \lambda_b w^2.		\label{LRTP}
\end{equation}
This leaves the following gauged $\U{}$ generator unbroken,
\begin{equation}
\T{Y}{} = \T{R}{3} + \frac{1}{2} \T{\BL}{},
\end{equation}
which can be identified as the SM hypercharge generator. In addition, Eq.~\eqref{LRvev} leaves four global $\U{}$ generators unbroken,
\begin{equation}\label{TCDGen}
\T{D}{} \equiv \frac{1}{2}\left( \T{X}{}-\T{\BL}{} \right)	\, , \quad	 \T{E}{} \equiv \T{F}{3} + \frac{1}{2} \T{Z}{} \, ,	\quad \T{G}{} \equiv \frac{1}{2} \left( \T{Z}{} + \T{\BL}{} \right)	\, , \quad \T{B}{} \, .				
\end{equation}
Thus, the VEV \eqref{LRvev} breaks the LR symmetry group \eqref{eq:LRgaugegroup} down to 
\begin{equation}
\SU{3}{C} \times \SU{2}{L} \times \U{Y}  \times \{ \U{D} \times \U{E} \times \U{G} \times \U{B} \}.
\end{equation}
which will also be the symmetry group of a SM-like effective model that is obtained when the heavy particles in the effective LR-symmetric model are integrated out. 

The scalar mass eigenstates in the minimum described by Eq.~\eqref{LRTP} are shown in Tab.~\ref{tab:LRBrokenMassStates}. Notice, in particular, that the scalar spectrum contains one massless complex d.o.f.~that is the ``global'' Goldstone boson from the breaking of the global part of the LR symmetry. We expect this to decouple at scales $\mu \ll w$, similarly to how $\tilde{\Phi}^I$ decouples for $\mu \ll v$.
\begin{table}[tbp]
  \begin{center}
    \begin{tabular}{llll}
     \toprule                     
       Fields    						& (Mass)$^2$ 											& 	$(Y,D,E)$			& Comment			\\      
      \midrule
       $\hh{L}{1}$						& $m_{h_2}^2 \equiv m_h^2 + \frac{1}{2} (\lambda_g - \lambda_f) w^2$		& 	$(-1/2,+1,0)$		&					\\
       $\hh{L}{2}$						& $m_{h_1}^2 \equiv m_h^2 - \frac{1}{2} \lambda_f w^2$					& 	$(+1/2,+1,0)$		&					\\
       $\Lr{1}{1}$						& $m_{r_2}^2 \equiv 2 \lambda_i w^2$										& 	$(-1,0,+1)$			&					\\
       $\mathrm{Re}[\Lr{2}{2}]$			& $m_{r_1}^2 \equiv 2 \lambda_b w^2$										& 	$(0,0,0)$			&					\\
       $\Lr{2}{1}$						& $0$													& 	$(-1,0,0)$			& Gauge Goldstone	\\
       $\mathrm{Im}[\Lr{2}{2}]$			& $0$													& 	$(0,0,0)$			& Gauge Goldstone	\\
       $\Lr{1}{2}$						& $0$													& 	$(0,0,+1)$			& Global Goldstone	\\
     \bottomrule
    \end{tabular}
    \caption{Mass eigenstates in $\hh{}{}$ and $\tilde{l}_\mathrm{R}$ after SSB of the LR symmetry by $\tilde{l}_\mathrm{R}$ VEV, the corresponding tree-level masses and $\U{}$ charges. All scalars are uncharged under the $\{\U{G} \times \U{B}\}$ global symmetry.}
    \label{tab:LRBrokenMassStates}
  \end{center}
\end{table}

The gauge fields after SSB of the LR symmetry are mixed and give rise to the massive and massless states shown in Tab.~\ref{tab:LRBrokenGaugeStates}. The massless states are the gauge fields of the unbroken gauge symmetries, and the massive gauge fields are the combinations
\begin{equation}
W'^{\pm}{}_\mu = \frac{1}{\sqrt{2}}\left( G_\mathrm{R}{}^1 \mp \i G_\mathrm{R}{}^2 \right)_\mu \, , \quad Z'^0{}_\mu = \cos{\theta_V} \, G_{\mathrm{\BL}}{}_\mu - \sin{\theta_V}\, G_\mathrm{R}{}^3_\mu
\end{equation}
with $\tan \theta_V = \g{R} / 2 \g{\BL} $. The hypercharge gauge coupling $\g{Y}$ becomes 
\begin{equation}
\g{Y} = \frac{2 \, \g{R} \, \g{\BL}}{\sqrt{4 \g{\BL}^2 + \g{R}^2}}.
\end{equation} 

\begin{table}[tbp]
  \begin{center}
    \begin{tabular}{llll}
     \toprule                     
       Fields    															& (Mass)$^2$ 				&  $Y$ 		& Comment								\\      
      \midrule
       $G_{\mathrm{C}}{}^{1,\dots, 8}_\mu$													& $0$						& 	$0$		& Gauge field of $\SU{3}{C}$	\\
       $G_{\mathrm{L}}{}^{1,2,3}_\mu$														& $0$						& 	$0$		& Gauge field of $\SU{2}{L}$				\\
       $s_{\theta_V} G_{\mathrm{\BL}}{}_\mu + c_{\theta_V} G_{\mathrm{R}}{}^3_\mu $ 						& $0$						& 	$0$		& Gauge field of $\U{Y}$		\\
       $W'^\pm{}_\mu$														& $m_{W'}^2 \equiv \frac{1}{4} \g{R}^2 w^2$	& 	$\pm 1$		&					\\
       $Z'^0{}_\mu$ 														& $m_{Z'}^2 \equiv \left( \g{\BL}^2 + \frac{1}{4} \g{R}^2 \right) w^2$	& 	$0$				&					\\
     \bottomrule
    \end{tabular}
    \caption{Gauge boson mass eigenstates and their hypercharges after the spontaneous LR symmetry breaking. All gauge bosons are uncharged under the global $\{\U{D}\times\U{E}\times\U{G}\times\U{B}\}$ group. Here, $c_{\theta_V} \equiv \cos \theta_V$ and $ s_{\theta_V}\equiv \sin \theta_V$.}
       \label{tab:LRBrokenGaugeStates}
  \end{center}
\end{table}

For general values of the Yukawa couplings in Eq.~\eqref{eq:LRYukawa}, many fermion fields become massive once $\langle \Lrt \rangle \neq 0$. However, in the approximation employed in this work, i.e. tree-level matching and one-loop RG evolution, most of the Yukawa couplings are zero and only one Dirac fermion becomes massive in the effective LR-symmetric model (shown in Tab.~\ref{tab:LRBrokenFermions}). This fermion is a Dirac state
\begin{equation}
\mathcal{D}_\mathrm{s}{}^c \equiv\left( \begin{array}{c}
D_\mathrm{L}^s{}^c 	\\
\QREd{1}{2}{c} 	
\end{array} \right)	
\end{equation}
and, when integrated out, leaves behind three $\SU{2}{L}$ doublet quarks and six $\SU{2}{L}$ singlet quarks that will make up the SM quark sector. 

\begin{table}[htbp]
  \begin{center}
    \begin{tabular}{llll}
     \toprule                     
       Fields    								& (Mass)$^2$ 												&  $(Y,D,E)$ 		& 	\\      
      \midrule
       $\mathcal{D}_{\mathrm{s}}{}^c $							& $ m^2_q \equiv \frac{1}{2} Y_\gamma^2 w^2 $						& 	$(+1/3,2/3,-1/3)$		&		\\
     \bottomrule
    \end{tabular}
    \caption{Massive quark field and its quantum numbers after the $\SU{2}{R}\times\U{\BL}$ breaking.}
    \label{tab:LRBrokenFermions}
  \end{center}
\end{table}

\subsection{Scenario II: Breaking directly to $\U{E.M.}$}\label{sec:breakingtoEM}

A second possible scenario is that in which both $\Lr{I}{R}$ and $\hh{L}{R}$ develop VEVs, which would directly trigger $\SU{2}{L}\times \SU{2}{R} \times \U{\BL} \rightarrow \U{E.M.}$. Although this case is certainly allowed by the model, and might also be triggered by RG running of effective Lagrangian parameters, there is no a priori reason that there will be any hierarchy in the VEVs of the fields consistent with the SM. This could lead to several massive gauge bosons with comparable masses which would immediately be in conflict with what is observed experimentally. So far, only the SM $W^\pm$ and $Z^0$ bosons have been observed, and the existing LHC bounds on extra gauge bosons \cite{wprimeresults,zprimeresults} would force an unnatural hierarchy which is precisely the problem one wants to avoid by means of radiative breaking. 

Nevertheless, let us explore the conditions under which such symmetry breaking might lead to an unbroken $\U{E.M.}$. The most general VEV setting, after accounting for gauge symmetries for $\Lr{I}{R}$ and $\hh{L}{R}$ would be:
\begin{equation} \label{eq:Minimum2VEV}
\langle \Lr{I}{R} \rangle=  \frac{1}{\sqrt{2}}
\begin{pmatrix}
w_1 & w_2 \\
0   & w_3
\end{pmatrix}   
\qquad
\langle\hh{L}{R}\rangle= \frac{1}{\sqrt{2}}
\begin{pmatrix}
v_1& 0 \\
0 & v_2
\end{pmatrix}.   
\end{equation}
However, not all of these minima lead to a $\U{E.M.}$ remaining gauge symmetry consistent with the proposed framework. To understand this, it is useful to think as if the VEVs are attained sequentially. Let us assume $\Lr{I}{R}$ gets its VEV first. From our previous analysis we know that in order to identify $\T{E.M.}{}= \T{L}{3} + \T{Y}{} $ we need $\hh{L}{R}$ to break down to two $\SU{2}{L}$ doublets with opposite hypercharge. This is only possible with $w_1=w_2=0$ (up to any symmetry transformation on $\langle \tilde{l}_\mathrm{R} \rangle$). Once $\Lr{I}{R}$ has obtained its VEV, there are physically different VEV settings in the Higgs bi-doublet we should explore separately:  $v_1 = v  \neq 0$ and  $v_2 =0$ (Case A),  $v_2 = v  \neq 0$ and $v_1=0$ (Case B), and $v_1 \neq 0$ and $v_2 \neq 0$ (Case C). In the following we will explore the details of Case A and B. Case C, where both Higgs VEVs are non-zero, is cumbersome and can be left for a future work. The reason for this is that for that case the masses in the scalar sector can not be obtained analytically and the type of analysis we will do in Sect.~\ref{sec:SA} is not feasible without further work. We will thus take a closer look at scenario I and the first two cases described above. 

\subsubsection{Case A}\label{sec:caseA}

Given the VEV setting
\begin{equation} \label{eq:Minimum2VEV_caseA}
\langle \Lr{I}{R} \rangle=
\begin{pmatrix}
0 & 0 \\
0   &  \frac{w}{\sqrt{2}}
\end{pmatrix}   \, ,
\qquad
\langle\hh{L}{R}\rangle=
\begin{pmatrix}
0 & 0 \\
0 &  \frac{v}{\sqrt{2}}
\end{pmatrix},   
\end{equation}
the extremal conditions become 
\begin{equation}
\left( m_h^2 + v^2 \lambda_a - \frac{1}{2} w^2 \lambda_f \right) v = 0 \, , \quad \left(m_R^2 + w^2 \lambda_b - \frac{1}{2} v^2 \lambda_f \right) w = 0.
\end{equation}
With $v \neq 0 $ and $w \neq 0$, there is only one unbroken gauge symmetry generator
\begin{equation}
\T{E.M.}{}=\T{L}{3} + \T{R}{3}+ \frac{1}{2} \T{\BL}{} 
\end{equation}
which corresponds exactly to the generator of $\U{E.M.}$ in our previous analysis with only $\Lr{I}{R}$ VEV. There are also two new $\U{}$ global symmetries in addition to $\{ \U{\Extraone} \times \U{\Extratwo} \}$ with generators
\begin{equation}
\T{V}{} = \T{X}{} - \T{\BL}{} + 4 \T{L}{3}   \, , \quad \T{W}{} = \T{F}{3}- \frac{1}{2}\T{\BL}{}
\end{equation}

The mass eigenstates in the scalar sector after this symmetry breaking are shown in Tab.~\ref{tab:LRBrokenMassStatesA}. In particular, there is one real state which obtains a mass of $\mathcal{O}(v)$ when $v \ll w$, which would be the candidate for the 125 GeV SM Higgs particle.

\subsubsection{Case B}\label{sec:caseB}

A second possible case follows from the VEV assignment

\begin{equation} \label{eq:Minimum2VEV_caseB}
\langle \Lr{I}{R} \rangle=
\begin{pmatrix}
0 & 0 \\
0   & \frac{w}{\sqrt{2}}
\end{pmatrix}   
\qquad
\langle\hh{L}{R}\rangle=
\begin{pmatrix}
 \frac{v}{\sqrt{2}} & 0 \\
0 & 0
\end{pmatrix}.   
\end{equation}
The extremal conditions in this case will be
\begin{equation}
\left( m_H^2 + v^2 \lambda_a + \frac{1}{2} w^2(-\lambda_f+ \lambda_g) \right) v = 0 \, , \quad \left( m_R^2 + w^2 \lambda_b + \frac{1}{2} v^2(-\lambda_f+ \lambda_g) \right) w = 0.
\end{equation}

The mass eigenstates in the scalar sector after this symmetry breaking are shown in Tab.~\ref{tab:LRBrokenMassStatesB}, where as in the previous case the spectrum contains a candidate for the SM Higgs particle. The VEV setting \eqref{eq:Minimum2VEV_caseB} leaves the same gauged $\U{E.M.}$ and global $\U{W}$ unbroken as the vev setting in Eq.~\eqref{eq:Minimum2VEV_caseA}. However, the $\U{V}$ is replaced by $\U{V'}$ which is generated by
\begin{equation}
\T{V'}{} = \T{X}{} -\T{\BL}{} -4 \T{L}{3} \,.
\end{equation}

\begin{table}[htbp]
  \begin{center}
       \begin{tabular}{llll}
     \toprule                     
       Fields    												& (Mass)$^2$ 													& Comment			\\      
      \midrule
       $s_\alpha \Lr{2}{1} - c_\alpha \hh{2}{1}$				& $\frac{1}{2} (v^2 + w^2) \lambda_g$							&					\\
       $\Lr{1}{1}$												& $\frac{1}{2} v^2 \lambda_g + 2 w^2 \lambda_i$				&					\\
       $\hh{1}{1}$												& $\frac{1}{2} w^2 \lambda_g + 2 v^2 \lambda_j$				&					\\
       $c_\eta \Re[\hh{2}{2}] + s_\eta \Re[\Lr{2}{2}]$			& $v^2 \lambda_a + w^2 \lambda_b + \sqrt{\ldots}$				& 					\\
       $ c_\eta  \Re[\Lr{2}{2}] - s_\eta \Re[\hh{2}{2}]$		& $v^2 \lambda_a + w^2 \lambda_b - \sqrt{\ldots}$				& 	 $\sim \mathcal{O}(v^2)$ for  $\tan \alpha \sim 0$	\\
       $\Lr{1}{2}$												& $0$															& Global Goldstone\\
       $\hh{1}{2}$												& $0$															& 	 Gauge Goldstone\\
       $\Im[\hh{2}{2}]$											& $0$															& Gauge Goldstone	\\
       $\Im[\Lr{2}{2}]$											& $0$															& Gauge Goldstone\\
       $c_\alpha \Lr{2}{1} + s_\alpha \hh{2}{1}$				& $0$															&  Gauge Goldstone 	\\
     \bottomrule
    \end{tabular}    \caption{Case A: Mass eigenstates in $\hh{}{}$ and $\tilde{l}_\mathrm{R}$ after SSB of the LR symmetry group to $\U{E.M.}$ and the corresponding tree-level masses. Here, $\sqrt{\ldots}=\sqrt{(v^2 \lambda_a - w^2 \lambda_b)^2+(v w \lambda_f)^2}$,  $c_\alpha = \cos \alpha, \, s_\alpha =\sin \alpha $ with $\tan \alpha = v / w$ and $c_\eta = \cos \eta, \, s_\eta =\sin \eta $ with $\eta$ being the corresponding mixing angle whose explicit form we omit for simplicity.}
   \label{tab:LRBrokenMassStatesA} 
  \end{center}
\end{table}

\begin{table}[htbp]
  \begin{center}
    \begin{tabular}{llll}
     \toprule                     
       Fields    												& (Mass)$^2$ 																							& Comment			\\      
      \midrule
       $c_\alpha \Re[\hh{1}{2}]+ s_\alpha \Re[\Lr{2}{1}]$						& $-\frac{1}{2} (v^2 + w^2) \lambda_g$																	&					\\
       $s_\alpha \Im[\Lr{2}{1}] - c_\alpha \Im[\hh{1}{2}]$							& $-\frac{1}{2} (v^2 + w^2) \lambda_g$																&					\\
       $\Lr{1}{1}$												& $-\frac{1}{2} v^2 \lambda_g + 2 w^2 \lambda_i$															&					\\
       $\hh{2}{2}$												& $-\frac{1}{2} w^2 \lambda_g + 2 v^2 \lambda_j$															&					\\
       $ c_\kappa  \Re[\hh{1}{1}] + s_\kappa  \Re[\Lr{2}{2}] $					& $v^2 \lambda_a + w^2 \lambda_b + \sqrt{\ldots}$			&					\\
       $ c_\kappa \Re[\Lr{2}{2}]  -  s_\kappa  \Re[\hh{1}{1}] $						& $v^2 \lambda_a + w^2 \lambda_b - \sqrt{\ldots}$			& $\sim \mathcal{O}(v^2)$ for  $\tan \alpha \sim 0$ \\
       $\Lr{1}{2}$												& $0$																								& Global Goldstone \\
       $\hh{2}{1}$												& $0$																								& Gauge Goldstone 	\\
       $\Im[\hh{1}{1}]$											& $0$																							& Gauge Goldstone 	\\
       $\Im[\Lr{2}{2}]$											& $0$																							& Gauge Goldstone\\
       $-s_\alpha \Re[\hh{1}{2}] +c_\alpha \Re[\Lr{2}{1}]$						& $0$																& Gauge Goldstone	\\
       $c_\alpha \Im[\Lr{2}{1}] + s_\alpha \Im[\hh{1}{2}] $						& $0$																	& 	Gauge Goldstone\\
     \bottomrule
    \end{tabular}
    \caption{Case B: Mass eigenstates in $\hh{}{}$ and $\tilde{l}_\mathrm{R}$ after SSB of the LR symmetry group directly to $\U{E.M.}$ and the corresponding tree-level masses. Here, $\sqrt{\ldots}=\sqrt{(v^2 \lambda_a - w^2 \lambda_b)^2+v^2 w^2(\lambda_g-\lambda_f)^2}$,  $c_\alpha = \cos \alpha, \, s_\alpha =\sin \alpha $ with $\tan \alpha = v / w$ and $c_\kappa = \cos \kappa, \, s_\kappa =\sin \kappa $ with $\kappa$ being the corresponding mixing angle whose explicit form we omit for simplicity.}
      \label{tab:LRBrokenMassStatesB}
  \end{center}
\end{table}

\section{Numerical results} \label{sec:Results}

The main question to answer for the proposed framework is whether for a consistent set of parameters of the trinification theory, the RG running in the effective LR-symmetric theory can trigger the radiative breaking of $\SU{2}{R} \times \U{\BL}$, and for what regions in parameter space this happens. In addition, we will explore under which circumstances we can get close to a realistic SM-like scalar sector, with a light $\SU{2}{L}$ scalar doublet with hypercharge $Y=+1/2$ remaining in the spectrum at lower energies (which potentially can induce EW symmetry breaking). The resulting low-scale mass spectrum after the radiative symmetry breaking will depend on our choice of initial parameters only, but the connection between the initial values of the parameters and the resulting mass spectrum is not obvious. In order to explore it, we implemented a parameter scanning framework using numerical integration of the RG equations together with a simulated annealing (SA) procedure to scan over the possible initial values of high-scale parameters.

We calculated one-loop $\beta$-functions for the effective LR-symmetric model using the package \texttt{pyr@te} \cite{Lyonnet:2013dna}, which are written in Appendix~\ref{sec:RGEs}.

\subsection{Parameter scan}

Effectively, we would like to explore a five-dimensional parameter subspace of the high-scale model $\lbrace \lambda_3, \varepsilon, \delta, g, y \rbrace$ assuming we have fixed the scale at which trinification is broken and imposed the constraints in Eq.~\eqref{eq:conditionsLam}. This is due to the fact that in the effective LR-symmetric model after tree-level matching, the $\beta$-functions only depend on those parameters as seen in Eq.~\eqref{eq:matchingcond}. Once a consistent set of high-scale model parameters is found, the matching can be performed and the RG equations can be numerically integrated yielding a scale dependence of the effective model parameters. The running starts from the matching scale $\mu_m$, which is chosen to be the trinification breaking VEV,
\begin{equation}
\mu_m = v_3,
\end{equation} 
since the heavy states in the trinification theory that we integrate out have masses of $\mathcal{O}(v_3)$. The running is then terminated at a lower scale $\mu_r$, which is defined as
\begin{equation}\label{eq:mur}
\mu_{r} =\sqrt{\frac{ |m_R^2(\mu_r)| + |m_h^2(\mu_r)|}{ 2}},
\end{equation}
since, at this scale, there are again states with masses of the same order as the renormalisation scale. These states then have to be integrated out before we can run down even further. Depending on the initial values at a high scale, $m_R^2$ may have run negative at this scale, thus triggering the radiative symmetry breaking we are looking for. However, we have to guarantee that, at the stopping scale $\mu_r$, the minimisation conditions for the VEV setting described in Eq.~\eqref{LRvev} are satisfied (i.e. that all the squared masses in Tab.~\ref{tab:LRBrokenMassStates} are positive). 

\subsubsection{Simulated annealing}\label{sec:SA}
In order to find viable parameter space points in the high-scale theory, we implemented the SA algorithm together with the numerical integration of $\beta$ functions. The SA is a method for estimating the global minimum of a given function $E(\lbrace p_i \rbrace)$ in a multi-dimensional parameter space $\lbrace p_i \rbrace$ \cite{Kirkpatrick671}. 

If we interpret the function $E(\lbrace p_i \rbrace)$ as the energy of a system whose physical state is defined by $\lbrace p_i \rbrace$, and imagine that the system is in thermal contact with a heat bath with temperature $T$, we can let this system approach its equilibrium state by employing the Metropolis algorithm. That is, we start with a random set of initial parameters, and propose random updates $\lbrace p_i \rbrace \rightarrow \lbrace p_j' \rbrace$ that are accepted with probability
\begin{equation}
P_\mathrm{acc}(\lbrace p_i \rbrace \rightarrow \lbrace p'_i \rbrace) = 
\begin{cases} 
1 &\mbox{if }  E(\lbrace p'_i \rbrace) < E( \lbrace p_i \rbrace) \\ 
\mathrm{e}^{(E( \lbrace p_i \rbrace )-E( \lbrace p'_i\rbrace )) / T} & \mbox{otherwise, }
\end{cases} \,.
\end{equation} 
Given a constant $T$, this procedure fulfils detailed balance w.r.t.~the canonical ensemble $\mathcal{P}(\lbrace p_i \rbrace ) \propto \mathrm{e}^{- E( \lbrace p_i \rbrace)/T}$, which in the limit $T\rightarrow \infty$ is a flat distribution where all $\lbrace p_i \rbrace$ are equally likely, while in the limit $T\rightarrow 0$ becomes highly peaked for the ground states of the system, i.e. the states $ \lbrace p_i \rbrace $ that minimise $E( \lbrace p_i \rbrace)$. 

The SA works by initialising the system at a large temperature, and then running the Metropolis algorithm while slowly (i.e.~adiabatically) decreasing the temperature until \mbox{$T \sim 0$}. In this way, $E(\lbrace p_i \rbrace)$ is minimised and the corresponding parameter space points $\lbrace p_i \rbrace$ are found. This procedure has the advantage of being easy to implement while also being less prone to get stuck in local minima compared to for example a gradient descent method since local energy barriers can be overcome by ``thermal fluctuations''. 

For the purpose of this work we defined 
\begin{equation}
E =  
\begin{cases} 
10 &\mbox{if } m_R^2(\mu) > 0 \, \forall \, \mu \,  \in \, (m_Z, \mu_{m})\\ 
\\
5 + \frac{\min(m^2_i)}{ \max(|m^2_i|)} &  \mbox{if } m^2_j < 0 \, \mbox{ for some } j\\ 
\\
2 \frac{\min(m^2_{h_i})}{\min(m_q^2, m^2_{Z'}, m^2_{W'})} + \frac{\min(m^2_{h_i})}{\max(m_q^2, m^2_{r_2}, m^2_{Z'}, m^2_{W'})} & \mbox{if } m_R^2(\mu_r) < 0 \mbox{ and } m^2_j(\mu_r) > 0 
\end{cases}
\\
\end{equation}
where $E = E( \lambda_3, \epsilon, \delta, g, y)$ and $m^2_i = (m_q^2, m^2_{h_i}, m^2_{R_i}, m^2_{Z'}, m^2_{W'})$ are the masses after radiative symmetry breaking evaluated at the scale $\mu_r$. Minimisation of this function guarantees that we find parameter space points where $m_R^2$ runs negative while also introducing a bias towards parameters that yield a light Higgs-like $\SU{2}{L}$ doublet.

\subsubsection{Choosing the trinification breaking scale}

\begin{figure}[htbp] 
  \centering
  \includegraphics[width=0.9\linewidth, natwidth=610,natheight=642]{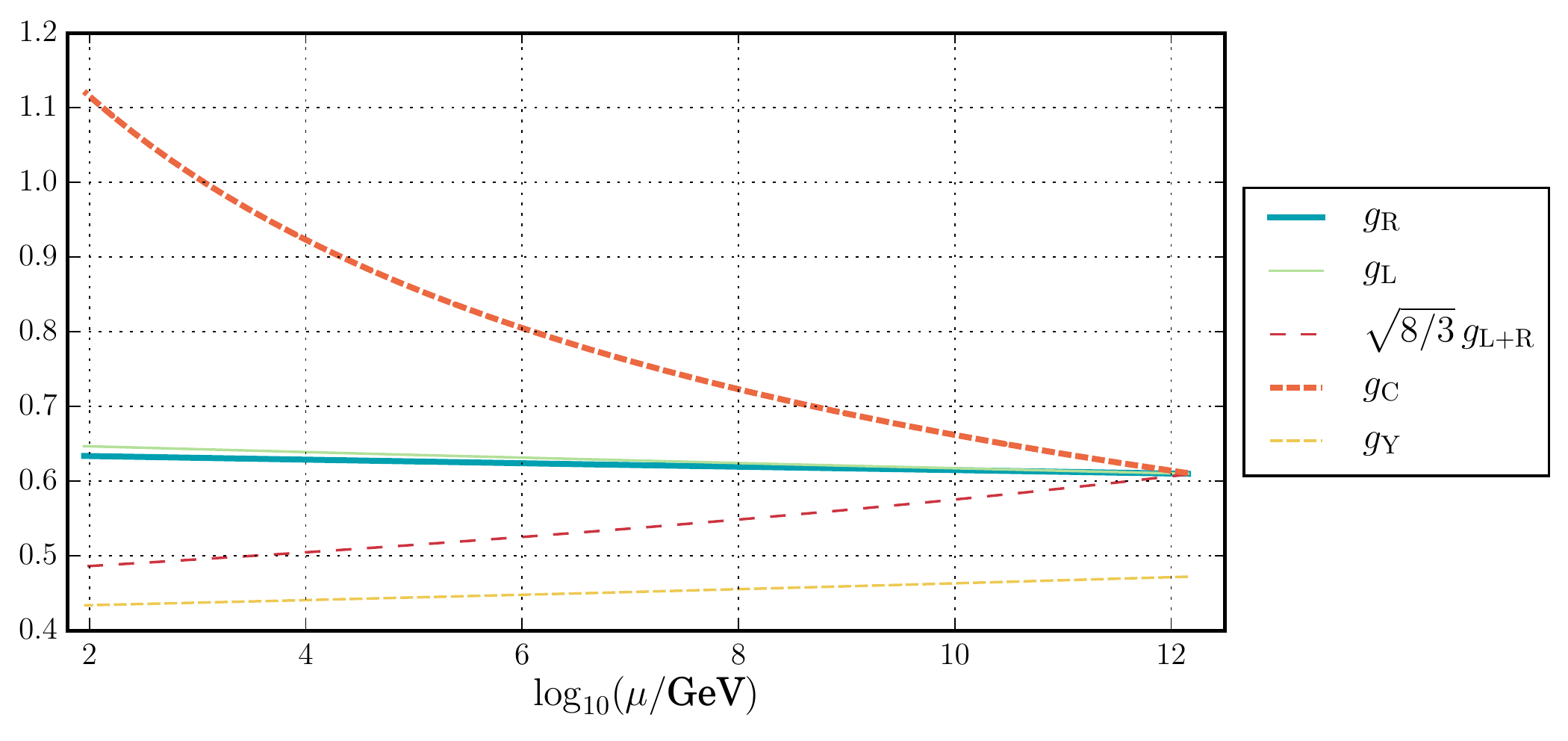}
  \caption{One-loop RG evolution of gauge couplings in the effective LR-symmetric model and matching condition for the hypercharge coupling $\g{Y}$ with $g_0=0.61$. A trinification breaking scale of $\mu_m \approx 10^{12.2}$ \gev leads to roughly SM values for $\g{L} \equiv \g{2}$, $\g{C}\equiv\g{3}$ and $\g{Y}\equiv\g{1}$ at $\mu \approx m_Z$.}
  \label{fig:gaugeplot}
\end{figure}

One of the free parameters of the proposed framework is the scale at which trinification symmetry is spontaneously broken. This scale, which is the starting point for all the successive symmetry breakings at low scales, is defined only by the trinification breaking VEV \eqref{eq:vevStructure}. In order to get an idea of what scales are sensible to explore, we integrated the one-loop RG equations for gauge couplings in the effective LR-symmetric model, an easy task due to the fact that at one-loop the $\beta$-functions only depend on the gauge couplings themselves. It was possible then to relate the trinification breaking scale to the measured values of the SM $\SU{2}{L}$, $\U{Y}$ and $\SU{3}{C}$ gauge couplings. We found that for a trinification breaking scale of $\mu_m=10^{12.2} \mbox{ GeV}$, the boundary condition of $g_0 = 0.61$ leads roughly to the SM values at $m_Z$. In Fig.~\ref{fig:gaugeplot} we show the result from integrating the gauge coupling $\beta$-functions from $\mu_m$ down to $m_Z$. Because the $\beta$-functions only depend on the gauge couplings at one-loop, this running is valid for any parameter space point with $g_0=0.61$ as boundary condition. We show in the same plot  
\begin{equation}
\g{Y}(\mu) \equiv \frac{2 \, \g{R} \, \g{\BL}}{\sqrt{4 \g{\BL}^2 + \g{R}^2}}
\end{equation}
which would be the matching condition for $g_1$, the hypercharge gauge coupling, as function of the scale $\mu$. Note however that in Fig.~\ref{fig:gaugeplot}, the running is performed all the way down to the EW scale $m_Z$, without decoupling the massive states at $\mu_r$. A more accurate calculation would implement this intermediate step, which would alter the slopes of the lines in Fig.~\ref{fig:gaugeplot} at scales below $\mu_r$. Therefore, this calculation should only serve as a very rough estimate of the numerical values of the matching scale (and the value of the trinification gauge coupling at this scale). 

\subsection{Regions of parameter space with radiative breaking}
\begin{figure}[htbp] 
  \centering
  \hspace{-10pt}
  \includegraphics[width=0.46\linewidth, height= 0.358 \linewidth, natwidth=610,natheight=642]{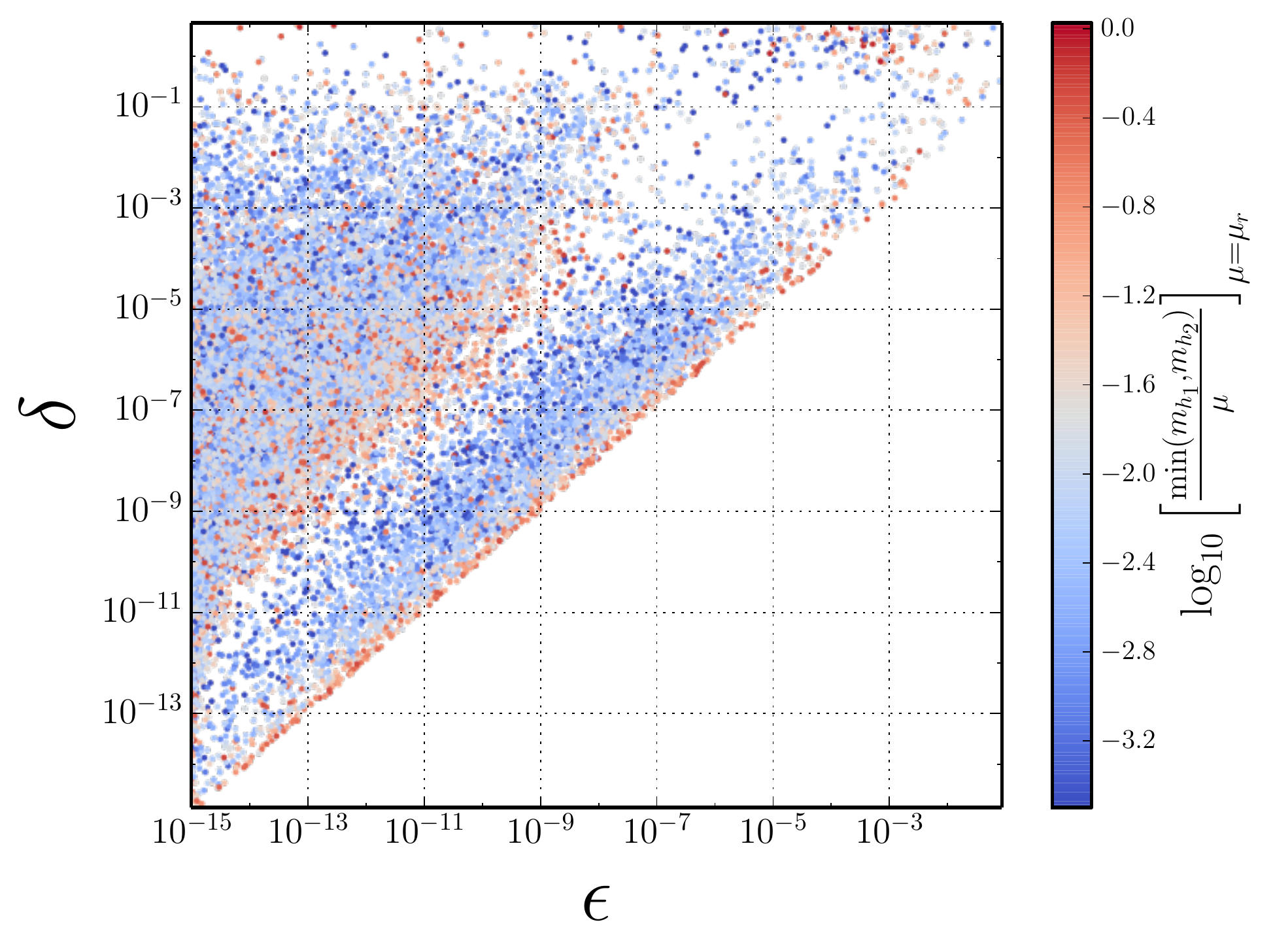} 
  \includegraphics[width=0.45\linewidth, natwidth=610,natheight=642]{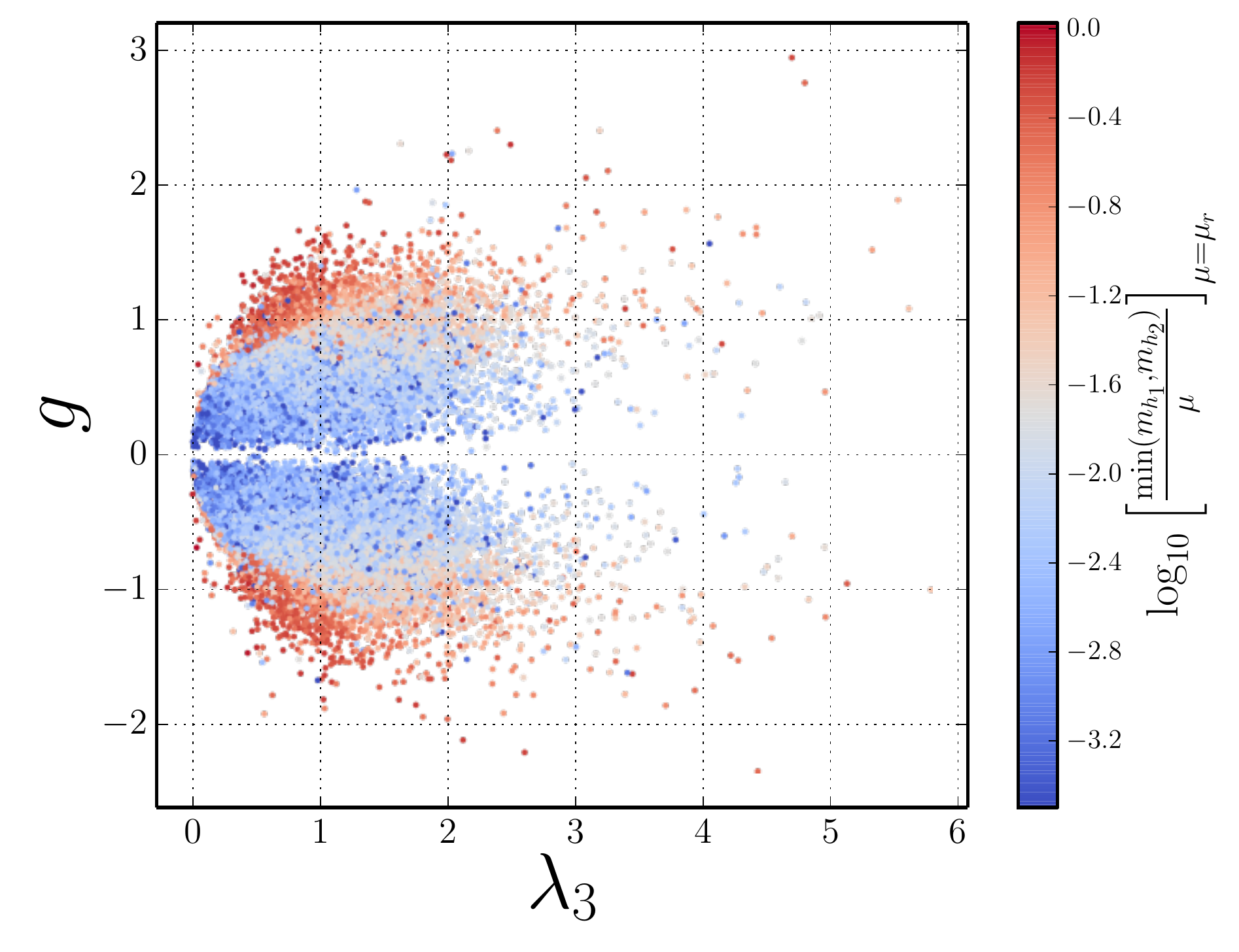} \\
  \includegraphics[width=0.45\linewidth, natwidth=610,natheight=642]{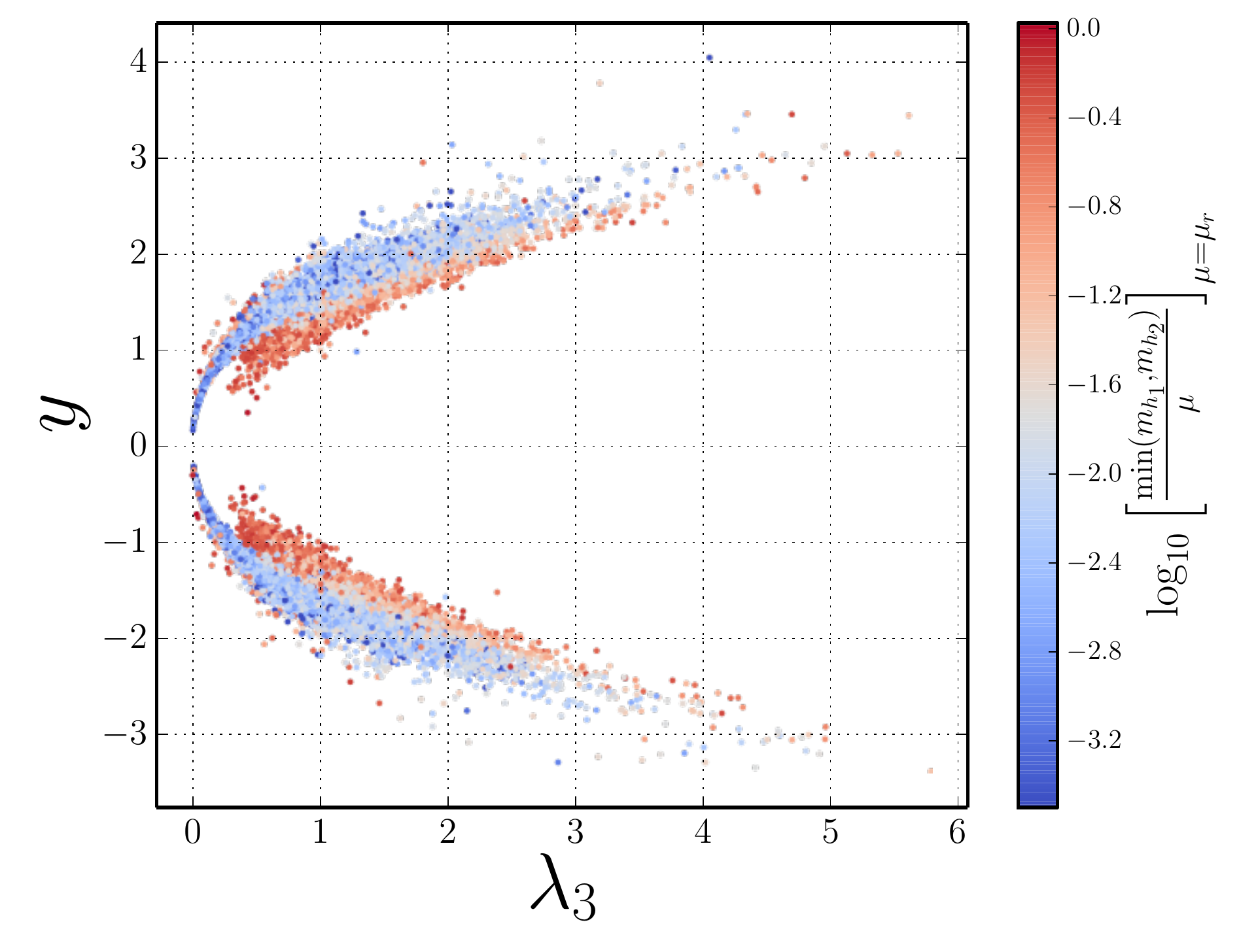} 
  \includegraphics[width=0.45\linewidth, natwidth=610,natheight=642]{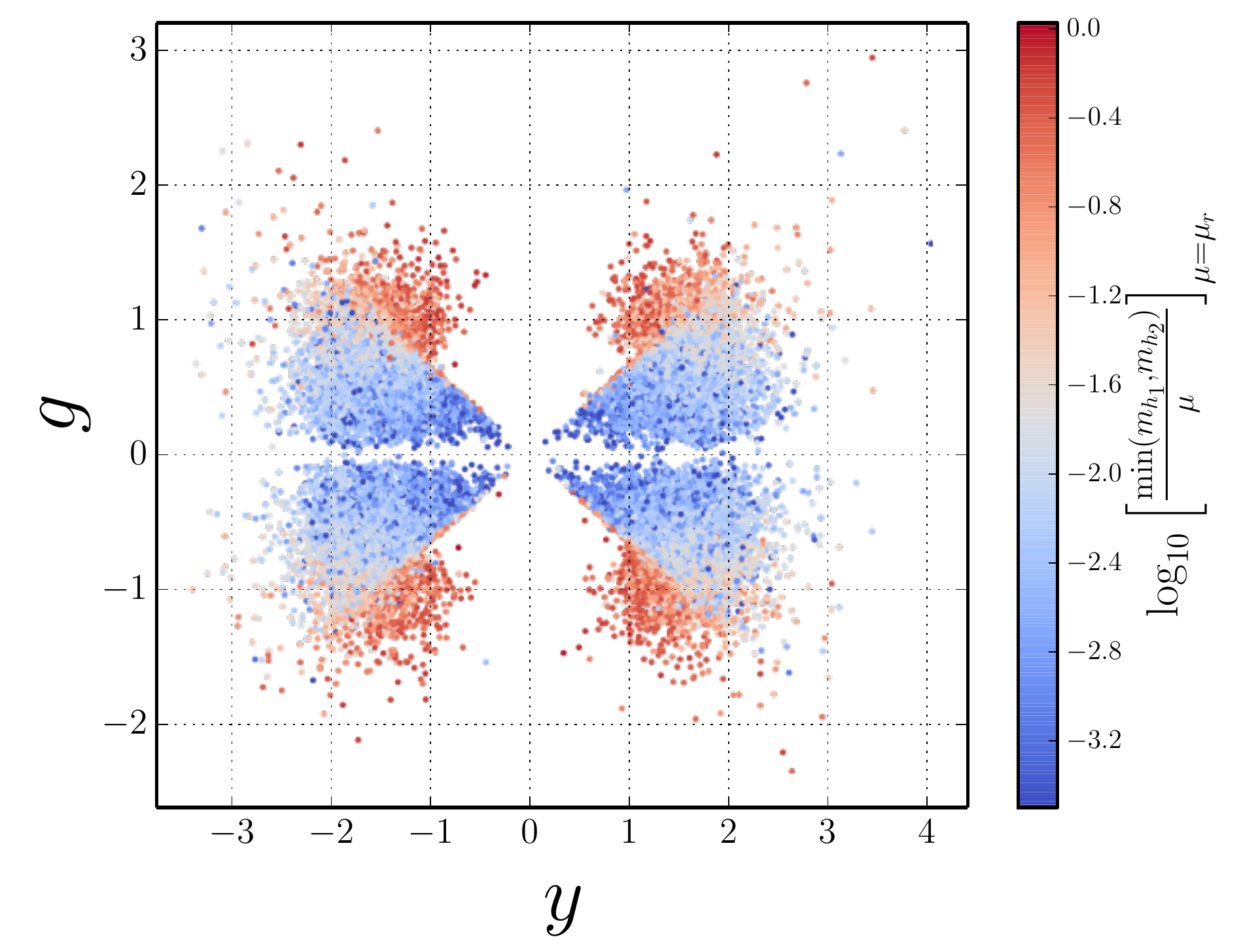} \\  
  \caption{Regions of the high-scale model parameter space with the radiative $\SU{2}{R} \times \U{\BL}$ symmetry breaking down to $\U{Y}$ 
  by RG evolution in the effective LR-symmetric model. The ranges for the scan were chosen such as to preserve the unitarity property at tree-level. The colours indicate the lightness of the lightest Higgs doublet compared to the renormalization scale at the stopping scale $\mu_r$. }
  \label{fig:allowedparameters}
\end{figure}

Using the framework described above we found $22081$ parameter space points by running our implementation of the SA algorithm allowing for high-scale parameters within the unitarity bounds and for a trinification breaking scale of $\mu_m=10^{12.2} \mbox{ GeV}$. Remarkably, we found that the considered model naturally contains large parameter 
space regions where $\SU{2}{R}\times\U{\BL}$ is radiatively broken down to $\U{Y}$ while a light Higgs doublet remains in the spectrum at the stopping scale $\mu_r$. In Fig.~\ref{fig:allowedparameters} we show the allowed regions in several slices of the high-scale parameter space. Most of the features of these regions can be explained by the structure of the mass-parameter $\beta$-functions in Eqs.~\eqref{eq:betamuR} and \eqref{eq:betamuh}. Our algorithm selects points where $m^2_R$ would have a positive $\beta$-function so that it could run to negative values at low scales. For example, we see that the allowed parameter space region always satisfies $\epsilon < \delta$ which translates into the inequality $m_R^2 < m_h^2$ at the matching scale. Although one can find points for which \eqref{eq:betamuR} is positive and $\epsilon > \delta$, for such points $m_h^2$ value runs negative before $m_R^2$ does. This would trigger an unwanted simultaneous breaking of $\SU{2}{L}$ and $\SU{2}{R}$ with the same VEV, as opposed to the desired situation where the VEV responsible for $\SU{2}{R}\times\U{\BL} \rightarrow \U{Y}$ is much larger than the Higgs VEV that triggers the EW symmetry breaking. 
\vspace{10pt}
 
\begin{figure}[htbp] 
  \centering
  \includegraphics[width=\linewidth, natwidth=610,natheight=642]{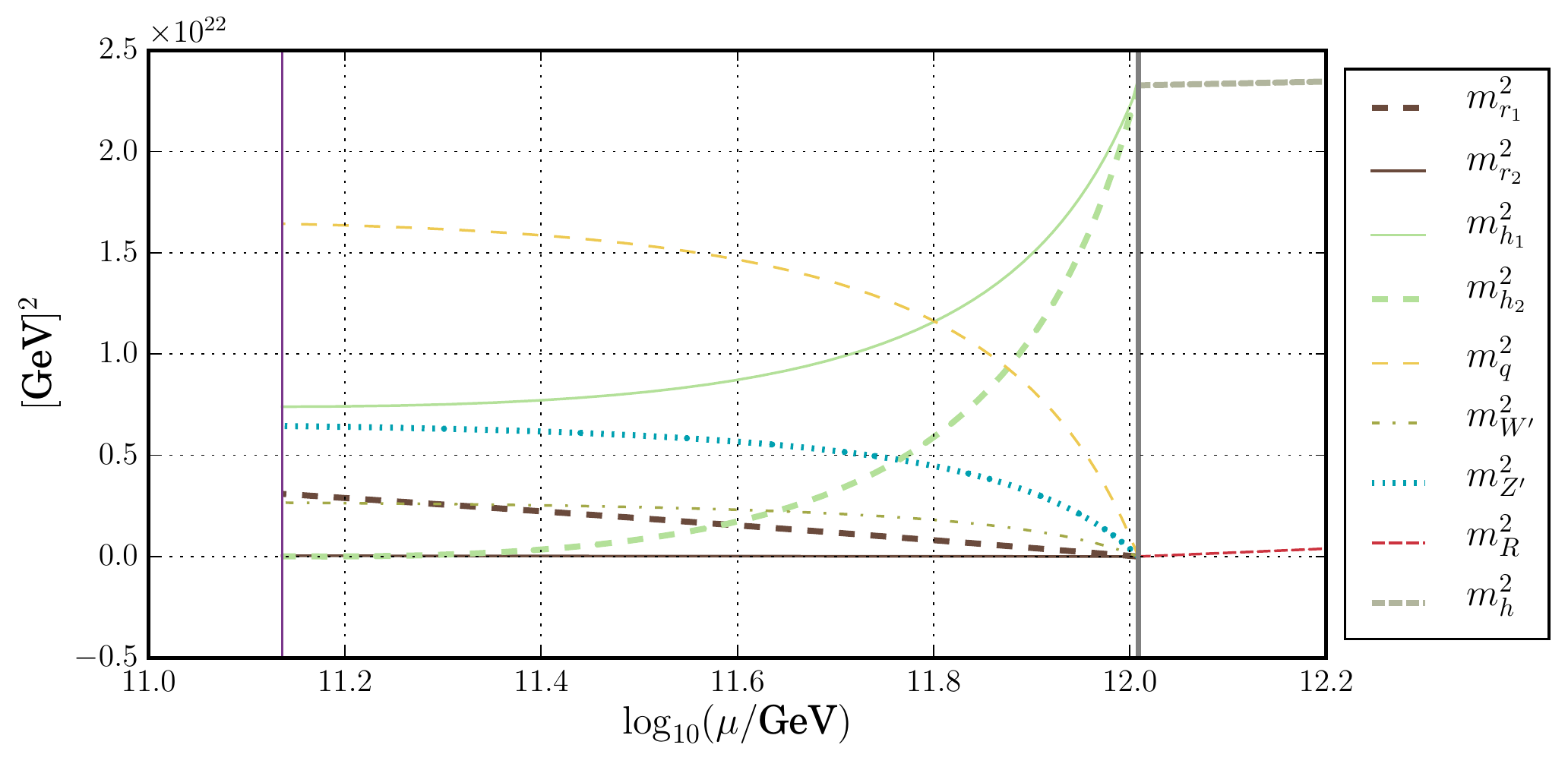}
  \caption{One-loop RG evolution of the mass parameters before and after the radiative $\SU{2}{R} \times \U{\BL}$ symmetry breaking down to $\U{Y}$ for an example point. The two vertical lines mark the scales at which $m_R^2$ first runs negative (right) and the scale $\mu_r$ at which the RG running is terminated (left), respectively.}
  \label{fig:massplot}
\end{figure}

We selected the most promising candidates from the results of the scan by requiring the maximal hierarchy between a light Higgs-like doublet and heavy exotic particles at the scale $\mu_r$ while having parameters within the perturbativity constraints. In Fig.~\ref{fig:massplot} we show the running of mass parameters before and after the radiative $\SU{2}{R} \times \U{\BL}$ symmetry breaking for a benchmark point satisfying those conditions. We find that it is possible to find some mass hierarchy at the symmetry breaking scale, with a Higgs-like scalar doublet coming from $\tilde{h}$ with masses up to two orders of magnitude lighter than the rest of the mass spectrum. In addition, we also observe that a complex scalar coming from $\Lrt$ prefers to have a small mass at $\mu_r$ (this is the dark solid curve in Fig.~\ref{fig:massplot}, corresponding to $m_{r_2}^2$ in Tab.~\ref{tab:LRBrokenMassStates}). This scalar is a singlet under $\SU{2}{L}$ while having unit hypercharge, meaning that it will have unit electric charge after EW symmetry breaking. At present, it is not clear to what extent this state accumulates a much larger mass when evolving from $\mu_r$ down to the EW scale.

Although the gauge couplings of the effective LR-symmetric model start with the same values due to the matching conditions and the $\mathbb{Z}_3$ symmetry in the high-scale model, the RG evolution induces a splitting as seen in Fig.~\ref{fig:gaugeplot}. It is interesting to note that although we did not impose the boundary condition $g_0=0.6$ for the SA algorithm, the allowed points in parameter space seem to be consistent with the boundary condition as seen in the upper right plot in Fig.~\ref{fig:allowedparameters}. We also note from Fig.~\ref{fig:gaugeplot} that there is an approximate relation $\g{L} \approx \g{R}$ that is exact at the matching scale, while a small splitting between $\g{L}$ and $\g{R}$ is generated in the low energy limit. This observation points towards an approximate $\mathbb{Z}_2$ symmetry between the $\SU{2}{L}$ and $\SU{2}{R}$ gauge groups. In fact, we can trace the origin of the radiative $\mathbb{Z}_2$ breaking to the scalar sector in the effective model, where the choice of keeping only $\tilde{l}_\mathrm{R}$ and $\tilde{h}$ leads to $\beta_{\g{L}} \neq \beta_{\g{R}}$. This is because only $\tilde{h}$ transforms under $\SU{2}{L}$ (while both $\tilde{l}_\mathrm{R}$ and $\tilde{h}$ transform under 
$\SU{2}{R}$).

This serves to prove that in the proposed framework it is possible to trigger the full symmetry breaking down to the SM gauge group by means of the trinification breaking VEV \eqref{eq:vevStructure} only, while at the same time generating a desired hierarchy at low energy scales. In other words, in the model proposed in this work, the RG evolution makes it possible to have a highly symmetric trinification model whose gauge group is naturally broken down to the SM gauge group.

For the alternative case of breaking directly to $\U{E.M.}$ discussed in Sect.~\ref{sec:breakingtoEM} we performed a similar analysis as we did above by preparing a SA scan to find parameter space points looking for a possibility for the radiative breaking to $\U{E.M.}$. For case A (discussed in Sect.~\ref{sec:caseA}) 6080 points were produced during three weeks where both $m_h^2$ and $m_R^2$ ran negative. However, none of the points showed positive squared masses in the scalar sector, i.e. no points were found where the desired vacuum was a minimum of the scalar potential. Similarly, for case B (discussed in Sect.~\ref{sec:caseB}) we run a SA scan that produced 32631 points, again with none of them showing stable minima with the desired radiative symmetry breaking. What this means is that as far as our analysis could tell, when $m_h^2$ and $m_R^2$ became negative through RG running, the true minimum of the scalar potential did not exhibit $\U{E.M.}$ as a remaining symmetry, thus making it unviable as a phenomenological model within the proposed framework. 
          
\section{Discussion and future work} \label{sec:Discussion}

\subsection{Fermion sector at one-loop}
In the present analysis, many Yukawa couplings (and also the Majorana mass $m_{\Phi^s}$) in the effective LR-symmetric model are zero simply due to the tree-level matching procedure. However, this will no longer be true once the matching and running are performed at a higher loop level. Although the full one-loop matching and two-loop running analysis is necessary to obtain precise numerical values that is planned for a future work, it is interesting to understand which diagrams will lead to non-vanishing matching conditions for some of the parameters in 
Eq.~\eqref{eq:LRYukawa}.

\begin{figure}[tbp]
  \centering
  \hspace{-10pt}
  \includegraphics[width=0.45\linewidth, natwidth=610,natheight=642]{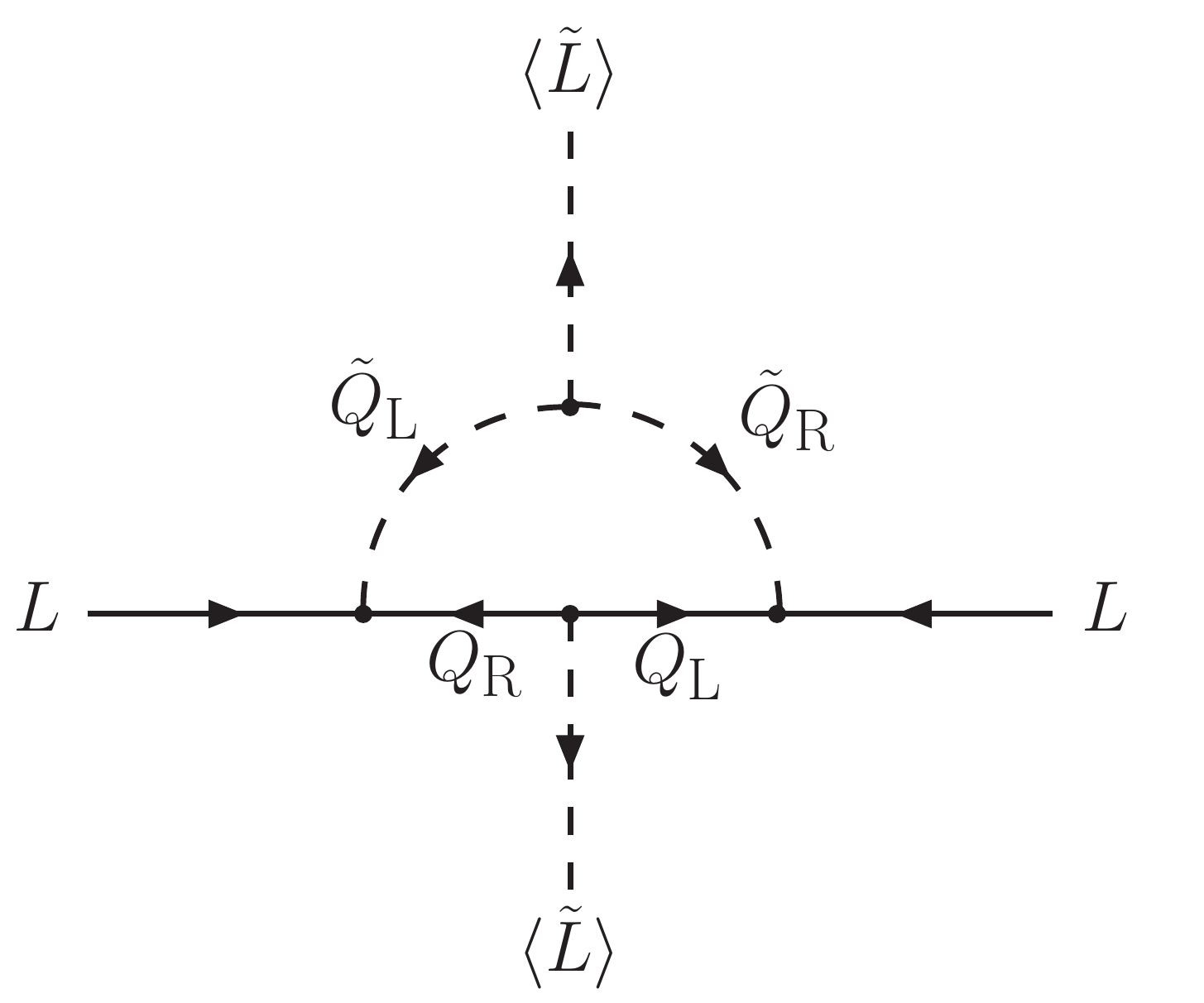} 
  \caption{A diagram in the trinification theory that constitutes the one-loop contribution to the matching onto the Majorana $\Phi^s$ mass, $m_{\Phi^s}$, in the effective 
  LR-symmetric model.}
  \label{fig:phi_loopmass}
\end{figure}

We first note that $m_{\Phi^s}$ receives a non-zero contribution from the diagram in Fig.~\ref{fig:phi_loopmass}, with the trinification VEV in Eq.~\eqref{eq:vevStructure}. From this we can estimate that $m_{\Phi^s}$ will be suppressed with respect to the trinification breaking scale $v_3$ as 
\begin{equation} \label{eq:mPhi_oneloop}
m_{\Phi^s}^{\mathrm{(1-loop)}} \sim \left( \frac{y^3}{(4 \pi)^2} \cdot \frac{\gamma}{v_3} \right) v_3.
\end{equation}
If the trinification Yukawa coupling $y$ and the scalar tri-linear coupling $\gamma$ are sufficiently large, it might be appropriate to integrate out $\Phi^s$ along with the heavy trinification-scale quarks in Tab.~\ref{tab:TriFermions}, so that it no longer appears in the effective LR-symmetric model. If instead $y$ and $\gamma$ are small, the suppression factor in Eq.~\eqref{eq:mPhi_oneloop} can easily be very small such that $m_{\Phi^s} \sim m_h , m_R$. At present, as was shown in Sect.~\ref{sec:Results}, we see no preference for neither large nor small values of $y$ and $\gamma$, and both possibilities thus remain open.  

\begin{figure}[H] 
  \centering
  \hspace{-10pt}
  \includegraphics[width=0.45\linewidth, natwidth=610,natheight=642]{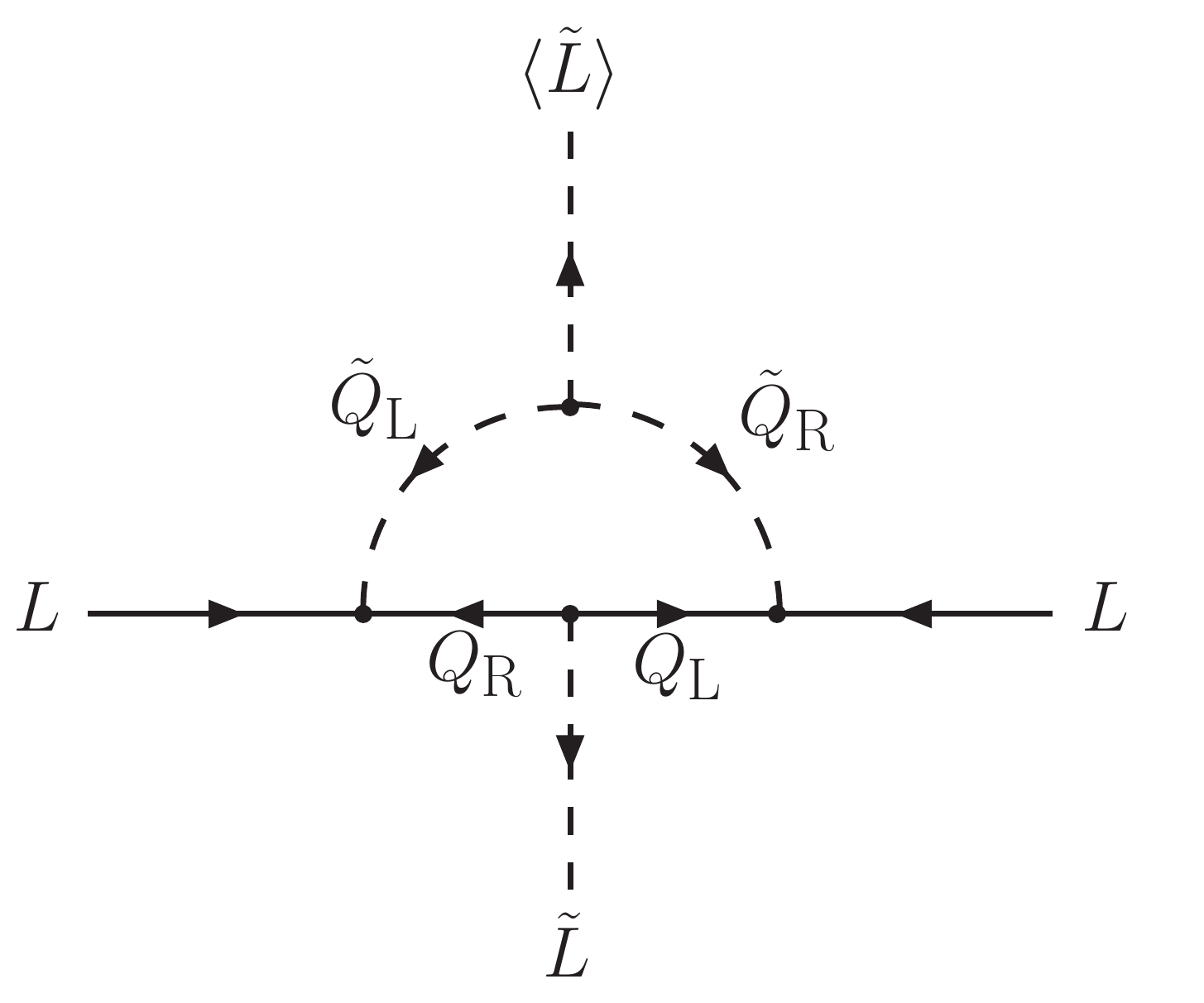} 
  \includegraphics[width=0.45\linewidth, natwidth=610,natheight=642]{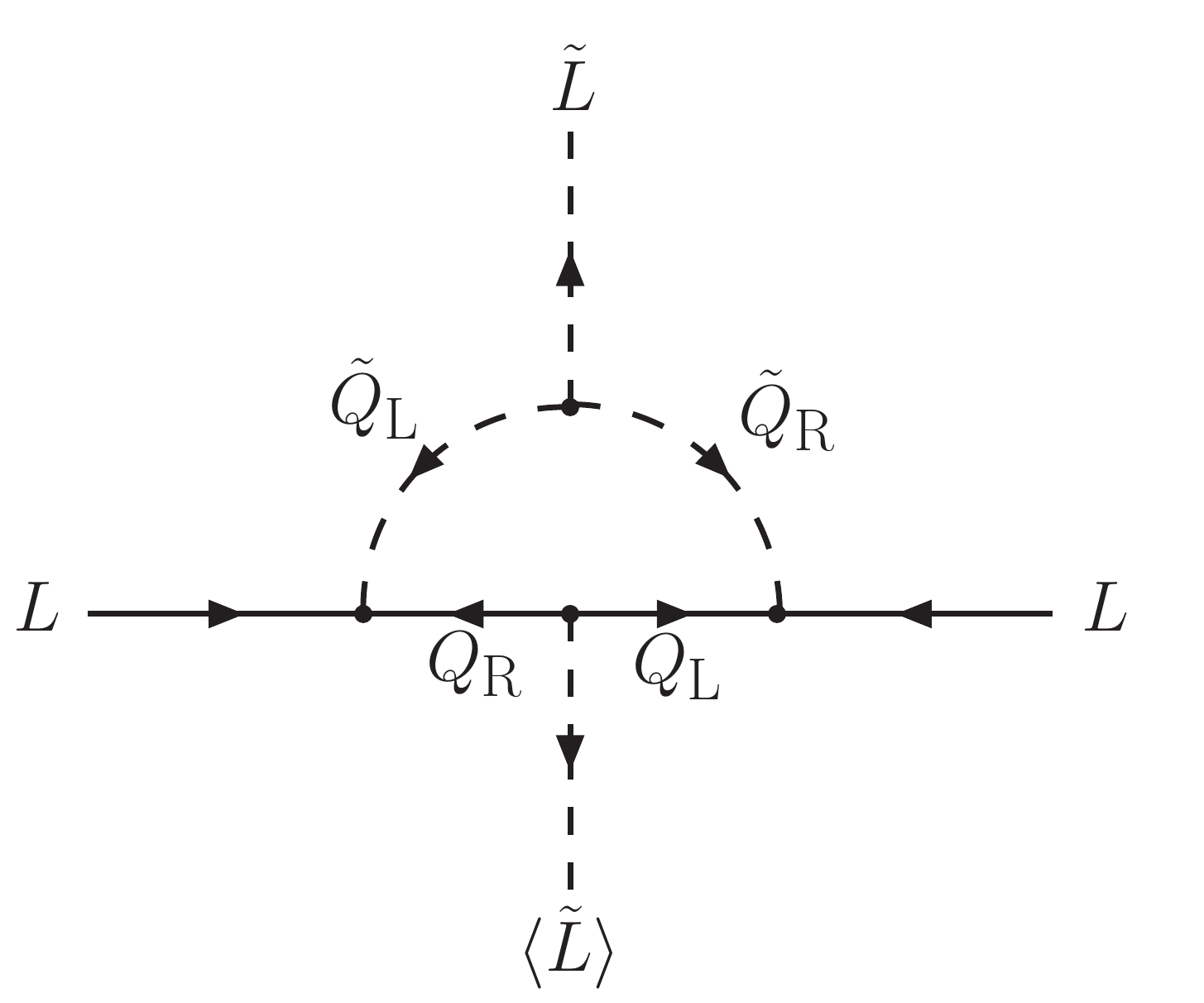} 
  \caption{These two diagrams give rise to Yukawa terms that, in turn, provide Dirac masses for two generations of right-handed neutrinos and $\Phi$'s, as well as for SM leptons 
  of the third generation.}
  \label{fig:loop_Yukawas}
\end{figure}

Next, we turn to the Yukawa interactions that are generated by the diagrams in Fig.~\ref{fig:loop_Yukawas}. These diagrams are of interest when the external leg $\tilde{L}$ corresponds to one of the two remaining scalars in the effective LR-symmetric model, namely $\tilde{h}$ and $\Lrt$. One can then show that the diagram is non-vanishing when the external fermion legs are such that the loop corresponds to Yukawa interactions of the types
\begin{equation} \label{eq:looped_yukawas}
\begin{aligned}
& \Lrs{I}{R} \,\, \Lrf{I}{R} \, \Phi^s + \mathrm{c.c.}\,,\\
& \Lrs{I}{R} \,\, \Lrf{s}{R} \, \Phi^I + \mathrm{c.c.}\,,\\
& \hhs{L}{R} \,\, \Llf{s}{L} \, \Lrf{s}{R}+ \mathrm{c.c.}\,,
\end{aligned}
\end{equation}
i.e. the Yukawa couplings $Y_\gamma$, $Y_\delta$ and $Y_\eta$ receive a non-zero contribution at one-loop. In particular, when $\langle \Lrt \rangle \neq 0$ the upper two interaction terms in Eq.~\eqref{eq:looped_yukawas} will provide masses to two generations of right-handed neutrinos. Note, however, that right-handed neutrinos receive a \textit{Dirac} mass that is formed together with two generations of $\Phi$. We can identify the last interaction term in Eq.~\eqref{eq:looped_yukawas} as containing the Yukawa term for SM leptons of the third generation.

\subsection{One-loop corrections to masses of light scalars} \label{sec:LightScalars_loop}

Although we have performed tree-level matching down to an effective theory, its important to understand whether 
the scalars $\hh{L}{R} , \,  \Lr{I}{R}$ can remain light as soon as the higher-order corrections are considered. In general, this may not be 
the case and thus it would no longer be justified to keep those light scalars in the low-energy theory. Even if the corrections 
would keep the scalars sufficiently light, the particular values for their masses also affect the RG flow potentially leading 
to different conclusions regarding the radiative LR-symmetry breaking. 
\begin{figure}[H] 
  \centering
  \begin{tabular}{p{0.5\textwidth} p{0.5\textwidth}}
  \vspace{0pt} \includegraphics[width=0.81\linewidth, natwidth=610,natheight=642]{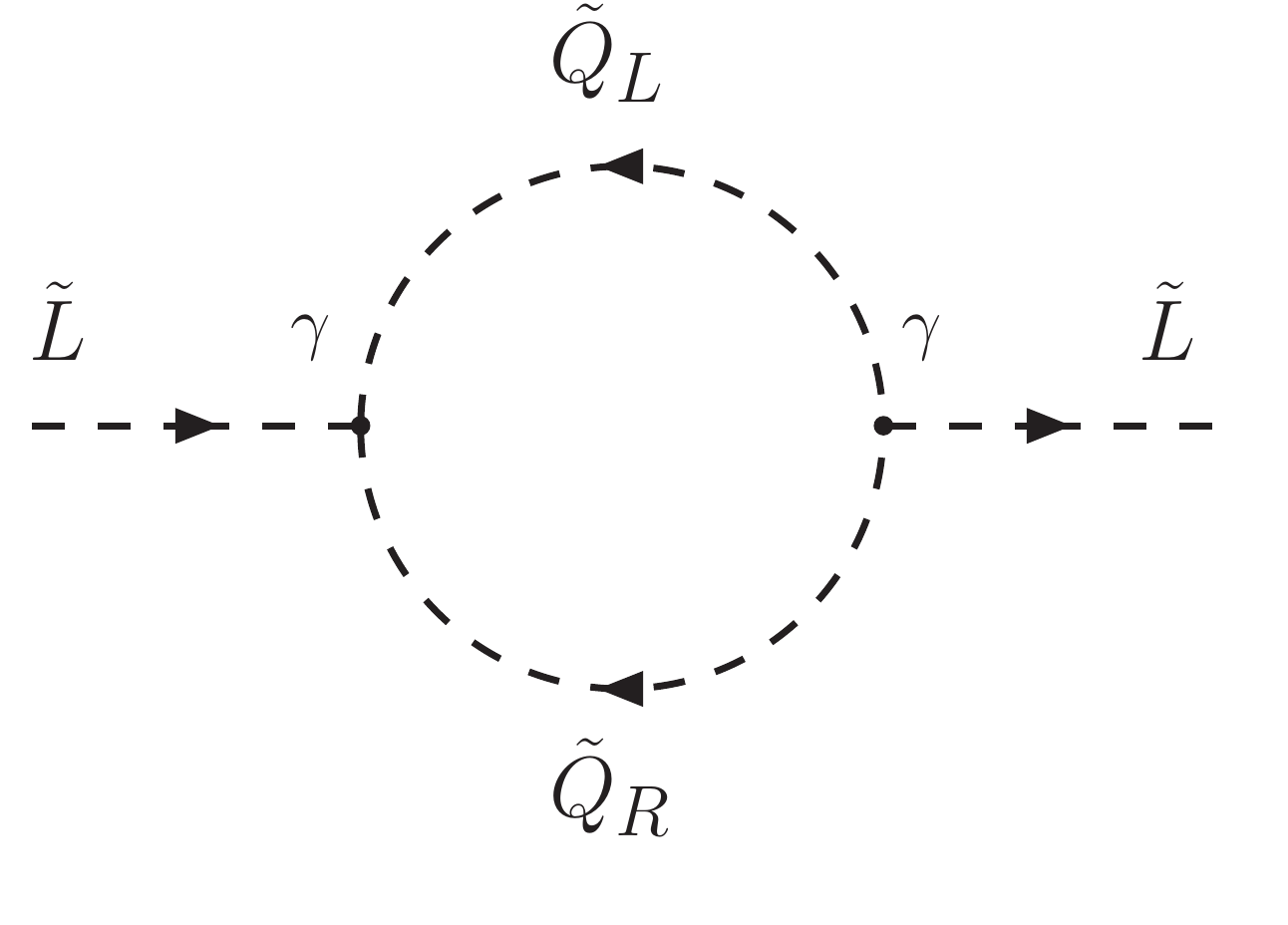}  &
  \vspace{0pt} \includegraphics[width=0.8\linewidth, natwidth=610,natheight=642]{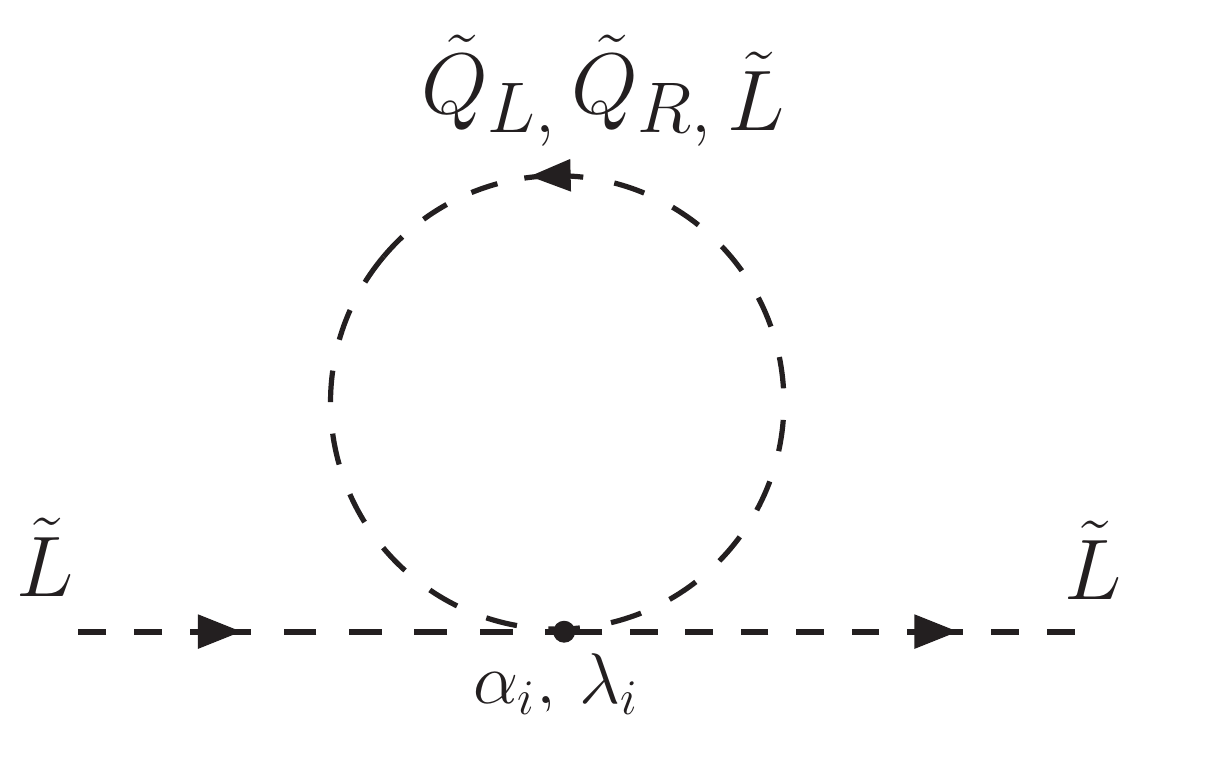} 
\end{tabular}
    \caption{Class of diagrams with a non-vanishing contribution to $m^2_h$ and $m_R^2$ in the trinification theory.}
  \label{fig:loopscalar}
\end{figure}

It turns out that the only non-vanishing one-loop contributions (in the zero external momenta approximation) 
to $m_h^2$ and $m_R^2$ come from the type of diagrams shown in Fig.~\ref{fig:loopscalar}.
Any other topology will either (i) be forbidden by the trinification symmetry (such as fermion loops), or 
(ii) vanish once the zero momentum approximation is taken into account. 

The tri-linear coupling $\gamma$ plays an important role in the one-loop matching conditions. Namely, it determines
the size of the one-loop corrections to $m_h^2$ and $m_R^2$ coming from the left panel in Fig.~\ref{fig:loopscalar}.
It thus suffices to make $\gamma/v_3 \ll 1 $ in order to keep those corrections small\footnote{The model presented here, 
while inspired by SUSY, is not supersymmetric. However, as long as SUSY is concerned, $\gamma/v_3 \ll 1$ condition
could be justified by $\gamma$ corresponding to a soft-SUSY breaking term whose natural values are around 
the SUSY breaking scale.}. For the second type of diagrams in Fig.~\ref{fig:loopscalar} (right), the corrections become functions of 
$\lambda_i$ but also $\alpha_i$ (the tree-level masses are only functions of $\lambda_i$). As long as the coloured 
scalars remain heavy, choosing appropriate values for $\alpha_i$ would give us enough freedom for $\hh{L}{R} , \,  \Lr{I}{R}$ 
to remain light. 


In the tree-level study of the model that has been presented here, we thus assume that the scalars are light leaving the specifics of 
the physical mechanism which appropriately tunes the values for $\gamma$ and $\alpha_i$ as well as the corresponding effect 
on the RG evolution for future work. However, we expect that the qualitative conclusions reached in this work would still be valid when accounting for the higher-order corrections due to the fact that, as discussed in section \ref{sec:Results}, the regions of the effective theory 
parameter space where the radiative LR-symmetry breaking takes place are quite broad.

\subsection{The low-energy LR-symmetric model with additional light Higgs-doublets}

We have shown that the proposed model brings an intriguing possibility of the radiative $\SU{2}{R} \times \U{\BL} \rightarrow \U{Y}$ breaking. To do this, we integrated out the maximal number of fields in the high scale trinification theory in order to make the effective LR-symmetric model as simple as possible. Although the radiative symmetry breaking in such a toy model is realized, we show in this section that in order to accurately generate all fermion masses in the SM, we need to consider the case where more scalars are present in the effective theory. This is due to the fact that some mass terms are forbidden by the global group that remains unbroken in the scenario considered in Sect.~\ref{sec:LR}. However, we note that the interplay between scalar mass parameters and their $\beta$-functions allowing $m_R^2$ to run negative, will still be present when integrating out fewer Higgs doublets. Although a more detailed study will be needed to find explicit regions of parameter space with the radiative symmetry breaking, we expect that the qualitative conclusions in Sect.~\ref{sec:Results} will not be changed.

\subsubsection{SM quarks}

\begin{table}[htbp]
  \begin{center}
    \begin{tabular}{cccc}
     \toprule                     
     Trinification  	& effective LR-symmetric model 							& $\U{E}$ 	& $\U{G}$ \\ 
     \midrule 	
     $\QL{2}{c}{2}$  	& $\QLE{2}{c}{2}$						& $-1/2$ 	& $-1/6$ \\
     $\QL{3}{c}{1}$  	& $\QLE{s}{c}{1}$ 						& $-1/2$ 	& $-2/3$ \\
     $\QL{3}{c}{2}$  	& $\QLE{s}{c}{2}$						& $-1/2$ 	& $-2/3$ \\
	 \midrule
     $\QRd{3}{1}{c}$ 	& $\QREd{s}{1}{c}$						& $+1/2$ 	& $+1/3$ \\
     $\QRd{3}{2}{c}$ 	& $\QREd{s}{2}{c}$ 						& $+1/2$ 	& $+1/3$ \\
     $\QRd{3}{3}{c}$ 	& $({D_\mathrm{R}^\dagger}^s)^3{}_c$ 	& $+1/2$ 	& $+5/6$ \\
     \bottomrule
    \end{tabular}
    \caption{Components in the trinification quark tri-triplets (and the corresponding fields in the effective LR-symmetric model) that should build up the left- and right-handed components of the lightest SM $u,\,d,\,s$ quarks. In the current realisation of the model, the mass terms for these quarks are forbidden by the global $\{\U{E} \times \U{G}\}$ group that is left unbroken by the two Higgs VEVs.} \label{tab:SMQuarks}
  \end{center}
\end{table}

With tree-level matching (and one-loop running) the only non-zero Yukawa coupling with SM particles is $Y_\zeta$ in Eq.~\eqref{eq:LRYukawa}. Arranging the $\tilde{h}$ and $\Lrt$ VEVs as in Eq.~\eqref{eq:Minimum2VEV}, we see that, through this term, $v_1$ gives masses to two up-type quarks which could be identified with the top and charm quarks of the SM. On the other hand, the VEV $v_2$ gives a mass to one down-type quark (which can be identified with the bottom quark) and also a contribution to the mass of the heavy down-type quark in Tab.~\ref{tab:LRBrokenFermions}.
However, there still remain one up-type and two down-type quarks that are massless at tree-level, which should be identified with the lightest \textit{u, d, s} quarks in the SM. The corresponding left-handed and right-handed components are shown in Tab.~\ref{tab:SMQuarks} along with their charges under the global group $\{\U{E} \times \U{G}\}$, which is unbroken by the Higgs VEVs in Eq.~\eqref{eq:Minimum2VEV}. Since none of the $Q_\mathrm{L}$ fields have the same global $\U{}$ charges as any of the $Q_\mathrm{R}^\dagger$ fields in Tab.~\ref{tab:SMQuarks}, it is not possible to generate Dirac masses for the \textit{u, d, s} quarks since those mass terms would violate $\{\U{E} \times \U{G}\}$. However, there is a simple way to accommodate these masses. For regions of parameter space with light Higgs doublets in the effective LR-symmetric model, their VEVs would break $\{\U{E} \times \U{G}\}$ enabling the radiative generation of the light quark mass terms.

\subsubsection{Colour singlet fermions}
As in the quark sector, many colour neutral fermion mass terms are forbidden by the global group that is left unbroken by the two Higgs VEVs. By looking at the global $\U{}$ charges of the components of the colour neutral fermions, we find that the only electrically charged and electrically neutral fermions that can participate in a fermion bilinear term are contained in
\begin{equation} \label{eq:FermMassComps}
\begin{aligned}
\psi_\mathrm{C} &= \left( \begin{array}{ccc|ccc}
\, \Lrf{s}{1} \,&\, \Llf{s}{1} \,&\, \HHf{2}{1}{2} \,\,&\,\, \HHf{s}{1}{2} \,&\, \HHf{s}{2}{1} \,&\, \Lrf{2}{1} \,
\end{array} \right)^T,			\\
\psi_\mathrm{N} &= \left( \begin{array}{ccccc|cccc}
\, \Lrf{s}{2} \,&\, \HHf{2}{1}{1} \,&\, \HHf{2}{2}{2} \,&\, \Phi^2 \,&\, \Llf{s}{2} \,\,&\,\, \Lrf{2}{2} \,&\, \HHf{s}{1}{1} \,&\, \HHf{s}{2}{2} \,&\, \Phi^s \,
\end{array} \right)^T,
\end{aligned}
\end{equation}
respectively. For all other colour neutral fermions, there is no fermion field with opposite $\U{}$ charges such that they together could form a mass term. In particular, note that $\psi_\mathrm{C,N}$ contain no first generation colour singlet fermions. 

Generally, we can then write the mass terms for the colour neutral fermions as
\begin{equation}
\frac{1}{2} \psi_{\mathrm{C}}^T \, M_\mathrm{C} \psi_\mathrm{C} + \frac{1}{2} \psi_{\mathrm{N}}^T \, M_\mathrm{N} \psi_\mathrm{N} + \mbox{c.c.}
\end{equation}
Upon demanding invariance under the global $\U{}$ groups (and of course $\U{E.M.}$), we find that the mass matrices $M_{\mathrm{C,N}}$ have the following structure: 
\begin{equation}
\begin{aligned}
M_\mathrm{C} &= \left( \begin{array}{ccc|ccc}
\,    	\,&\, \bullet	\,&\, \bullet	\,\,&\,\,     	\,&\,     	\,&\,    		\\
\bullet	&    		&    		&    		&    		&    		\\	
\bullet	&    		&    		&    		&    		&    		\\	\hline

    	&    		&    		&    		& \bullet	& \bullet	\\
    	&    		&    		& \bullet	&    		&    		\\	
    	&    		&    		& \bullet	&    		&    	

\end{array}\right),							\qquad											
M_\mathrm{N} &= \left( \begin{array}{ccccc|cccc}
			\,&\, \bullet	\,&\, \bullet	\,&\, \bullet	\,&\, \bullet	\,\,&\,\, 			\,&\,			\,&\,			\,&\,			\\
\, \bullet	\,&\,			&			&			&			&			&			& 			&			\\	
\bullet	&			&			&			&			&			&			& 			&			\\	
\bullet	& 			& 			&			& 			& 			&			& 			&			\\		
\bullet	&			&			& 			&			&			&			& 			&			\\	\hline
		
		&			&			& 			&			\,\,&\,\, \star		\,&\, \bullet	\,&\, \bullet	\,&\, \bullet	\\
		&			&			& 			&			& \bullet	& \star		& \bullet	& \bullet	\\
		&			&			& 			&			& \bullet	& \bullet	& \star		& \bullet	\\
		&			&			& 			&			& \bullet	& \bullet	& \bullet	& \star				

\end{array}\right),
\end {aligned} 
\end{equation}
where entries marked with a `$\bullet$' (`$\star$') denote the Dirac (Majorana) contributions that are allowed to be non-zero. These would amount to two electrically charged massive Dirac fermions (which we would have to identify with the $\tau$-lepton and muon in the SM), one massive electrically neutral Dirac fermion, and four electrically neutral Weyl fermions that receive both Dirac and Majorana mass contributions. 

The tri-triplet $\LL{i}{l}{r}$ contains twelve electrically charged Weyl fermions, meaning that we will still have eight electrically charged Weyl fermions whose mass terms are forbidden by the unbroken global group. Out of the remaining fifteen electrically neutral components in $\LL{i}{l}{r}$, at least nine Weyl fermions are necessarily massless. To make more components in $\LL{i}{l}{r}$ massive (to evade obvious inconsistencies with phenomenology), we have to include more Higgs doublets in the effective LR-symmetric model originating from the high-scale trinification theory. 

\subsection{CKM mixing with additional Higgs doublets}

In the previous section, we have seen that in order to explain the observed fermionic mass spectra, more components of $\tilde{L}$ should be kept in the effective LR-symmetric 
model. By looking at the CSS mass spectra, we see that this will be possible when
\begin{equation}
\lambda_{2,3,4} \ll \lambda_{1} \, ,
\end{equation}
in which case the fields
\begin{equation}
\Sl{I}{L}{R} \, , \quad \Sl{I}{3}{R} \, , \quad \Sl{I}{L}{3} \, , \quad \Sl{3}{L}{R} \quad ,
\end{equation}
would all remain in the effective LR-symmetric model (see Tab.~\ref{tab:CSSMassStates}). In this case, the CSS potential exhibits an approximate $\mathrm{O}(54)$ symmetry. Although this increase in the number of scalar fields would lead to a substantial increase in complexity in the low-energy theory, we are confident that the radiative $\SU{2}{R} \times \U{\BL} \rightarrow \U{Y}$ breaking will still be present as mentioned above.
In this section, we show in a straightforward tree-level analysis, that the structure of the SM Cabibbo-Kobayashi-Maskawa (CKM) mixing matrix in the Cabibbo form emerges as a consequence of $\SU{3}{F}$ if VEVs are strategically placed in the tri-doublet $\Sl{I}{L}{R}$. These VEVs would be allowed at lower scales when keeping the extra fields in the effective theory after SSB of trinification. 

Consider the following VEV setting:
\begin{equation}
\langle \Sl{1}{L}{R} \rangle = \tfrac{1}{\sqrt{2}} \begin{pmatrix}
h_1 & 0  \\
0 & h_2
\end{pmatrix}	\, , \quad 
\langle \Sl{2}{L}{R} \rangle = \tfrac{1}{\sqrt{2}} \begin{pmatrix}
h_3 & 0  \\
0 & 0
\end{pmatrix}
\end{equation}
In terms of the full trinification tri-triplet $\tilde{L}$, this means that
\begin{equation} 
  \label{eq:Minimum}
\langle \tilde{L}^1\rangle = \tfrac{1}{\sqrt{2}}
\begin{pmatrix}
h_1 & 0 & 0 \\
0 & h_2 & 0 \\
0 & 0     & 0
\end{pmatrix} \,,
\qquad
\langle \tilde{L}^2\rangle = \tfrac{1}{\sqrt{2}}
\begin{pmatrix}
h_3 & 0    & 0 \\
0     & 0    & 0 \\
0     & w    & 0
\end{pmatrix}\,,
\qquad
\langle \tilde{L}^3\rangle = \tfrac{1}{\sqrt{2}}
\begin{pmatrix}
0 & 0 & 0 \\
0 & 0 & 0 \\
0 & 0 & v_3
\end{pmatrix} \,.
\end{equation}
where we have also indicated the trinification and LR symmetry breaking VEVs $v_3$ and $w$. The VEV setting \eqref{eq:Minimum} leaves the group
\begin{equation}
\SU{3}{C} \times \U{E.M.}  \times \left\{ \U{P} \times \U{B} \right\}
\end{equation}
unbroken, where the global $\U{P}$ is generated by
\begin{equation}
\T{P}{} \equiv \frac{1}{\sqrt{3}} \left( \T{R}{8}+\T{F}{8} \right).
\end{equation}
For a consistency with the SM, we have to impose the following hierarchy between the above VEVs
\begin{eqnarray}
\label{VEV-hierarchy}
v \gg w \gg h_{1,2,3} \sim 10^2 - 10^3 \, \rm{GeV} \,,
\end{eqnarray}
such that $h_{1,2,3}$ would correspond to the SM-breaking Higgs VEVs. In the gauge sector, at tree level one recovers one massless 
(photon) state
\begin{eqnarray}
A_\mu=\frac{1}{2\sqrt{2}} \Big[{G_{\rm L}}^8_\mu + {G_{\rm R}}^8_\mu - \sqrt{3} ({G_{\rm L}}^3_\mu + {G_{\rm R}}^3_\mu) \Big] \,,
\end{eqnarray}
$W^\pm$ bosons
\begin{eqnarray}
W^\pm_\mu = {G_{\rm L}}^{1,2}_\mu \,, \qquad m_W^2 \simeq \frac18 \, g^2\sum_i h_i^2 \,,
\end{eqnarray}
and the $Z^0$ boson
\begin{eqnarray}
Z^0_\mu=\frac{1}{2\sqrt{10}} \Big[5{G_{\rm L}}^3_\mu - 3{G_{\rm R}}^3_\mu + \sqrt{3} ({G_{\rm L}}^8_\mu + {G_{\rm R}}^8_\mu) \Big] \,, \qquad
m_Z^2 \simeq \frac{2}{10} \, g^2\sum_i h_i^2 \,,
\end{eqnarray}
in a rough consistency with the SM. Besides, at $w$ scale one finds heavy ${W'}^\pm$ and ${Z'}^0$ bosons
\begin{eqnarray}
&& {W'}^\pm_\mu = {G_{\rm R}}^{1,2}_\mu \,, \qquad m_{W'}^2 \simeq \frac18 \, g^2 w^2 \,, \\
&& {Z'}^0_\mu = \frac{1}{\sqrt{10}} \Big[2{G_{\rm R}}^3_\mu + \sqrt{3} ({G_{\rm L}}^8_\mu + {G_{\rm R}}^8_\mu) \Big] \,, \qquad 
m_{Z'}^2 \simeq \frac{5}{16} \,g^2 w^2 \, ,
\end{eqnarray}
which can be recognised as the heavy vector states in Tab.~\ref{tab:LRBrokenGaugeStates}. The other nine gauge bosons corresponding 
to broken (by $v_3$) generators of $[\SU{3}{}]^2 \to [\SU{2}{}]^2$ in trinification get masses at the GUT scale $\mu_{m} \sim v_3$ (see Tab.~\ref{tab:LRBrokenGaugeStates}).

In the quark sector with $Q^i_\mathrm{L}=\{u_\mathrm{L}^i,\,d_\mathrm{L}^i,\,D_\mathrm{L}^i\}$, we obtain three weak-singlet down-type quarks $D^i=\{D,\,S,\,B\}$ 
that acquire large (Dirac) tree-level masses
\begin{eqnarray}
m_B\simeq \tfrac{1}{\sqrt{2}}yw \,, \qquad m_D \simeq m_S \simeq \tfrac{1}{\sqrt{2}}y v \,, 
\end{eqnarray}
and hence decouple from the SM. The other three down-type states $d^i=\{d,\,s,\,b\}$ remain light
\begin{eqnarray}
m_d=0 \,, \qquad m_s \simeq m_b \simeq \tfrac{1}{\sqrt{2}} y h_2 \,, 
\end{eqnarray}
and could thus be identified with masses of down, strange and bottom quarks of the SM, respectively, such that 
there is no tree-level splitting between $s$ and $b$ quarks, and $d$-quark is massless. Interestingly enough,
all the down-type quarks $d^i$ and $D^i$ practically do not mix with each other to the leading order
in small $h_i/v,\, h_i/w$ and $w/v$ ratios. Note, while it is possible to introduce a non-zero tree-level 
splitting between $s$ and $b$ quarks by imposing more VEVs in neutral components of two Higgs doublets, 
a non-zero $d$-quark mass can only acquire a non-zero value by an unnaturally small VEV in a neutral component of a third Higgs doublet, 
but we do not consider this situation here. All the physical up-type quarks emerge as mixures of trinification up-type quarks 
$u^i=\{u^1,\,u^2,\,u^3\}$ remain light
\begin{eqnarray}
&& u=\frac{u^1 h_1 + u^2 h_3}{\sqrt{h_1^2+h_3^2}}\,, \qquad
c=\frac{-u^2 h_1 + u^1 h_3}{\sqrt{h_1^2+h_3^2}}\,, \qquad t = u^3 \,, \\
&& m_u=0 \,, \qquad m_c \simeq m_t \simeq \tfrac{1}{\sqrt{2}}y \sqrt{h_1^2+h_3^2} \,, 
\end{eqnarray}
which could be identified with masses of up $u$, charm $c$ and top $t$ quarks of the SM, respectively. Again, 
in the considering scenario, there is no tree-level splitting between $c$ and $t$, and it can not be generated at tree 
level by imposing any additional VEVs. The observed substantial charm-top and strange-bottom splittings can be 
in principle generated radiatively by (i) RG runnings of the corresponding Yukawa couplings which will have different 
slopes as long as trinification symmetry is broken, and by (ii) higher-loop effects which may modify the starting values for 
the Yukawa couplings at the matching scale. The quark CKM mixing aquires an approximate 
Cabbibo form already at tree level
\begin{eqnarray}
V^{\rm CKM} \simeq 
\begin{pmatrix}
\cos\theta_C & \sin\theta_C & 0 \\
-\sin\theta_C & \cos\theta_C & 0 \\
0 & 0 & 1
\end{pmatrix} \,, \qquad \tan\theta_C = \frac{h_1}{h_3} \,,
\end{eqnarray}
which is a remarkable feature of the family symmetry, while small observed distortions of the Cabbibo mixing could only be generated at a higher-loop level. Non-unitarity corrections to the quark CKM mixing are also suppressed by small $h_i/v,\, h_i/w$ and $w/v$ ratios. This means that phenomenological constraints on those corrections could be important for setting lower limits on hierarchies between 
the trinification symmetry breaking scales.

\section{Conclusions}  \label{sec:Conclusions}

In this work we have introduced a GUT based on the trinification gauge group. By introducing a global $\SU{3}{F}$ family symmetry, our model resolves some of the issues with previous attempts to work with gauge trinification-based models while also considerably reduces the number of free parameters. 

We found that SSB of the trinification symmetry can be triggered by the VEV of only one component of a scalar $\mathbf{27}$-plet and that the minimum is, in a large part of the parameter space, the global one. We found that radiative breaking of gauge (i.e. $\SU{2}{R} \times  \U{\BL}$) and global symmetries, that are not present in the SM, was possible in the effective LR-symmetric model that is left after SSB of trinification. We did so by studying the most simple scenario (two light scalar multiplets remaining in the effective theory) where the mass-squared parameter for a scalar field charged under such symmetries could become negative by means of the RG evolution. By implementing a parameter scan algorithm using simulated annealing, we were able to efficiently scan the parameter space of the trinification theory and found regions where the radiative breaking happens in the chosen 
effective LR-symmetric model. 

We also explored under which circumstances the high-scale theory might reproduce the masses and hierarchies of the SM at lower energies. We found that the simple scenario used to understand the radiative symmetry breaking needs to be extended in order to get for example CKM mixing and masses for all SM fermions. By having more light scalar multiplets present in the effective theory, their VEVs could break the remaining global symmetries which forbid the necessary mass terms in the low-energy theory. We also show that if such fields are present, the proposed model has good potential to result in a realistic quark mass spectrum resembling the SM one, while keeping the ingredients necessary to trigger the radiative breaking shown in this work. It is clear then that future studies should include one-loop matching, two-loop RG running and the extra scalar multiplets in the effective theory. Although we have shown the feasibility of radiative breaking and the possibility to explain the hierarchies in mass parameters for the proposed model, in order to offer a complete consistency with the SM such extended study needs to be performed in future work.

\acknowledgments

The authors would like to thank N.-E.~Bomark, E.~Corrigan, W.~Porod, M.~Sampaio and F.~Staub for insightful discussions 
during the development of this work. A. P. M. is supported by the FCT grant SFRH/BPD/97126/2013 and partially 
by the H2020-MSCA-RISE-2015 Grant agreement No StronGrHEP-690904, and by the CIDMA  project UID/MAT/04106/2013. 
A. P. M. also acknowledges the THEP group at Lund University for all hospitality and significant support provided for the development 
of this work. J. E. C.-M., R. P. and J. W. acknowledge the warm hospitality of the Gr@v group at Aveiro University.
The work by J. E. C.-M. was supported by the Crafoord Foundation. R. P. and J. W. were partially supported by 
the Swedish Research Council, contract number 621-2013-428.

\appendix

\section{ RG equations for the LR-symmetric theory} \label{sec:RGEs}

In this appendix we list the one-loop $\beta$-functions for the LR symmetric theory described in Section \ref{sec:LR}. The convention we will follow is that for a given coupling $g$, the $\beta$function is defined as  
\begin{equation}
			 \beta_g = \frac{\d g}{\d t} 
\end{equation}
\noindent where $t=\mbox{log}(\mu)$ with $\mu$ the renormalisation scale.

   \subsection{Gauge couplings}

					\begin{align}
						(4\pi)^2\beta_{\g{R}} &= - \frac{2}{3} \g{R}^{3}\\\nonumber
						\\ 
						(4\pi)^2\beta_{\g{L}}  &= - \g{L}^{3}\\\nonumber
						\\ 
						(4\pi)^2\beta_{\g{\BL}} &= \frac{124}{9} \g{\BL}^{3}\\\nonumber
						\\ 
						(4\pi)^2\beta_{\g{C}}  &= - \frac{19}{3} \g{C}^{3}
					\end{align}

	        \subsection{Yukawa couplings }

	   	\begin{align}
	   	&\begin{aligned}
		(4\pi)^2\beta_{Y_{\alpha}} =
		&\left( \frac{7}{2}  |Y_{\alpha}|^2 + |Y_{\beta}|^2 + 3 |Y_{\gamma}|^2 + 2 |Y_{\delta}|^2 - \frac{9}{4} \g{R}^{2} - 3 \g{\BL}^{2} \right) Y_{\alpha}
	        \end{aligned}
		\\ \nonumber
		\\
		&\begin{aligned}
		(4\pi)^2\beta_{Y_{\beta}} =
		&\left( |Y_{\alpha}|^2 + 3  |Y_{\beta}|^2 + 3 |Y_{\gamma}|^2 + |Y_{\epsilon}|^2 - \frac{9}{4} \g{R}^{2} - 3 \g{\BL}^{2}       \right) Y_{\beta} 
		\end{aligned}		
		\\ \nonumber
		\\
	        &\begin{aligned}
		(4\pi)^2\beta_{Y_{\gamma}} =
		&\left( |Y_{\alpha}|^2 + |Y_{\beta}|^2 + \frac{11}{2}  |Y_{\gamma}|^2 +  |Y_{\zeta}|^2 - 8 \g{C}^{2} - \frac{9}{4} \g{R}^{2}   - \frac{5}{3} \g{\BL}^{2}      \right) Y_{\gamma}
		\end{aligned}  
	       	\\ \nonumber
	  	\\
		&\begin{aligned}
		(4\pi)^2\beta_{Y_{\delta}} =
		&\left( 2  |Y_{\alpha}|^2 +\frac{7}{2} |Y_{\delta}|^2 +|Y_{\epsilon}|^2  + 6 |Y_{\zeta}|^2 - \frac{9}{4} \g{L}^{2}   - \frac{9}{4} \g{R}^{2} \right) Y_{\delta} 
	       \end{aligned}
		\\ \nonumber
		\\
		&\begin{aligned}
		(4\pi)^2\beta_{Y_{\epsilon}} =
		&\left( |Y_{\beta}|^2 + |Y_{\delta}|^2 +3  |Y_{\epsilon}|^2 + 6 |Y_{\zeta}|^2 - \frac{9}{4} \g{L}^{2} - \frac{9}{4} \g{R}^{2} - 6 \g{\BL}^{2}    \right) Y_{\epsilon}
	        \end{aligned}		
	        \\ \nonumber
		\\
	        &\begin{aligned}
		(4\pi)^2\beta_{Y_{\zeta}} =
		&\left(\frac{1}{2}  |Y_{\gamma}|^2 +  |Y_{\delta}|^2 +|Y_{\epsilon}|^2 + 8 |Y_{\zeta}|^2 - 8 \g{C}^{2}  - \frac{9}{4} \g{L}^{2}  - \frac{9}{4} \g{R}^{2} - \frac{2}{3} \g{\BL}^{2} \right) Y_{\zeta}
		\end{aligned}
		\end{align}

	      \subsection{Scalar masses}

			\begin{align} \label{eq:betamuh}
					(4\pi)^2\beta_{m^2_h} = & \left(20 \lambda_{a} - 8 \lambda_{j} + 2 |Y_{\delta}|^2 + 2 |Y_{\epsilon}|^2 + 12 |Y_{\zeta}|^2 - \frac{9}{2} \g{L}^{2} - \frac{9}{2} \g{R}^{2} \right) m^2_h \\ \nonumber
				& + 4\left( \lambda_{g} +2 \lambda_{f} \right) m^2_R - 4 |Y_{\delta}|^2 m_{\Phi^s}^2 
			\end{align}

			\begin{align}\label{eq:betamuR}
					(4\pi)^2\beta_{m^2_R} = & \left( 20 \lambda_{b} - 8 \lambda_{i} + 2 |Y_{\alpha}|^2 +2 |Y_{\beta}|^2 + 6|Y_{\gamma}|^2 - \frac{9}{2} \g{R}^{2}- 6\g{\BL}^{2}  \right) m^2_R \\ \nonumber
				& + 4\left(\lambda_{g} +2 \lambda_{f} \right) m^2_h - 4  |Y_{\alpha}|^2 m_{\Phi^s}^2
			\end{align}
			
			\subsection{Singlet Fermion mass}

			\begin{align}
					(4\pi)^2\beta_{m_{\Phi^s}} =
				& \,\, 4 (|Y_{\alpha}|^2 + |Y_{\delta}|^2)  m_{\Phi^s}
			\end{align}

			\subsection{Quartic couplings}
		\begin{align}	
			&\begin{aligned}
					(4\pi)^2\beta_{\lambda_{a}} = & \,\, 32 \lambda_{a}^{2} + 4 \lambda_{f}^{2} + 2 \lambda_{g}^{2} + 4 \lambda_{f} \lambda_{g} + 16 \lambda_{j}^{2} - 16 \lambda_{a} \lambda_{j}   \\ 
					&  + 4 \left( |Y_{\epsilon}|^2 + |Y_{\delta}|^2 + 6 |Y_{\zeta}|^2 \right) \lambda_{a} 				\\ 
					& -2 \left( |Y_{\epsilon}|^4 +|Y_{\delta}|^4 +6 |Y_{\zeta}|^4	 \right)							\\
					& - 9 \left( \g{L}^{2}	+ \g{R}^{2}	\right) \lambda_{a}												\\ 
					& + \frac{9}{8}  \g{L}^{4} + \frac{3}{4} \g{L}^{2} \g{R}^{2} + \frac{9}{8} \g{R}^{4}  			
			\end{aligned}
			\\ \nonumber
			\\
			&\begin{aligned}
					(4\pi)^2\beta_{\lambda_{b}} = & \,\, 32 \lambda_{b}^{2} + 4 \lambda_{f}^{2} + 2 \lambda_{g}^{2} +4 \lambda_{f} \lambda_{g} + 16 \lambda_{i}^{2}  - 16 \lambda_{b} \lambda_{i}	\\ 
					& + 4 \left( |Y_{\alpha}|^2 + |Y_{\beta}|^2 + 3 |Y_{\gamma}|^2 \right) \lambda_{b}					\\ 
					& - 2 \left( |Y_{\alpha}|^4 + |Y_{\beta}|^4 + 3 |Y_{\gamma}|^4 \right)  											\\ 
					& - 3 \left( 3 \g{R}^{2} + 4  \g{\BL}^{2} \right) \lambda_{b}										\\ 
					& + \frac{9}{8} \g{R}^{4}  + 3  \g{R}^{2} \g{\BL}^{2} + 6 \g{\BL}^{4} 
			\end{aligned}
			\\ \nonumber
			\\
			&\begin{aligned}
					(4\pi)^2\beta_{\lambda_{f}} = & \,\, 4 \lambda_f^2 + 2 \lambda_g^2 + 4 \left(5 \lambda_f + 2 \lambda_g \right) \left( \lambda_a + \lambda_b \right) - 8 \left( \lambda_f + \lambda_g \right) \left( \lambda_i + \lambda_j \right) \\ 
					& + 2 \left( |Y_{\alpha}|^2 +|Y_{\beta}|^2 + 3 |Y_{\gamma}|^2 + |Y_{\delta}|^2 + |Y_{\epsilon}|^2 + 6 |Y_{\zeta}|^2 \right)	\lambda_{f} \\ 
					& - 4 \left( |Y_{\alpha}|^2 |Y_{\delta}|^2 + |Y_{\beta}|^2 |Y_{\epsilon}|^2  + 3 |Y_{\gamma}|^2 |Y_{\zeta}|^2 \right)  \\ 
					& - 3 \left( \frac{3}{2} \g{L}^{2} + 3 \g{R}^{2} + 2 \g{\BL}^{2} \right) \lambda_f + \frac{9}{4} \g{R}^{4} 
			\end{aligned}
			    	\end{align}
				\begin{align}
			&\begin{aligned}
				(4\pi)^2\beta_{\lambda_{g}}= & \,\, \lambda_{g}( 2 |Y_{\alpha}|^2+  2 |Y_{\beta}|^2 +  6 |Y_{\gamma}|^2+ 2 |Y_{\delta}|^2 + 2 |Y_{\epsilon}|^2 +12 |Y_{\zeta}|^2  )  \\ 
				& - \frac{9}{2} \lambda_{g} \g{L}^{2} + \frac{1}{2} |Y_{\alpha}|^2 |Y_{\delta}|^2 + \frac{5}{2}|Y_{\beta}|^2 |Y_{\epsilon}|^2+ \frac{15}{2} |Y_{\gamma}|^2 |Y_{\zeta}|^2\\ 
				&+8 \lambda_{g} \lambda_{i} + 8 \lambda_{g} \lambda_{j} +8 \lambda_{f} \lambda_{g} +4 \lambda_{b} \lambda_{g} +4 \lambda_{a} \lambda_{g}  \\ 
				&- 9 \lambda_{g} \g{R}^{2} - 6 \lambda_{g} \g{\BL}^{2} + 4 \lambda_{g}^{2}  
			\end{aligned}
			\\ \nonumber
			\\
			&\begin{aligned}
					(4\pi)^2\beta_{\lambda_{i}} = & - 16 \lambda_{i}^{2} + \lambda_{g}^{2} +24 \lambda_{b} \lambda_{i} 	\\ 
					& + 4 \left(|Y_{\alpha}|^2 +3|Y_{\beta}|^2+ 3 |Y_{\gamma}|^2 \right) \lambda_i		\\ 
					& - \frac{1}{8} \left( |Y_{\alpha}|^4 + 5 |Y_{\beta}|^4 +3 |Y_{\gamma}|^4 \right)			\\ 
					& - 3 \left(3 \g{R}^{2}  + 4 \g{\BL}^{2}\right) \lambda_i + 3 \g{\BL}^{2} \g{R}^{2}
			\end{aligned}
		     	\\ \nonumber
			\\
			&\begin{aligned}
					(4\pi)^2\beta_{\lambda_{j}} = & - 16 \lambda_{j}^{2} +\lambda_{g}^{2} +24 \lambda_{a} \lambda_{j} \\
					& + 4 \left( |Y_{\delta}|^2 +|Y_{\epsilon}|^2 + 6 |Y_{\zeta}|^2 \right) \lambda_{j}  \\ 
					& - \frac{5}{8} \left( |Y_{\delta}|^4 +|Y_{\epsilon}|^4 + 6 |Y_{\zeta}|^4 \right) 	\\ 
					& - 9 \left( \g{R}^{2} + \g{L}^{2} \right) \lambda_{j} - \frac{3}{2} \g{L}^{2} \g{R}^{2}
			\end{aligned}
		         \end{align}

\newpage

\bibliographystyle{JHEP}
\bibliography{bib}

\providecommand{\href}[2]{#2}\begingroup\raggedright\begin{thebibliography}{10}

\bibitem{Aad:2012tfa}
{\scshape ATLAS} collaboration, G.~Aad et~al., \emph{{Observation of a new
  particle in the search for the Standard Model Higgs boson with the ATLAS
  detector at the LHC}},
  \href{http://dx.doi.org/10.1016/j.physletb.2012.08.020}{\emph{Phys. Lett.}
  {\bf B716} (2012) 1--29}, [\href{http://arxiv.org/abs/1207.7214}{{\tt
  1207.7214}}].

\bibitem{Chatrchyan:2012xdj}
{\scshape CMS} collaboration, S.~Chatrchyan et~al., \emph{{Observation of a new
  boson at a mass of 125 GeV with the CMS experiment at the LHC}},
  \href{http://dx.doi.org/10.1016/j.physletb.2012.08.021}{\emph{Phys. Lett.}
  {\bf B716} (2012) 30--61}, [\href{http://arxiv.org/abs/1207.7235}{{\tt
  1207.7235}}].

\bibitem{original}
A.~De~R\'ujula, H.~Georgi and S.~L. Glashow{\emph{, in Fifth Workshop on Grand
  Unification edited by K. Kang, H. Fried, and F. Frampton} (1984) }.

\bibitem{Babu:1985gi}
K.~S. Babu, X.-G. He and S.~Pakvasa, \emph{Neutrino masses and proton decay
  modes in $\mathrm{SU}(3)\times\mathrm{SU}(3)\times\mathrm{SU}(3)$
  trinification}, \href{http://dx.doi.org/10.1103/PhysRevD.33.763}{\emph{Phys.
  Rev. D} {\bf 33} (Feb, 1986) 763--772}.

\bibitem{Lazarides:1993uw}
G.~Lazarides and C.~Panagiotakopoulos, \emph{{MSSM from SUSY Trinification}},
  \href{http://dx.doi.org/10.1016/0370-2693(94)00925-2}{\emph{Phys. Lett.} {\bf
  B336} (1994) 190--193}, [\href{http://arxiv.org/abs/hep-ph/9403317}{{\tt
  hep-ph/9403317}}].

\bibitem{Lazarides:1994px}
G.~Lazarides and C.~Panagiotakopoulos, \emph{{MSSM and large $\tan \beta$ from
  SUSY Trinification}},
  \href{http://dx.doi.org/10.1103/PhysRevD.51.2486}{\emph{Phys. Rev.} {\bf D51}
  (1995) 2486--2488}, [\href{http://arxiv.org/abs/hep-ph/9407286}{{\tt
  hep-ph/9407286}}].

\bibitem{Kim:2003cha}
J.~E. Kim, \emph{{$Z_3$ orbifold construction of $SU(3)^3$ GUT with $\sin^2
  \theta^0_W = \frac{3}{8}$}},
  \href{http://dx.doi.org/10.1016/S0370-2693(03)00567-7}{\emph{Phys. Lett.}
  {\bf B564} (2003) 35--41}, [\href{http://arxiv.org/abs/hep-th/0301177}{{\tt
  hep-th/0301177}}].

\bibitem{Willenbrock:2003ca}
S.~Willenbrock, \emph{{Triplicated Trinification}},
  \href{http://dx.doi.org/10.1016/S0370-2693(03)00419-2}{\emph{Phys. Lett.}
  {\bf B561} (2003) 130--134}, [\href{http://arxiv.org/abs/hep-ph/0302168}{{\tt
  hep-ph/0302168}}].

\bibitem{Carone:2004rp}
C.~D. Carone and J.~M. Conroy, \emph{{Higgsless GUT Breaking and
  Trinification}},
  \href{http://dx.doi.org/10.1103/PhysRevD.70.075013}{\emph{Phys. Rev.} {\bf
  D70} (2004) 075013}, [\href{http://arxiv.org/abs/hep-ph/0407116}{{\tt
  hep-ph/0407116}}].

\bibitem{Hetzel:2015cca}
J.~Hetzel, \emph{{Phenomenology of a left-right-symmetric model inspired by the
  trinification model}}.
\newblock PhD thesis, Inst. Appl. Math., Heidelberg, 2015.
\newblock \href{http://arxiv.org/abs/1504.06739}{{\tt 1504.06739}}.

\bibitem{Gursey:1975ki}
F.~Gursey, P.~Ramond and P.~Sikivie, \emph{{A universal gauge theory model
  based on $E_6$}},
  \href{http://dx.doi.org/10.1016/0370-2693(76)90417-2}{\emph{Phys. Lett.} {\bf
  B60} (1976) 177--180}.

\bibitem{Achiman:1978vg}
Y.~Achiman and B.~Stech, \emph{{Quark-Lepton Symmetry and Mass Scales in an E6
  Unified Gauge Model}},
  \href{http://dx.doi.org/10.1016/0370-2693(78)90584-1}{\emph{Phys. Lett.} {\bf
  B77} (1978) 389--393}.

\bibitem{Shafi:1978gg}
Q.~Shafi, \emph{{$E_6$ as a Unifying Gauge Symmetry}},
  \href{http://dx.doi.org/10.1016/0370-2693(78)90248-4}{\emph{Phys. Lett.} {\bf
  B79} (1978) 301--303}.

\bibitem{Barbieri:1981yy}
R.~Barbieri, D.~V. Nanopoulos and A.~Masiero, \emph{{Hierarchical Fermion
  Masses in E6}},
  \href{http://dx.doi.org/10.1016/0370-2693(81)90589-X}{\emph{Phys. Lett.} {\bf
  B104} (1981) 194--198}.

\bibitem{Stech:2003sb}
B.~Stech and Z.~Tavartkiladze, \emph{{Fermion masses and coupling unification
  in $E_6$: Life in the desert}},
  \href{http://dx.doi.org/10.1103/PhysRevD.70.035002}{\emph{Phys. Rev.} {\bf
  D70} (2004) 035002}, [\href{http://arxiv.org/abs/hep-ph/0311161}{{\tt
  hep-ph/0311161}}].

\bibitem{Stech:2010tv}
B.~Stech, \emph{{Neutrino Properties from $E_6 \times SO(3) \times Z_2$}},
  \href{http://dx.doi.org/10.1002/prop.201000034}{\emph{Fortsch. Phys.} {\bf
  58} (2010) 692--698}, [\href{http://arxiv.org/abs/1003.0581}{{\tt
  1003.0581}}].

\bibitem{King:2005my}
S.~F. King, S.~Moretti and R.~Nevzorov, \emph{{Exceptional Supersymmetric
  Standard Model}},
  \href{http://dx.doi.org/10.1016/j.physletb.2005.12.070}{\emph{Phys. Lett.}
  {\bf B634} (2006) 278--284}, [\href{http://arxiv.org/abs/hep-ph/0511256}{{\tt
  hep-ph/0511256}}].

\bibitem{King:2005jy}
S.~F. King, S.~Moretti and R.~Nevzorov, \emph{{Theory and Phenomenology of an
  Exceptional Supersymmetric Standard Model}},
  \href{http://dx.doi.org/10.1103/PhysRevD.73.035009}{\emph{Phys. Rev.} {\bf
  D73} (2006) 035009}, [\href{http://arxiv.org/abs/hep-ph/0510419}{{\tt
  hep-ph/0510419}}].

\bibitem{King:2007uj}
S.~F. King, S.~Moretti and R.~Nevzorov, \emph{{Gauge Coupling Unification in
  the Exceptional Supersymmetric Standard Model}},
  \href{http://dx.doi.org/10.1016/j.physletb.2007.04.061}{\emph{Phys. Lett.}
  {\bf B650} (2007) 57--64}, [\href{http://arxiv.org/abs/hep-ph/0701064}{{\tt
  hep-ph/0701064}}].

\bibitem{Braam:2010sy}
F.~Braam, A.~Knochel and J.~Reuter, \emph{{An Exceptional SSM from $E_6$
  Orbifold GUTs with intermediate LR symmetry}},
  \href{http://dx.doi.org/10.1007/JHEP06(2010)013}{\emph{JHEP} {\bf 06} (2010)
  013}, [\href{http://arxiv.org/abs/1001.4074}{{\tt 1001.4074}}].

\bibitem{Athron:2008np}
P.~Athron, S.~F. King, D.~J. Miller, S.~Moretti, R.~Nevzorov, S.~F. King
  et~al., \emph{{The Constrained E$_6$SSM}},
  \href{http://arxiv.org/abs/0810.0617}{{\tt 0810.0617}}.

\bibitem{King:2008qb}
S.~F. King, R.~Luo, D.~J. Miller and R.~Nevzorov, \emph{{Leptogenesis in the
  Exceptional Supersymmetric Standard Model: Flavour dependent lepton
  asymmetries}},
  \href{http://dx.doi.org/10.1088/1126-6708/2008/12/042}{\emph{JHEP} {\bf 12}
  (2008) 042}, [\href{http://arxiv.org/abs/0806.0330}{{\tt 0806.0330}}].

\bibitem{Athron:2009ue}
P.~Athron, S.~F. King, D.~J. Miller, S.~Moretti and R.~Nevzorov,
  \emph{{Predictions of the Constrained Exceptional Supersymmetric Standard
  Model}}, \href{http://dx.doi.org/10.1016/j.physletb.2009.10.051}{\emph{Phys.
  Lett.} {\bf B681} (2009) 448--456},
  [\href{http://arxiv.org/abs/0901.1192}{{\tt 0901.1192}}].

\bibitem{Athron:2009bs}
P.~Athron, S.~F. King, D.~J. Miller, S.~Moretti and R.~Nevzorov, \emph{{The
  Constrained Exceptional Supersymmetric Standard Model}},
  \href{http://dx.doi.org/10.1103/PhysRevD.80.035009}{\emph{Phys. Rev.} {\bf
  D80} (2009) 035009}, [\href{http://arxiv.org/abs/0904.2169}{{\tt
  0904.2169}}].

\bibitem{Athron:2011wu}
P.~Athron, S.~F. King, D.~J. Miller, S.~Moretti and R.~Nevzorov, \emph{{LHC
  Signatures of the Constrained Exceptional Supersymmetric Standard Model}},
  \href{http://dx.doi.org/10.1103/PhysRevD.84.055006}{\emph{Phys. Rev.} {\bf
  D84} (2011) 055006}, [\href{http://arxiv.org/abs/1102.4363}{{\tt
  1102.4363}}].

\bibitem{Hall:2010ix}
J.~P. Hall, S.~F. King, R.~Nevzorov, S.~Pakvasa, M.~Sher, R.~Nevzorov et~al.,
  \emph{{Novel Higgs Decays and Dark Matter in the $E_6$SSM}},
  \href{http://dx.doi.org/10.1103/PhysRevD.83.075013}{\emph{Phys. Rev.} {\bf
  D83} (2011) 075013}, [\href{http://arxiv.org/abs/1012.5114}{{\tt
  1012.5114}}].

\bibitem{Athron:2012sq}
P.~Athron, S.~F. King, D.~J. Miller, S.~Moretti and R.~Nevzorov,
  \emph{{Constrained Exceptional Supersymmetric Standard Model with a Higgs
  Near 125 GeV}},
  \href{http://dx.doi.org/10.1103/PhysRevD.86.095003}{\emph{Phys. Rev.} {\bf
  D86} (2012) 095003}, [\href{http://arxiv.org/abs/1206.5028}{{\tt
  1206.5028}}].

\bibitem{Nevzorov:2012hs}
R.~Nevzorov, \emph{{$E_6$ inspired supersymmetric models with exact custodial
  symmetry}}, \href{http://dx.doi.org/10.1103/PhysRevD.87.015029}{\emph{Phys.
  Rev.} {\bf D87} (2013) 015029}, [\href{http://arxiv.org/abs/1205.5967}{{\tt
  1205.5967}}].

\bibitem{Nevzorov:2013tta}
R.~Nevzorov and S.~Pakvasa, \emph{{Exotic Higgs decays in the $E_6$ inspired
  SUSY models}},
  \href{http://dx.doi.org/10.1016/j.physletb.2013.11.050}{\emph{Phys. Lett.}
  {\bf B728} (2014) 210--215}, [\href{http://arxiv.org/abs/1308.1021}{{\tt
  1308.1021}}].

\bibitem{Nevzorov:2013ixa}
R.~Nevzorov, \emph{{Quasifixed point scenarios and the Higgs mass in the $E_6$
  inspired supersymmetric models}},
  \href{http://dx.doi.org/10.1103/PhysRevD.89.055010}{\emph{Phys. Rev.} {\bf
  D89} (2014) 055010}, [\href{http://arxiv.org/abs/1309.4738}{{\tt
  1309.4738}}].

\bibitem{Nevzorov:2015sha}
R.~Nevzorov and A.~W. Thomas, \emph{{$E_6$ inspired composite Higgs model}},
  \href{http://dx.doi.org/10.1103/PhysRevD.92.075007}{\emph{Phys. Rev.} {\bf
  D92} (2015) 075007}, [\href{http://arxiv.org/abs/1507.02101}{{\tt
  1507.02101}}].

\bibitem{Athron:2015vxg}
P.~Athron, D.~Harries, R.~Nevzorov and A.~G. Williams, \emph{{$E_6$ Inspired
  SUSY Benchmarks, Dark Matter Relic Density and a 125 GeV Higgs}},
  \href{http://arxiv.org/abs/1512.07040}{{\tt 1512.07040}}.

\bibitem{King:2016wep}
S.~F. King and R.~Nevzorov, \emph{{750 GeV Diphoton Resonance from Singlets in
  an Exceptional Supersymmetric Standard Model}},
  \href{http://dx.doi.org/10.1007/JHEP03(2016)139}{\emph{JHEP} {\bf 03} (2016)
  139}, [\href{http://arxiv.org/abs/1601.07242}{{\tt 1601.07242}}].

\bibitem{Athron:2012pw}
P.~Athron, D.~Stockinger and A.~Voigt, \emph{{Threshold Corrections in the
  Exceptional Supersymmetric Standard Model}},
  \href{http://dx.doi.org/10.1103/PhysRevD.86.095012}{\emph{Phys. Rev.} {\bf
  D86} (2012) 095012}, [\href{http://arxiv.org/abs/1209.1470}{{\tt
  1209.1470}}].

\bibitem{Kawamura:2013rj}
Y.~Kawamura and T.~Miura, \emph{{Classification of Standard Model Particles in
  $E_6$ Orbifold Grand Unified Theories}},
  \href{http://dx.doi.org/10.1142/S0217751X13500553}{\emph{Int. J. Mod. Phys.}
  {\bf A28} (2013) 1350055}, [\href{http://arxiv.org/abs/1301.7469}{{\tt
  1301.7469}}].

\bibitem{Rizzo:2012rf}
T.~G. Rizzo, \emph{{Gauge Kinetic Mixing in the $E_6$SSM}}, {\emph{Phys.Rev.}
  {\bf D85} (2012) 055010}, [\href{http://arxiv.org/abs/1201.2898}{{\tt
  1201.2898}}].

\bibitem{Reuter:2010nx}
J.~Reuter and D.~Wiesler, \emph{{Distorted mass edges at LHC from
  supersymmetric leptoquarks}},
  \href{http://dx.doi.org/10.1103/PhysRevD.84.015012}{\emph{Phys. Rev.} {\bf
  D84} (2011) 015012}, [\href{http://arxiv.org/abs/1010.4215}{{\tt
  1010.4215}}].

\bibitem{Gross:1984dd}
D.~J. Gross, J.~A. Harvey, E.~J. Martinec and R.~Rohm, \emph{{The Heterotic
  String}}, \href{http://dx.doi.org/10.1103/PhysRevLett.54.502}{\emph{Phys.
  Rev. Lett.} {\bf 54} (1985) 502--505}.

\bibitem{Cremmer:1979uq}
E.~Cremmer, J.~Scherk and J.~H. Schwarz, \emph{{Spontaneously Broken N=8
  Supergravity}},
  \href{http://dx.doi.org/10.1016/0370-2693(79)90654-3}{\emph{Phys. Lett.} {\bf
  B84} (1979) 83--86}.

\bibitem{Ma:1986we}
E.~Ma, \emph{{Particle Dichotomy and Left-Right Decomposition of $E_6$
  Superstring Models}},
  \href{http://dx.doi.org/10.1103/PhysRevD.36.274}{\emph{Phys. Rev.} {\bf D36}
  (1987) 274}.

\bibitem{Ma:1995xk}
E.~Ma, \emph{{Neutrino masses in an extended gauge model with $E_6$ particle
  content}}, \href{http://dx.doi.org/10.1016/0370-2693(96)00524-2}{\emph{Phys.
  Lett.} {\bf B380} (1996) 286--290},
  [\href{http://arxiv.org/abs/hep-ph/9507348}{{\tt hep-ph/9507348}}].

\bibitem{Kim:2004pe}
J.~E. Kim, \emph{{Trinification with $\sin^2 \theta_W = \frac{3}{8}$ and seesaw
  neutrino mass}},
  \href{http://dx.doi.org/10.1016/j.physletb.2004.04.017}{\emph{Phys. Lett.}
  {\bf B591} (2004) 119--126}, [\href{http://arxiv.org/abs/hep-ph/0403196}{{\tt
  hep-ph/0403196}}].

\bibitem{Cauet:2010ng}
C.~Cauet, H.~Pas, S.~Wiesenfeldt, C.~Cauet, H.~Pas and S.~Wiesenfeldt,
  \emph{{Trinification, the Hierarchy Problem and Inverse Seesaw Neutrino
  Masses}}, \href{http://dx.doi.org/10.1103/PhysRevD.83.093008}{\emph{Phys.
  Rev.} {\bf D83} (2011) 093008}, [\href{http://arxiv.org/abs/1012.4083}{{\tt
  1012.4083}}].

\bibitem{Sayre:2006ma}
J.~Sayre, S.~Wiesenfeldt and S.~Willenbrock, \emph{{Minimal trinification}},
  \href{http://dx.doi.org/10.1103/PhysRevD.73.035013}{\emph{Phys. Rev.} {\bf
  D73} (2006) 035013}, [\href{http://arxiv.org/abs/hep-ph/0601040}{{\tt
  hep-ph/0601040}}].

\bibitem{Huang:2001bm}
C.-S. Huang, J.~Jiang, T.-j. Li and W.~Liao, \emph{{N=2 six-dimensional
  supersymmetric $E_6$ breaking}},
  \href{http://dx.doi.org/10.1016/S0370-2693(02)01335-7}{\emph{Phys. Lett.}
  {\bf B530} (2002) 218--226}, [\href{http://arxiv.org/abs/hep-th/0112046}{{\tt
  hep-th/0112046}}].

\bibitem{Georgi:1979md}
H.~Georgi, \emph{{Towards a Grand Unified Theory of Flavor}},
  \href{http://dx.doi.org/10.1016/0550-3213(79)90497-8}{\emph{Nucl. Phys.} {\bf
  B156} (1979) 126--134}.

\bibitem{Stech:2014tla}
B.~Stech, \emph{{Trinification Phenomenology and the structure of Higgs
  Bosons}}, \href{http://dx.doi.org/10.1007/JHEP08(2014)139}{\emph{JHEP} {\bf
  08} (2014) 139}, [\href{http://arxiv.org/abs/1403.2714}{{\tt 1403.2714}}].

\bibitem{Hetzel:2015bla}
J.~Hetzel and B.~Stech, \emph{{Low-energy phenomenology of trinification: an
  effective left-right-symmetric model}},
  \href{http://dx.doi.org/10.1103/PhysRevD.91.055026}{\emph{Phys. Rev.} {\bf
  D91} (2015) 055026}, [\href{http://arxiv.org/abs/1502.00919}{{\tt
  1502.00919}}].

\bibitem{lee2008hom4ps}
T.~Lee, T.~Li and C.~Tsai, \emph{Hom4{PS}-2.0: a software package for solving
  polynomial systems by the polyhedral homotopy continuation method},
  {\emph{Computing} {\bf 83} (2008) 109--133}.

\bibitem{Burgess:1998kh}
C.~P. Burgess, \emph{{A Goldstone boson primer}},  in \emph{{11th Summer School
  and Symposium on Nuclear Physics (NuSS 98): Effective Theories of Matter (1)
  Seoul, Korea, June 23-27, 1998}}, 1998.
\newblock \href{http://arxiv.org/abs/hep-ph/9812468}{{\tt hep-ph/9812468}}.

\bibitem{Burgess:1998ku}
C.~P. Burgess, \emph{{Goldstone and Pseudo-Goldstone Bosons in Nuclear,
  Particle and Condensed-Matter Physics}},
  \href{http://dx.doi.org/10.1016/S0370-1573(99)00111-8}{\emph{Phys. Rept.}
  {\bf 330} (2000) 193--261}, [\href{http://arxiv.org/abs/hep-th/9808176}{{\tt
  hep-th/9808176}}].

\bibitem{Dedes:1995sb}
A.~Dedes, A.~B. Lahanas and K.~Tamvakis, \emph{{Radiative electroweak symmetry
  breaking in the MSSM and low-energy threshold}},
  \href{http://dx.doi.org/10.1103/PhysRevD.53.3793}{\emph{Phys. Rev.} {\bf D53}
  (1996) 3793--3807}, [\href{http://arxiv.org/abs/hep-ph/9504239}{{\tt
  hep-ph/9504239}}].

\bibitem{Gamberini:1989jw}
G.~Gamberini, G.~Ridolfi and F.~Zwirner, \emph{{On Radiative Gauge Symmetry
  Breaking in the Minimal Supersymmetric Model}},
  \href{http://dx.doi.org/10.1016/0550-3213(90)90211-U}{\emph{Nucl.Phys.} {\bf
  B331} (1990) 331--349}.

\bibitem{Carena:1993ag}
M.~Carena, S.~Pokorski and C.~E.~M. Wagner, \emph{{On the unification of
  couplings in the minimal supersymmetric Standard Model}},
  \href{http://dx.doi.org/10.1016/0550-3213(93)90161-H}{\emph{Nucl. Phys.} {\bf
  B406} (1993) 59--89}, [\href{http://arxiv.org/abs/hep-ph/9303202}{{\tt
  hep-ph/9303202}}].

\bibitem{wprimeresults}
{\scshape ATLAS} collaboration, M.~Aaboud et~al., \emph{{Search for new
  resonances in events with one lepton and missing transverse momentum in $pp$
  collisions at $\sqrt{s} = 13$ TeV with the ATLAS detector}},
  \href{http://arxiv.org/abs/1606.03977}{{\tt 1606.03977}}.

\bibitem{zprimeresults}
{\scshape ATLAS} collaboration, \emph{{Search for new phenomena in the dilepton
  final state using proton-proton collisions at $\sqrt{s} = 13$ TeV with the
  ATLAS detector}}, .

\bibitem{Lyonnet:2013dna}
F.~Lyonnet, I.~Schienbein, F.~Staub and A.~Wingerter, \emph{{PyR@TE:
  Renormalization Group Equations for General Gauge Theories}},
  \href{http://dx.doi.org/10.1016/j.cpc.2013.12.002}{\emph{Comput. Phys.
  Commun.} {\bf 185} (2014) 1130--1152},
  [\href{http://arxiv.org/abs/1309.7030}{{\tt 1309.7030}}].

\bibitem{Kirkpatrick671}
S.~Kirkpatrick, C.~D. Gelatt and M.~P. Vecchi, \emph{Optimization by simulated
  annealing},
  \href{http://dx.doi.org/10.1126/science.220.4598.671}{\emph{Science} {\bf
  220} (1983) 671--680},
  [\href{http://arxiv.org/abs/http://science.sciencemag.org/content/220/4598/671.full.pdf}{{\tt
  http://science.sciencemag.org/content/220/4598/671.full.pdf}}].

\end{thebibliography}\endgroup

\end{document}